\documentclass[preprint,prd,showpacs,showkeys,byrevtex,footinbib,eqsecnum,unsortedaddress]{revtex4}
\usepackage{amsfonts}
\usepackage{amsmath}
\usepackage{amssymb}

\begin{document}

\title{Infrared Behavior of High-Temperature QCD}
\author{A. Abada}
\email{a.abada@uaeu.ac.ae}
\affiliation{Physics Department, United Arab Emirates University, P.O.B. 17551, Al Ain,
United Arab Emirates}
\altaffiliation{On leave from: D\'{e}partement de Physique, ENS, BP 92 Vieux Kouba, 16050
Alger, Algeria.}
\author{K. Bouakaz}
\email{bouakazk@caramail.com}
\affiliation{D\'{e}partement de Physique, Ecole Normale Sup\'{e}rieure, BP 92 Vieux
Kouba, 16050 Alger, Algeria}
\author{O. Azi}
\email{Azi@mpikg-golm.mpg.de}
\affiliation{Max Planck Institut F\"{u}r Kolloid und Grenzfl\"{a}chenforschung
(MPIKG-Golm), Abteilung Theorie, D-14424 Postdam, Germany}

\begin{abstract}
The damping rate $\gamma _{t}\left( p\right)$ of on-shell transverse gluons
with ultrasoft momentum $p$ is calculated in the context of
next-to-leading-order hard-thermal-loop-summed perturbation of
high-temperature QCD. It is obtained in an
expansion to second order in $p$. The first coefficient is recovered but that of
order $p^{2}$ is found divergent in the infrared. Divergences from
light-like momenta do also occur but are circumvented. Our result and method
are critically discussed, particularly regarding a Ward identity obtained in
the literature. When enforcing the equality between $\gamma _{t}\left(
0\right)$ and $\gamma _{l}\left( 0\right)$, a rough estimate of the
magnetic mass is obtained. Carrying a similar calculation in the context of
scalar quantum electrodynamics shows that the early ultrasoft-momentum
expansion we make has little to do with the infrared sensitivity of the
result.
\end{abstract}

\date{\today}
\maketitle

\section{Introduction}

At high temperature $T$, the properties of the quark-gluon plasma are
described in the context of massless QCD with a small (running) coupling constant $g$,
which implies there is a hierarchy of scales $g^{n}T$, $n$ a nonnegative
integer \cite{aurenche-gelis-kobes-petitgirard1}. The quasiparticle spectra
have been determined to lowest order in the coupling
\cite{kalash-klimo,klimov1,klimov2,weldon1,weldon2}. There are transverse gluons
and longitudinal ones (plasmons), ordinary quarks and plasminos
\cite{braaten}. In \cite{kalash-klimo} was recognized the problem of gauge
dependence of naive one-loop-order dispersion relations for slow-moving
particles. A systematic reorganization of the perturbation became necessary
and a proposal was made in \cite{BP1,BP2,BP3,pisarski1,pisarski2,
pisarski3,frenkel-taylor}. It consists in summing the so-called hard thermal
loops (HTL) into dressed propagators and vertices \cite{le bellac}.

Hard thermal loops form a lowest-order approximation of massless QCD at
high temperature describing slow-moving quasiparticles. They have been
encapsulated in a generating-functional formalism
\cite{action1,action2,action3,action4,action5} and once a connection to the
eikonal of a Chern-Simons gauge theory was made
\cite{chern-simons1,chern-simons2,chern-simons3}, a hydrodynamic approach
\cite{hydro1,hydro2,hydro3} showed that they are essentially classical. This
suggests that `true' quantum effects would effectively arise at
next-to-leading order in HTL-summed perturbation, and one question to ask is
whether this perturbation is consistent and reliable. More particularly, is
a next-order term really a correction or of the same magnitude as the
one considered? Also, does the HTL-summed perturbation take account of and
describe all possible effects of the theory? Indeed, note that the
functionals in the literature \cite{hydro1,hydro2,hydro3} generate
\textit{only} the hard thermal loops and \textit{not} the \textit{full} quantum
field theory organized in a HTL-summed perturbation. Finally, are actual
calculations safe from potential infinities?

With this in perspective, many quantities have been examined in
next-to-leading order calculations. In general, finite and consistent
results are obtained when fast-moving particles are involved
\cite{baier-nakkagawa-niegawa-redlich,pisarski1,lebedev-smilga1,
lebedev-smilga2,burgess-marini,rebhan1,braaten-thoma,thoma-gyulassy,
altherr-petitgirard-del rio gaztelurrutia,baier-kobes},
but difficulties arise when slow-moving quasiparticles are dealt with. For
example, the determination of the production rate of soft real photons from
a hot quark-gluon plasma \cite{baier-peigne-schiff,aurenche-becherrawy-petitgirard}
has revealed that the hard thermal loops introduce non-cancellable collinear
divergences when external momenta are light-like. It is shown in another approach
\cite{aurenche-gelis-kobes-petitgirard1} that it is not even the hard thermal
loops that dominate at leading order. The problem of collinear divergences
is looked into in \cite{flechsig-rebhan} and an improved gauge-invariant
effective action is proposed that takes into account asymptotic thermal
masses. Further work \cite{aurenche-gelis-zaraket,aurenche-gelis-kobes-zaraket,
aurenche-gelis-kobes-petitgirard1,aurenche-gelis-kobes-petitgirard2}
on the production of photons from a hot quark-gluon plasma using this
improved action shows that (i) two and three-loop graphs are of the same
magnitude as the lowest order term; (ii) though infrared divergences
associated with ultrasoft transverse gluons cancel in graphs with what is
termed `abelian topography', due to a sort of a finite-temperature
Kinoshita-Lee-Nauenberg theorem, the result depends strongly on the introduced magnetic mass.

Regarding the infrared sector more specifically, there are strong
indications that here too, problems may arise when pushing the HTL-summed
perturbation to next-to-leading order. Indeed, it is well known that at
lowest order $gT$, static chromoelectric fields get screened but not the
static chromomagnetic ones. These are believed to get so in nonabelian gauge
theories at next-to-leading order $g^{2}T$\cite{linde1,linde2,gross-pisarski-yaffe},
the so-called magnetic scale. It is also believed that the determination of
the chromomagnetic screening length, the inverse of the magnetic mass $\mu$,
is nonperturbative \cite{braaten-nieto1,rebhan3}. Therefore, it is quite possible that non-screened
static chromomagnetic fields spoil some, if not all, next-to-leading-order
calculations. For example, the works \cite{rebhan2,braaten-nieto1,rebhan3}
discuss next-to-leading-order nonabelian Debye screening in HTL-summed
perturbation and it is explicitly shown that the results depend strongly on
the type of infrared regularization used.

There also seems to be difficulties in the infrared when dealing with
next-to-leading-order dynamic quantities. The first non-trivial such
quantities to investigate are the slow-moving quasiparticle damping rates.
Indeed, though they come from one-loop calculations with fully dressed
propagators and vertices, they only involve the imaginary part of the
one-loop dressed self-energies, something that simplifies matters
considerably. The first work to cite is \cite{BPgamt} which determines the
damping rate $\gamma _{t}(0)$ for non-moving transverse gluons and finds the
finite and positive result: 
\begin{equation}
\gamma _{t}(0)=0.088N_{c}\,g^{2}T, \label{gamt-0}
\end{equation}
where $N_{c}$ is the number of colors. Another result is that of the damping
rates $\gamma _{\pm }(0)$ for non-moving quarks found independently in
\cite{kobes-kunstatter-mak} and \cite{braaten-pisarski (quarks)} which read: 
\begin{equation}
\gamma _{\pm }(0)=a_{0}C_{f}\,g^{2}T, \label{gamq-0}
\end{equation}
where $C_{f}=\left( N_{c}^{2}-1\right) /2N_{c}$ and $a_{0}$ is a finite
positive numerical constant that depends on $N_{c}$ and the number of
flavors $N_{f}$. However, in contrast with these finite results, the damping
rates for on-shell slow-moving quasiparticles with momentum $\mathbf{p}$
estimated in \cite{pisarski4} indicate the presence of a logarithmic
divergence in $g$: 
\begin{equation}
\gamma (p)\sim -ag^{2}Tv(p)\ln g, \label{gam glu qua-nonzero p-pisarski}
\end{equation}
where $a$ is a positive numerical constant, $v(p)$ the quasiparticle group
velocity and $p=\left\vert \mathbf{p}\right\vert$. Note that if one takes
the limit $p\rightarrow 0$, one cannot recover neither the finite result
(\ref{gamt-0}) nor (\ref{gamq-0}). Another estimate is done in
\cite{flechsig-rebhan-schulz} and a result similar to (\ref{gam glu qua-nonzero
p-pisarski}) is found: 
\begin{equation}
\gamma (p)\sim -\frac{g^{2}N_{c}T}{4\pi }v(p)\ln g.
\label{gam soft p-flechsig-rebhan-schulz}
\end{equation}

The infrared problem in the damping rates is more emphasized in \cite{AAB,AA},
where a direct and explicit calculation of the damping rate $\gamma_{l}(0)$
for non-moving longitudinal gluons in the sole framework of the
HTL-summed perturbation shows that this latter is infrared sensitive: 
\begin{equation}
\gamma _{l}(0)=\left( \frac{250}{27\pi ^{2}\bar{\mu}^{2}}-2.471908492\right) 
\frac{g^{2}N_{c}T}{24\pi }, \label{gaml-0}
\end{equation}
where $\bar{\mu}$ is the magnetic mass $\mu $, taken as an infrared
regulator, in units of the electric thermal mass. This peculiar result
contrasts with the expectation that at zero momentum, there must be no
difference between longitudinal and transverse gluons, and so the two
corresponding damping rates ought to be equal \cite{BPgamt}, a statement
emphasized in \cite{dirks-niegawa-okano}. Also, another direct and explicit
calculation \cite{ABD1} of the damping rates $\gamma _{\pm }(p)$ for
ultrasoft \cite{aurenche-gelis-zaraket--reinbach-schulz} quarks suggests
that these too bear infrared divergences.

The purpose of the present article is to determine the damping rate $\gamma
_{t}(p)$ for a transverse gluon with an ultrasoft momentum $\mathbf{p}$. The
aim behind is an additional attempt to understand the infrared behavior of
massless Quantum Chromodynamics at high temperature. The calculation is done
in the strict context of next-to-leading order HTL-summed perturbation
theory as developed in \cite{BP1,BP2,BP3,pisarski1,
pisarski2,pisarski3,frenkel-taylor}. In the back of
our mind is the persistent question whether this scheme is reliable and
efficient enough when it comes to describe the properties of the quark-gluon
plasma, most importantly beyond lowest order. We will seek an expression for 
$\gamma _{t}(p)$ in an expansion in powers of the ultrasoft external
momentum $p$. We recover the first, finite and positive, coefficient
(\ref{gamt-0}) found in \cite{BPgamt}. All coefficients of odd powers of $p$
vanish. We determine explicitly the coefficient of $p^{2}$ and find it
sensitive in the infrared, see (\ref{gamma_t--final}). This finding relies
on the way the calculation is performed and much of the discussion we carry
orbits around this point. This may be felt slightly unfortunate a situation,
but it should be seen as an additional reflection of the degree of
difficulty to carry analytic estimates of physical quantities beyond lowest
order directly in QCD.

In the next section, we give the definition of the damping rate in the
context of HTL-summed perturbation. The problem amounts to calculating the
imaginary part of the transverse-gluon one-loop self-energy with fully
HTL-dressed propagators and vertices. We carry this calculation in section
three. The analytic results of this section have already been reported in 
\cite{AAT} and so, this part of the work is not new; we merely give here
substantially more details regarding the intermediary steps. The evaluation
of the analytic integrals is carried out in section four. It necessitates
the disentanglement of the infrared-sensitive pieces from the finite ones,
quite delicate a task. We choose the magnetic mass $\mu$ as the infrared
regulator. It also requires the handling of potential divergences coming
from light-like momenta. We explicitly show that these latter can be
circumvented without having recourse to the improved HTL-summation of
\cite{flechsig-rebhan}, whereas the infrared divergences persist. All of this
section is new.

Section five is devoted to a detailed critical discussion of the result we
obtain and the method we use. First we explain why it is admissible to
regulate the infrared sector using the magnetic mass $\mu$, and assuming
that all of massless QCD is in the HTL summation scheme, we impose the
equality between $\gamma _{l}(0)$ and $\gamma _{t}(0)$ to obtain a rough
estimate of $\mu$. Various orders of magnitude are discussed and we come to
the conclusion that the HTL-summed perturbation may be useful well into the
quark-gluon-plasma phase. Next we address the issue of the equality that we
do not obtain in \cite{AA} between the damping rates for the longitudinal
and transverse non-moving gluons, in view of the work \cite{dirks-niegawa-okano}
in which a Slavnov-Taylor identity for the gluon
polarization tensor in Coulomb gauge is derived and, when formally applied
to the next-to-leading order self-energy, equality is found. We show that
this identity, though important in itself, is not of direct help to us. We
then compare our result with other estimates in the literature,
\cite{pisarski4,flechsig-rebhan-schulz}, and argue that there is no contradiction
between what we find and these works: the regions of the external momentum
$p$ are not the same. To indicate that our method of calculation has little
to do with the occurrence of infrared sensitive coefficients in ultrasoft
physical quantities, we carry a similar calculation in the context of scalar
quantum electrodynamics where it is possible to avoid the early expansion in
powers of the external ultrasoft momentum. We show that results with early
expansion and without are the same. We conclude the discussion with few
final comments.

One appendix contains the small-momentum expansion of the gluon on-shell
energies, the residue and cut functions involved in the spectral densities
of the dressed propagators, together with their first and second
derivatives. We should mention that only the first few terms of the
expansions are displayed; in actual use, many additional terms are needed. A
second appendix gives the results of the terms contributing to the imaginary
part of the transverse-gluon one-loop HTL-dressed self-energy we have not
detailed in the main text.

\section{Gluon damping rate in HTL-summed perturbation}

The calculation is performed in the imaginary-time formalism where the
euclidean momentum of the gluon is $P^{\mu }=(p_{0},\mathbf{p})$ with
$P^{2}=p_{0}^{2}+p^{2}$ and the bosonic Matsubara frequency$\ p_{0}=2\pi nT$, 
$n$ an integer. In this formalism, real-time quantities are obtained via the
analytic continuation $p_{0}=-i\omega+0^{+}$, where $\omega$ is the energy
of the gluon. A momentum is said to be soft if both $\omega$ and $p$ are
of order $gT$; it is said to be hard if one is or both are of order $T$. The
three-momentum $\mathbf{p}$ of the on-shell gluon is said to be
\textit{ultrasoft} if $p$ is much smaller than $gT$, say of the order $g^{2}T$ and
smaller. The calculation is carried in the strict Coulomb gauge $\xi _{C}=0$
where the separation between the longitudinal and transverse components is
straightforward. The HTL results we quote in this section can all be found
in \cite{BP1,BP2,BP3,pisarski1,pisarski2,pisarski3,frenkel-taylor,le bellac}.

In the strict Coulomb gauge, the dressed gluon propagator $^{\ast }\Delta
_{\mu \nu }\left( P\right)$ has a simple structure. It is given by $^{\ast
}\Delta _{00}\left( P\right) =\,^{\ast }\Delta _{l}\left( P\right) $,
$^{\ast }\Delta _{0i}\left( P\right) =0$ and $^{\ast }\Delta _{ij}\left(
P\right) =\left( \delta _{ij}-\hat{p}_{i}\hat{p}_{j}\right) \,^{\ast }\Delta
_{t}\left( P\right) $ with: 
\begin{equation}
^{\ast }\Delta _{l}(P)=\frac{1}{p^{2}-\delta \Pi _{l}(P)};\qquad ^{\ast
}\Delta _{t}(P)=\frac{1}{P^{2}-\delta \Pi _{t}(P)}\,,
\label{gluon propagator}
\end{equation}
where $\delta \Pi _{l}(P)$ and $\delta \Pi _{t}(P)$ are hard thermal loops
given by: 
\begin{equation}
\delta \Pi _{l}(P)=3m_{g}^{2}Q_{1}\left( \frac{ip_{0}}{p}\right) ;\qquad
\delta \Pi _{t}(P)=\frac{3}{5}m_{g}^{2}\left[ Q_{3}\left( \frac{ip_{0}}{p}
\right) -Q_{1}\left( \frac{ip_{0}}{p}\right) -\frac{5}{3}\right] .
\label{delta pi_l-t}
\end{equation}
The $Q_{n}$ are Legendre functions of the second kind and $m_{g}=\sqrt{N_{c}
+N_{f}/2}\,gT/3$ to lowest order is the gluon thermal mass. The poles
of $^{\ast }\Delta _{t(l)}(-i\omega ,\mathbf{p})$ determine the transverse
(longitudinal) gluon dispersion relation. These write: 
\begin{equation}
\ln \frac{\omega _{t}+k}{\omega _{t}-k}=\frac{2k\left( -2k^{2}\omega
_{t}^{2}+2k^{4}+3m_{g}^{2}\omega _{t}^{2}\right) }{3m_{g}^{2}\,\omega
_{t}\left( \omega _{t}^{2}-k^{2}\right) };\qquad \ln \frac{\omega _{l}+k}
{\omega _{l}-k}=\frac{2k\left( 3m_{g}^{2}+k^{2}\right) }{3m_{g}^{2}\,\omega
_{l}}. \label{dispersion relations}
\end{equation}
At this lowest order $gT$ in the dispersion relations, the on-shell gluon
energies are real and no damping occurs. To get the damping rates to their
lowest order $g^{2}T$ in HTL-summed perturbation, one has to include in the
dispersion relations the contribution from the next-to-leading-order
self-energy $^{\ast }\Pi $ which has a more complicated structure than that
of the corresponding hard thermal loop. It satisfies the less restrictive
identity $P_{\mu }P_{\nu }\,^{\ast }\Pi _{\mu \nu }(P)=0$, which means
$^{\ast }\Pi $ depends on three scalar functions. The advantage of the strict
Coulomb gauge is that only two of these are relevant, namely: 
\begin{equation}
^{\ast }\Pi _{l}(P)\equiv \,^{\ast }\Pi _{00}(P);\qquad ^{\ast }\Pi
_{t}(P)\equiv \frac{1}{2}\left( \delta _{ij}-\hat{p}_{i}\hat{p}_{j}\right)
\,^{\ast }\Pi _{ij}(P). \label{pi star_l,t}
\end{equation}
The transverse-gluon dispersion relation reads: 
\begin{equation}
-\Omega _{t}^{2}+p^{2}-\delta \Pi _{t}(-i\Omega _{t},p)-\,^{\ast }\Pi
_{t}(-i\Omega _{t},p)=0. \label{transverse dispersion relation}
\end{equation}
The transverse-gluon damping rate is defined by $\gamma _{t}(p)=
-\mathrm{Im}\Omega _{t}\left( p\right) $. It is $g$-times smaller than
$\omega _{t}(p)$, and so we have to lowest order: 
\begin{equation}
\gamma _{t}(p)=\left. \frac{\mathrm{Im}\,^{\ast }\Pi _{t}(-i\omega ,p)}
{2\omega +\partial _{\omega }\delta \Pi _{t}(-i\omega ,p)}\right\vert
_{\omega =\omega _{t}(p)+i0^{+}}, \label{definition1 gamma_t}
\end{equation}
where $\partial _{\omega }$ stands for $\partial /\partial \omega $. Using
the expression of $\delta \Pi _{t}$ in (\ref{delta pi_l-t}), we have for
soft momenta: 
\begin{equation}
\gamma _{t}(p)=\left. \frac{1}{2}\left[ 1-\frac{1}{10}\left( \frac{p}{m_{g}}
\right) ^{2}+\dots\right] \, \mathrm{Im}\,^{\ast }\Pi _{t}(-i\omega
,p)\right\vert _{\omega =\omega _{t}(p)+i0^{+}}. \label{definition2 gamma_t}
\end{equation}
Hence, determining $\gamma _{t}(p)$ amounts to calculating the imaginary
part of the next-to-leading-order transverse-gluon self-energy.

In HTL-summed perturbation, the next-to-leading-order gluon self-energy in
the strict Coulomb gauge is given by \cite{BPgamt}: 
\begin{align}
^{\ast }\Pi ^{\mu \nu }(P)& =-\frac{g^{2}N_{c}}{2}\mathrm{Tr}_{\mathrm{soft}}
\left[ \,^{\ast }\Gamma ^{\mu \nu \lambda \sigma }(P,-P,K,-K)\,^{\ast
}\Delta _{\lambda \sigma }(K)\right. \notag \\
& +\left. \,^{\ast }\Gamma ^{\sigma \mu \lambda }(-Q,P,-K)\,^{\ast }\Delta
_{\lambda \lambda ^{\prime }}(K)\,^{\ast }\Gamma ^{\lambda ^{\prime }\nu
\sigma ^{\prime }}(-K,P,-Q)\,^{\ast }\Delta _{\sigma ^{\prime }\sigma }(Q) 
\right] . \label{definition pi star}
\end{align}
$K$ is the loop momentum, $Q=P-K$ and $\mathrm{Tr}\equiv T\underset{k_{0}}
{\sum }\int \dfrac{d^{3}k}{(2\pi )^{3}}$ with $k_{0}=2n\pi T$. The subscript
`soft' means that only soft values of $K$ are allowed in the integrals,
which implies that both propagators and vertices must be dressed; this is
already indicated in (\ref{definition pi star}). The dressed vertices 
$^{\ast }\Gamma $ are of the form: 
\begin{equation}
^{\ast }\Gamma ^{(n)}=\Gamma ^{(n)}+\delta \Gamma ^{(n)}\,;\qquad n=3,4,
\label{dressed vertex}
\end{equation}
where $\Gamma ^{3(4)}$ is the tree three (four)-gluon vertex and $\delta
\Gamma ^{3(4)}$ is the corresponding hard thermal loop. We have: 
\begin{align}
\delta \Gamma ^{\mu \nu \lambda }(-Q,P,-K)& =3m_{g}^{2}\int \frac{d\Omega
_{s}}{4\pi }\frac{S^{\mu }S^{\nu }S^{\lambda }}{PS\,}\left( \frac{iq_{0}}{QS}
-\frac{ik_{0}}{KS}\right) ;  \label{3 gluon HTL vertex} \\
\delta \Gamma ^{\mu \nu \lambda \sigma }(P,-P,K,-K)& =3m_{g}^{2}\int \frac{
d\Omega _{s}}{4\pi }\frac{S^{\mu }S^{\nu }S^{\lambda }S^{\sigma }}{PS\,KS}
\left( \frac{ip_{0}-ik_{0}}{PS-KS}-\frac{ip_{0}+ik_{0}}{PS+KS}\right) ,
\label{4 gluon HTL vertex}
\end{align}
with $S\equiv (i,\mathbf{\hat{s}})$ and $\Omega _{s}$ the solid angle of
$\mathbf{\hat{s}}$.

To get an expression for the imaginary part of $^{\ast }\Pi _{t}$, the
next-to-leading-order transverse-gluon self-energy, we have first to perform
the above angular integrals, then perform the Matsubara sum, then
analytically continue to real gluon energies $p_{0}=-i\omega +0^{+}$ and
implement the on-shell condition. Last is to find a way to perform the
integration over the gluon loop three-momentum $\mathbf{k}$. However, it is
technically difficult to follow the above sequence of operations. Especially
the integration over $\Omega _{s}$ in (\ref{3 gluon HTL vertex}) and (\ref{4
gluon HTL vertex}). To render this latter feasible, we expand $^{\ast }\Pi $
in powers of the external gluon momentum $p/m_{g}$, a small parameter when
$p$ is ultrasoft. The Matsubara sum is performed using the spectral
representations of the different quantities involved. The subsequent
integration over $\Omega _{k}$ is analytically done and the remaining
integrals have to be performed numerically, once infrared-sensitive pieces,
if any, are extracted. The rationale behind this procedure is as follows.
Expecting infrared sensitivity, we introduce from the outset an infrared
regulator which remains fixed for the rest of the calculation. It can be
viewed as physically representing the magnetic scale $g^{2}T$, or better the
magnetic mass $\mu$. The integration $\int_{0}^{+\infty }dk$ is therefore
replaced by $\int_{\mu }^{+\infty }dk$. This means that the loop momentum $k$
is always greater than $\mu $. As we said, we always regard in this work $p$
as ultrasoft, i.e., $p<\mu \leq k$. This condition allows the expansion of
quantities function of $q=\left\vert \mathbf{p}-\mathbf{k}\right\vert $ in
powers of $p$, quantities like $1/QS$ and $^{\ast }\Delta (Q)$. We have
already discussed the range of validity of this expansion in \cite{ABD1} and
will come back to this issue later in section five.

\section{one-loop HTL-summed gluon self-energy}

From now on, we set $m_{g}=1$. This will simplify writing the intermediary
steps as well as the expressions we obtain. From (\ref{definition pi star})
and the structure of the gluon propagator $^{\ast }\Delta _{\mu \nu }$ given
in the text before (\ref{gluon propagator}), we have the following explicit
expression for $^{\ast }\Pi _{t}$: 
\begin{align}
^{\ast }\Pi _{t}(P)& =-\frac{g^{2}N_{c}}{4}\left( \delta _{ij}-\hat{p}_{i}
\hat{p}_{j}\right) T\sum_{k_{0}}\int \frac{d^{3}k}{\left( 2\pi \right) ^{3}}
\left[ \,^{\ast }\Gamma ^{ij00}(P,-P,K,-K)\,^{\ast }\Delta _{l}(K)\right. 
\notag \\
& \hspace{-0.5in}+\,^{\ast }\Gamma ^{ijmn}(P,-P,K,-K)\left( \delta _{mn}-
\hat{k}_{m}\hat{k}_{n}\right) \,^{\ast }\Delta _{t}(K)  \notag \\
& \hspace{-0.5in}+\,^{\ast }\Gamma ^{0i0}(-Q,P,-K)\,^{\ast }\Delta
_{l}(K)\,^{\ast }\Gamma ^{0j0}(-K,P,-Q)\,^{\ast }\Delta _{l}(Q)  \notag \\
& \hspace{-0.5in}+\,^{\ast }\Gamma ^{0im}(-Q,P,-K)\left( \delta _{mn}-\hat{k}
_{m}\hat{k}_{n}\right) \,^{\ast }\Delta _{t}(K)\,^{\ast }\Gamma
^{nj0}(-K,P,-Q)\,^{\ast }\Delta _{l}(Q)  \notag \\
& \hspace{-0.5in}+\,^{\ast }\Gamma ^{mi0}(-Q,P,-K)\,^{\ast }\Delta
_{l}(K)\,^{\ast }\Gamma ^{0jn}(-K,P,-Q)\left( \delta _{nm}-\hat{q}_{n}\hat{q}
_{m}\right) \,^{\ast }\Delta _{t}(Q)  \notag \\
& \hspace{-0.5in}+\left. \,^{\ast }\Gamma ^{mir}(-Q,P,-K)\left( \delta _{rs}-
\hat{k}_{r}\hat{k}_{s}\right) \,^{\ast }\Delta _{t}(K)\,^{\ast }\Gamma
^{sjn}(-K,P,-Q)\left( \delta _{nm}-\hat{q}_{n}\hat{q}_{m}\right) \,^{\ast
}\Delta _{t}(Q)\right] \hspace{-2pt}.  \label{pi star explicit}
\end{align}
There are six contributions: two from the four-gluon $(4g)$ vertex and four
from the three-gluon ($(3g$) vertex. Each contribution has to be calculated
separately. We will show the details of how we get the $3gll$-contribution
where the two propagators are both longitudinal. We will comment afterwards
on how to work out the other contributions.

\subsection{Solid-angle integrals}

Using the expression of the dressed $3g$-vertex, we have: 
\begin{align}
^{\ast }\Pi _{t3gll}(P)& =\frac{g^{2}N_{c}}{8\pi ^{2}}T\sum_{k_{0}}\int 
\frac{d^{3}k}{4\pi }\left[ \left( \mathbf{p}-2\mathbf{k}\right) ^{2}-\left( 
\mathbf{p}-2\mathbf{k}\right) \hspace{-2pt}.\mathbf{\hat{p}}^{2}\right. 
\notag \\
& \hspace{-0.82in}+6\int \frac{d\Omega _{s}}{4\pi }\frac{\left( \mathbf{p}-2
\mathbf{k}\right) \hspace{-2pt}.\mathbf{\hat{s}}-\left( \mathbf{p}-2\mathbf{k
}\right) \hspace{-2pt}.\mathbf{\hat{p}\,\hat{s}}.\mathbf{\hat{p}}}{PS}\left( 
\frac{ik_{0}}{KS}-\frac{iq_{0}}{QS}\right)  \notag \\
& \hspace{-0.82in}+\left. 9\int \frac{d\Omega _{s_{1}}}{4\pi }\int \frac{
d\Omega _{s_{2}}}{4\pi }\frac{\mathbf{\hat{s}}_{1}.\mathbf{\hat{s}}_{2}-
\mathbf{\hat{s}}_{1}.\mathbf{\hat{p}}\,\mathbf{\hat{s}}_{2}.\mathbf{\hat{p}}
}{PS_{1}\,PS_{2}}\left( \hspace{-1pt}\frac{ik_{0}}{KS_{1}}-\frac{iq_{0}}{
QS_{1}}\hspace{-1pt}\right) \left( \hspace{-1pt}\frac{ik_{0}}{KS_{2}}-\frac{
iq_{0}}{QS_{2}}\hspace{-1pt}\right) \hspace{-2pt}\right] \,^{\ast }\Delta
_{l}(K)\,^{\ast }\Delta _{l}(Q)\hspace{-1pt}. \label{pi star t3gll-1}
\end{align}
We first work out the contribution that involves no integration over $\Omega
_{s}$. We have $\left( \mathbf{p}-2\mathbf{k}\right) ^{2}-\left( \mathbf{p}-2
\mathbf{k}\right) \hspace{-2pt}.\mathbf{\hat{p}}\hspace{1pt}^{2}=4k^{2}\sin
^{2}\psi $ with $\psi =\left( \mathbf{\hat{p}},\mathbf{\hat{k}}\right) $. To
integrate over the solid angle $\Omega _{k}$, we expand:
\begin{equation}
^{\ast }\Delta _{l(t)}(q_{0},q)=\left[ 1-p\cos \psi \,\partial _{k}+\frac{
p^{2}}{2}\left( \frac{\sin ^{2}\psi }{k}\partial _{k}+\cos ^{2}\psi
\,\partial _{k}^{2}\right) +\dots\right] \,^{\ast }\Delta _{l(t)}(q_{0},k)\,,
\label{expansion delta star}
\end{equation}
where $\partial _{k}$ stands for $\partial /\partial k$. The solid-angle
integrals become straightforward and we find: 
\begin{align}
& \hspace{-1in}T\sum_{k_{0}}\int \frac{d^{3}k}{4\pi }\left[ \left( \mathbf{p}
-2\mathbf{k}\right) ^{2}-\left( \mathbf{p}-2\mathbf{k}\right) \hspace{-2pt}.
\mathbf{\hat{p}}\hspace{1pt}^{2}\right] \,^{\ast }\Delta _{l}(K)\,^{\ast
}\Delta _{l}(Q)  \notag \\
\hspace{1in}& =\frac{8}{3}T\sum_{k_{0}}\int_{\mu }^{+\infty }k^{4}dk\,^{\ast
}\Delta _{l}(K)\left[ 1+\frac{p^{2}}{2}\left( \frac{2}{k}\partial _{k}+\frac{
1}{2}\partial _{k}^{2}\right) +...\right] \,^{\ast }\Delta _{l}(q_{0},k).
\label{first term in 3gll before spectral decomp.}
\end{align}

Next we work out the term involving one single integral over $\Omega _{s}$.
It is sufficient to concentrate on the piece with $ik_{0}/KS$ because the
other one, that involving $iq_{0}/QS$, is in fact equal to the first one. To
get a manageable expression for the solid-angle integral, we use the
following expansion: 
\begin{equation}
\dfrac{1}{PS}=\dfrac{1}{ip_{0}}\left[ 1-\dfrac{\mathbf{p}.\mathbf{\hat{s}}}{
ip_{0}}-\dfrac{\mathbf{p}.\mathbf{\hat{s}}^{2}}{p_{0}^{2}}+...\right] \,.
\label{expansion-1/PS}
\end{equation}
Also, it is best here to measure the solid angle $\Omega _{s}\equiv (\theta
,\varphi )$ with respect to $\mathbf{\hat{k}}$ such that $\theta =\left( 
\mathbf{\hat{k}},\mathbf{\hat{s}}\right) $. We can then write: 
\begin{align}
& \hspace{-0.6in}\int \frac{d\Omega _{s}}{4\pi }\frac{\left( \mathbf{p}-2
\mathbf{k}\right) \hspace{-2pt}.\mathbf{\hat{s}}-\left( \mathbf{p}-2\mathbf{k
}\right) \hspace{-2pt}.\mathbf{\hat{p}}\,\mathbf{\hat{s}}.\mathbf{\hat{p}}}{
PS}\left( \frac{ik_{0}}{KS}-\frac{iq_{0}}{QS}\right)  \notag \\
\hspace{0.6in}& =-4k\sin \psi \frac{ik_{0}}{ip_{0}}\int \frac{d\Omega _{s}}{
4\pi }\frac{\sin \psi \cos \theta +\cos \psi \sin \theta \sin \varphi }{
ik_{0}+k\cos \theta }\left[ 1-\dfrac{\mathbf{p}.\mathbf{\hat{s}}}{ip_{0}}-
\dfrac{\mathbf{p}.\mathbf{\hat{s}}^{2}}{p_{0}^{2}}+...\right] .
\label{one solid angle in 3gll-1st expression}
\end{align}
The angular integrals are performed using the relation $\mathbf{\hat{p}}.
\mathbf{\hat{s}}=\cos \psi \cos \theta -\sin \psi \sin \theta \sin \varphi $.
We get:
\begin{align}
\int \frac{d\Omega _{s}}{4\pi }\frac{\left( \mathbf{p}-2\mathbf{k}\right) 
\hspace{-2pt}.\mathbf{\hat{s}}-\left( \mathbf{p}-2\mathbf{k}\right) \hspace{
-2pt}.\mathbf{\hat{p}}\,\mathbf{\hat{s}}.\mathbf{\hat{p}}}{PS}\left( \frac{
ik_{0}}{KS}-\frac{iq_{0}}{QS}\right) & =-2\frac{ik_{0}}{ip_{0}}\left(
1-x^{2}\right) \left[ 2\left( 1-\frac{ik_{0}}{k}Q_{0k}\right) \right.  \notag
\\
& \hspace{-2.5in}+\frac{p}{ip_{0}}x\left[ 3\frac{ik_{0}}{k}+\left( 1+3\frac{
k_{0}^{2}}{k^{2}}\right) Q_{0k}\right] -\frac{p^{2}}{p_{0}^{2}}\left[ \frac{2
}{3}\left( 1-2x^{2}\right) +\left( 1-5x^{2}\right) \frac{k_{0}^{2}}{k^{2}}
\right.  \notag \\
& \hspace{-2.5in}\left. -\left. \frac{ik_{0}}{k}\left( 1-3x^{2}+\left(
1-5x^{2}\right) \frac{k_{0}^{2}}{k^{2}}\right) Q_{0k}\right] +\dots\right] ,
\label{one solid angle in 3gll-2nd expression}
\end{align}
where $x=\cos \psi $ and $Q_{0k}=Q_{0}(\frac{ik_{0}}{k})$. We put this
expression back inside the integral over $d^{3}k$ and use (\ref{expansion
delta star}) to perform the integral over $\Omega _{k}$. We get: 
\begin{align}
& \hspace{-0.2in}6\,T\hspace{-2pt}\sum_{k_{0}}\hspace{-2pt}\int \frac{d^{3}k
}{4\pi }\hspace{-2pt}\int \frac{d\Omega _{s}}{4\pi }\frac{\left( \mathbf{p}-2
\mathbf{k}\right) \hspace{-2pt}.\mathbf{\hat{s}}-\left( \mathbf{p}-2\mathbf{k
}\right) \hspace{-2pt}.\mathbf{\hat{p}}\,\mathbf{\hat{s}}.\mathbf{\hat{p}}}{
PS}\left( \frac{ik_{0}}{KS}-\frac{iq_{0}}{QS}\right) \hspace{1pt}^{\ast }
\hspace{-2pt}\Delta _{l}(K)\,^{\ast }\Delta _{l}(Q)  \notag \\
& =-16\,T\sum_{k_{0}}\int_{\mu }^{+\infty }k^{2}dk\,^{\ast }\Delta _{l}(K)
\frac{ik_{0}}{ip_{0}}\left[ \left( 1-\frac{ik_{0}}{k}Q_{0k}\right) +\frac{
p^{2}}{5p_{0}^{2}}\left[ -1+\frac{ik_{0}}{k}Q_{0k}\right. \right.  \notag \\
& \left. \hspace{-2pt}+\hspace{-3pt}\left. \frac{ip_{0}}{2}\hspace{-2pt}
\left( \hspace{-2pt}3\frac{ik_{0}}{k}\hspace{-1pt}+\hspace{-2pt}\hspace{-1pt}
\left( \hspace{-1pt}\hspace{-1pt}1+3\frac{k_{0}^{2}}{k^{2}}\right) Q_{0k}
\hspace{-2pt}\right) \hspace{-2pt}\hspace{-1pt}\partial _{k}+p_{0}^{2}
\hspace{-1pt}\hspace{-2pt}\left( 1-\frac{ik_{0}}{k}Q_{0k}\right) \hspace{-2pt
}\left( \hspace{-2pt}\hspace{-1pt}\frac{2}{k}\partial _{k}+\hspace{-1pt}
\frac{1}{2}\partial _{k}^{2}\hspace{-1pt}\hspace{-1pt}\right) \hspace{-2pt}
\right] +...\right] \hspace{0.5pt}^{\ast }\Delta _{l}(q_{0},\hspace{-1pt}k).
\label{one solid angle in 3gll before matsubara}
\end{align}

Next we look at the terms with two solid-angle integrals. They are actually
equal to 
\begin{equation*}
-18\,T\sum_{k_{0}}\int \frac{d^{3}k}{4\pi }\,^{\ast }\Delta _{l}(K)\,^{\ast
}\Delta _{l}(Q)\int \hspace{-1pt}\frac{d\Omega _{s_{1}}}{4\pi }\int \hspace{
-1pt}\frac{d\Omega _{s_{2}}}{4\pi }\frac{\mathbf{\hat{s}}_{1}.\mathbf{\hat{s}
}_{2}-\mathbf{\hat{s}}_{1}.\mathbf{\hat{p}}\,\mathbf{\hat{s}}_{2}.\mathbf{
\hat{p}}}{PS_{1}\,PS_{2}}\left( \frac{k_{0}^{2}}{KS_{1}\,KS_{2}}-\frac{
k_{0}q_{0}}{KS_{1}\,QS_{2}}\right) ,
\end{equation*}
and we first look at the piece containing $k_{0}^{2}/KS_{1}\,KS_{2}$; the
other one will require more labor as we will see. The additional difficulty
here is the coupling between the two solid-angle integrals introduced by the
dot product $\mathbf{\hat{s}}_{1}.\mathbf{\hat{s}}_{2}$. In order to be able
to carry forward, it is best to write $\mathbf{\hat{s}}_{1}.\mathbf{\hat{s}}
_{2}-\mathbf{\hat{s}}_{1}.\mathbf{\hat{p}}\,\mathbf{\hat{s}}_{2}.\mathbf{
\hat{p}}$ explicitly in trigonometric terms, which will automatically
decouple the double solid-angle integral into the following form: 
\begin{align}
\int \frac{d\Omega _{s_{1}}}{4\pi }\int \frac{d\Omega _{s_{2}}}{4\pi }\frac{
\mathbf{\hat{s}}_{1}.\mathbf{\hat{s}}_{2}-\mathbf{\hat{s}}_{1}.\mathbf{\hat{p
}}\,\mathbf{\hat{s}}_{2}.\mathbf{\hat{p}}}{PS_{1}\,PS_{2}}\frac{k_{0}^{2}}{
KS_{1}\,KS_{2}}& =k_{0}^{2}\left[ \sin ^{2}\psi \left[ \int \frac{d\Omega
_{s}}{4\pi }\frac{\cos \theta }{PS\,KS}\right] ^{2}\right.  \notag \\
& \hspace{-3in}+\left. \cos ^{2}\psi \left[ \int \frac{d\Omega _{s}}{4\pi }
\frac{\sin \theta \sin \varphi }{PS\,KS}\right] ^{2}+2\cos \psi \sin \psi
\int \frac{d\Omega _{s}}{4\pi }\frac{\cos \theta }{PS\,KS}\int \frac{d\Omega
_{s}}{4\pi }\frac{\sin \theta \sin \varphi }{PS\,KS}\right] .
\label{two solid angle in 3gll-1st expression}
\end{align}
The problem reduces therefore to carrying out only a single solid-angle
integral each time, which is straightforward using (\ref{expansion-1/PS}).
We have the following intermediary results: 
\begin{align}
\int \frac{d\Omega _{s}}{4\pi }\frac{\cos \theta }{PS\,KS}& =\frac{1}{ip_{0}k
}\left[ 1-\frac{ik_{0}}{k}Q_{0k}+\frac{px}{ip_{0}}\frac{ik_{0}}{k}\left( 1-
\frac{ik_{0}}{k}Q_{0k}\right) \right.  \notag \\
& \left. -\frac{p^{2}}{p_{0}^{2}}\left[ \frac{1}{3}+\frac{1-3x^{2}}{2}\frac{
k_{0}^{2}}{k^{2}}-\frac{ik_{0}}{2k}\left( 1-x^{2}+\left( 1-3x^{2}\right) 
\frac{k_{0}^{2}}{k^{2}}\right) Q_{0k}\right] +...\right] ;
\label{cos theta/PS KS}
\end{align}
\begin{align}
\int \frac{d\Omega _{s}}{4\pi }\frac{\sin \theta \sin \varphi }{PS\,KS}& =
\frac{\sin \psi }{ip_{0}k}\left[ \frac{p}{2ip_{0}}\left( \frac{ik_{0}}{k}
+\left( 1+\frac{k_{0}^{2}}{k^{2}}\right) Q_{0k}\right) \right.  \notag \\
& \hspace{1in}\left. +\frac{p^{2}x}{p_{0}^{2}}\left( \frac{2}{3}+\frac{
k_{0}^{2}}{k^{2}}-\frac{ik_{0}}{k}\left( 1+\frac{k_{0}^{2}}{k^{2}}\right)
Q_{0k}\right) +...\right] ,  \label{sin theta sin phi/PS KS}
\end{align}
whereas we have $\int \frac{d\Omega _{s}}{4\pi }\frac{\sin \theta \cos
\varphi }{PS\,KS}=0\,$. Putting these back into (\ref{two solid angle in
3gll-1st expression}) and organizing the products and squares in powers of $p$,
we get: 
\begin{align}
\int \frac{d\Omega _{s_{1}}}{4\pi }\int \frac{d\Omega _{s_{2}}}{4\pi }\frac{
\mathbf{\hat{s}}_{1}.\mathbf{\hat{s}}_{2}-\mathbf{\hat{s}}_{1}.\mathbf{\hat{p
}}\,\mathbf{\hat{s}}_{2}.\mathbf{\hat{p}}}{PS_{1}\,PS_{2}}\frac{k_{0}^{2}}{
KS_{1}\,KS_{2}}& =-\frac{(1-x^{2})k_{0}^{2}}{p_{0}^{2}k^{2}}\left[ \left( 1-
\frac{ik_{0}}{k}Q_{0k}\right) ^{2}\right.  \notag \\
& \hspace{-3.11in}+\frac{px}{ip_{0}}\left( 1-\frac{ik_{0}}{k}Q_{0k}\right)
\left( 3\frac{ik_{0}}{k}+\left( 1+3\frac{k_{0}^{2}}{k^{2}}\right)
Q_{0k}\right) -\frac{p^{2}}{p_{0}^{2}}\left[ \frac{x^{2}}{4}\left( \frac{
ik_{0}}{k}+\left( 1+\frac{k_{0}^{2}}{k^{2}}\right) Q_{0k}\right) ^{2}\right.
\notag \\
& \hspace{-0.12in}\hspace{-3in}\left. +\left. \hspace{-2pt}\left( 1-\frac{
ik_{0}}{k}Q_{0k}\right) \hspace{-2pt}\left( \frac{2}{3}-\frac{4}{3}x^{2}
\hspace{-2pt}+\hspace{-2pt}\left( 1-7x^{2}\right) \frac{k_{0}^{2}}{k^{2}}
\hspace{-2pt}-\hspace{-2pt}\left( 1-4x^{2}+\hspace{-2pt}\left(
1-7x^{2}\right) \frac{k_{0}^{2}}{k^{2}}\right) \frac{ik_{0}}{k}Q_{0k}\right) 
\hspace{-2pt}\right] +...\hspace{-2pt}\right] \hspace{-2pt}
\label{two solid angle in 3gll-2st expression}
\end{align}
We put this expression back under $\int d^{3}k$ and use (\ref{expansion
delta star}) to perform the integral over the solid angle $\Omega _{k}$. We
get:
\begin{align}
& \hspace{-0.3in}18\,T\sum_{k_{0}}\int \frac{d^{3}k}{4\pi }\,^{\ast }\Delta
_{l}(K)\,^{\ast }\Delta _{l}(Q)\int \frac{d\Omega _{s_{1}}}{4\pi }\int \frac{
d\Omega _{s_{2}}}{4\pi }\frac{\mathbf{\hat{s}}_{1}.\mathbf{\hat{s}}_{2}-
\mathbf{\hat{s}}_{1}.\mathbf{\hat{p}}\,\mathbf{\hat{s}}_{2}.\mathbf{\hat{p}}
}{PS_{1}\,PS_{2}}\frac{k_{0}^{2}}{KS_{1}\,KS_{2}}  \notag \\
\hspace{-1in}& =\hspace{-2pt}-12T\hspace{-2pt}\sum_{k_{0}}\frac{k_{0}^{2}}{
p_{0}^{2}}\int_{\mu }^{+\infty }\hspace{-2pt}dk\,^{\ast }\Delta _{l}(K)\left[
\hspace{-2pt}\left( 1-\frac{ik_{0}}{k}Q_{0k}\right) ^{2}\hspace{-2pt}-\frac{
p^{2}}{5p_{0}^{2}}\left[ \hspace{-2pt}\frac{1}{4}\left( \frac{ik_{0}}{k}
+\left( 1+\frac{k_{0}^{2}}{k^{2}}\right) Q_{0k}\right) ^{2}\right. \right. 
\notag \\
& +\left( 1-\frac{ik_{0}}{k}Q_{0k}\right) \left( 2-2\frac{k_{0}^{2}}{k^{2}}
-\left( 1-2\frac{k_{0}^{2}}{k^{2}}\right) \frac{ik_{0}}{k}Q_{0k}\right)
-ip_{0}\left( 1-\frac{ik_{0}}{k}Q_{0k}\right)  \notag \\
& \left. \hspace{-2pt}\times \left. \hspace{-2pt}\hspace{-2pt}\left( 3\frac{
ik_{0}}{k}+\hspace{-2pt}\left( \hspace{-1pt}1+3\frac{k_{0}^{2}}{k^{2}}
\right) Q_{0k}\hspace{-1pt}\right) \hspace{-2pt}\partial _{k}\hspace{-2pt}-
\hspace{-2pt}\frac{p_{0}^{2}}{2}\hspace{-2pt}\left( 1-\frac{ik_{0}}{k}Q_{0k}
\hspace{-1pt}\right) ^{2}\hspace{-2pt}\left( \frac{4}{k}\partial
_{k}+\partial _{k}^{2}\right) \hspace{-1pt}\right] \hspace{-2pt}\hspace{-2pt}
+...\right] \hspace{0.5pt}^{\ast }\Delta _{l}(q_{0},k).\hspace{-22pt}
\label{two solid angle term in 3gll}
\end{align}
Note that generally, odd powers of $p$ do not cancel in the intermediary
steps until the integration over $\Omega _{k}$ is performed.

The piece in the double solid-angle integral in (\ref{pi star t3gll-1})
containing $k_{0}q_{0}/KS_{1}\hspace{1pt}QS_{2}$ deserves more attention.
Because of the presence of both $\mathbf{k}$ and $\mathbf{q}$, it is most
suitable to measure the solid angles with respect to $\mathbf{\hat{p}}$.
Then we have:
\begin{align}
\int \frac{d\Omega _{s_{1}}}{4\pi }\int \frac{d\Omega _{s_{2}}}{4\pi }\frac{
\mathbf{\hat{s}}_{1}.\mathbf{\hat{s}}_{2}-\mathbf{\hat{s}}_{1}.\mathbf{\hat{p
}}\,\mathbf{\hat{s}}_{2}.\mathbf{\hat{p}}}{PS_{1}\,PS_{2}}\frac{k_{0}q_{0}}{
KS_{1}\,QS_{2}}& =k_{0}q_{0}\left[ \int \frac{d\Omega _{s}}{4\pi }\frac{\sin
\theta _{0}\cos \varphi _{0}}{PS\hspace{1pt}KS}\int \frac{d\Omega _{s}}{4\pi 
}\frac{\sin \theta _{0}\cos \varphi _{0}}{PS\hspace{1pt}QS}\right.  \notag \\
& \hspace{-1in}+\left. \int \frac{d\Omega _{s}}{4\pi }\frac{\sin \theta
_{0}\sin \varphi _{0}}{PS\hspace{1pt}KS}\int \frac{d\Omega _{s}}{4\pi }\frac{
\sin \theta _{0}\sin \varphi _{0}}{PS\hspace{1pt}QS}\right] ,
\label{two solid angle with 1/KS QS-first}
\end{align}
where we have added the subscript zero to indicates that the solid angle
$\Omega _{s}\equiv \left( \theta _{0},\varphi _{0}\right) $ is measured with
respect to $\mathbf{\hat{p}}$. Each solid-angle integral can be carried out
separately using the following identities:
\begin{equation*}
\cos \theta _{0}=\cos \psi \cos \theta -\sin \psi \sin \theta \sin \varphi 
\hspace{1pt};
\end{equation*}
\begin{equation*}
\sin \theta _{0}\sin \varphi _{0}=\sin \psi \cos \theta +\cos \psi \sin
\theta \sin \varphi \hspace{1pt};
\end{equation*}
\begin{equation}
\sin \theta _{0}\cos \varphi _{0}=\sin \theta \cos \varphi \hspace{1pt}.
\label{cos theta_zero etc--cos theta etc}
\end{equation}
We get: 
\begin{equation}
\int \frac{d\Omega _{s}}{4\pi }\frac{\sin \theta _{0}\cos \varphi _{0}}{PS
\hspace{1pt}KS}=0\hspace{1pt}, \label{solid-angle integral wrt p-hat-cos phi_0}
\end{equation}
using (\ref{cos theta_zero etc--cos theta etc}) and the fact that $\int 
\frac{d\Omega _{s}}{4\pi }\frac{\sin \theta \cos \varphi }{PS\,KS}=0$. The
other integral in $1/KS$ is:
\begin{align}
\int \frac{d\Omega _{s}}{4\pi }\frac{\sin \theta _{0}\sin \varphi _{0}}{PS
\hspace{1pt}KS}& =\frac{\sin \psi }{ip_{0}k}\left[ 1-\frac{ik_{0}}{k}Q_{0k}+
\frac{px}{2ip_{0}}\left[ 3\frac{ik_{0}}{k}+\left( 1+3\frac{k_{0}^{2}}{k^{2}}
\right) Q_{0k}\right] \right.  \notag \\
& \hspace{-1.2in}\left. -\frac{p^{2}}{2p_{0}^{2}}\left[ \frac{2}{3}\left(
1-2x^{2}\right) +\left( 1-5x^{2}\right) \frac{k_{0}^{2}}{k^{2}}-\left(
1-3x^{2}+\left( 1-5x^{2}\right) \frac{k_{0}^{2}}{k^{2}}\right) \frac{ik_{0}}{
k}Q_{0k}\right] +...\right] . \label{solid-angle integral wrt p-hat-sin phi_0}
\end{align}
We have similar results for the integrals with $1/QS$ in (\ref{two solid
angle with 1/KS QS-first}): we need only to replace $K$ by $Q$ and the angle 
$\psi =(\mathbf{\hat{p}},\mathbf{\hat{k}})$ by $\chi =(\mathbf{\hat{p}},
\mathbf{\hat{q}})$. We then multiply the two expressions and put back the
result under $\int d^{3}k/4\pi $. But here, the integral over $\Omega _{k}$
is not straightforward yet because the integrand still depends on $q$ and
the angle $\left( \mathbf{\hat{p}},\mathbf{\hat{q}}\right) $. Hence a
further expansion is necessary, but instead of expanding $Q_{0q}$ directly,
it is most suitable as we will explain why shortly to express first $Q_{0k}$
and $Q_{0q}$ in terms of $^{\ast }\Delta _{l}^{-1}(K)$ and $^{\ast }\Delta
_{l}^{-1}(Q)$ respectively, using (\ref{gluon propagator}) and(\ref{delta
pi_l-t}). We have:
\begin{equation}
\,^{\ast }\Delta _{lK}^{-1}=k^{2}+3\left( 1-\frac{ik_{0}}{k}Q_{0k}\right) ,
\label{delta-l--1 versus Qo}
\end{equation}
and a similar result when replacing $K$ by $Q$. We have written $^{\ast
}\Delta _{lK}$ for $^{\ast }\Delta _{l}(K)$ for short. The reason behind
this preliminary replacement is to eliminate the occurrence of products of
more than two functions necessitating a spectral decomposition. Such
undesirable products complicate unnecessarily the subsequent extraction of
the imaginary part. After the replacement (\ref{delta-l--1 versus Qo}) is
done and apparent simplifications made, we obtain: 
\begin{align}
18T\sum_{k_{0}}\hspace{-2pt}\int \hspace{-2pt}\frac{d^{3}\hspace{-1pt}k}
{4\pi }\,^{\ast }\Delta _{lK}\,^{\ast }\Delta _{lQ}\hspace{-2pt}\int \hspace{
-2pt}\frac{d\Omega _{s_{1}}}{4\pi }\hspace{-2pt}\int \hspace{-2pt}\frac{
d\Omega _{s_{2}}}{4\pi }\frac{\mathbf{\hat{s}}_{1}.\mathbf{\hat{s}}_{2}-
\mathbf{\hat{s}}_{1}.\mathbf{\hat{p}}\,\mathbf{\hat{s}}_{2}.\mathbf{\hat{p}}
}{PS_{1}\,PS_{2}}\frac{k_{0}q_{0}}{KS_{1}\,QS_{2}}& =2T\sum_{k_{0}}\int 
\frac{d^{3}\hspace{-1pt}k}{4\pi }\,^{\ast }\Delta _{lK}  \notag \\
& \hspace{-4.5in}\times \frac{\sin \psi \sin \chi }{p_{0}^{2}}\left[
-k_{0}q_{0}kq-\frac{pxq_{0}}{p_{0}}\left( 3+k^{2}+3k_{0}^{2}\right) q-\frac{
p^{2}}{2p_{0}^{2}}\left[ -2\frac{k_{0}q_{0}}{k}\left( 1-5x^{2}\right.
\right. \right.  \notag \\
& \hspace{-0.5in}\left. \hspace{-4in}\left. \left. +\left( 1-3x^{2}\right)
k^{2}+\left( 1-5x^{2}\right) k_{0}^{2}\right) q+\frac{xy}{2}\left(
3+k^{2}+3k_{0}^{2}\right) \left( 3+q^{2}+3q_{0}^{2}\right) \right] +...
\right] \,^{\ast }\Delta _{lQ}\hspace{1pt},
\label{two solid angle with 1/KS QS-second}
\end{align}
where $y=\cos \chi $. We only need now a final expansion of functions of $q$
in powers of $p$. We use the additional relations: 
\begin{equation}
q\cos \chi =p-k\cos \psi ;\qquad q\sin \chi =-k\sin \psi \hspace{1pt},
\label{cos sin chi versus cos sin psi}
\end{equation}
and the following two useful expansions: 
\begin{equation}
q=k\left( 1-x\frac{p}{k}+\frac{1}{2}\left( 1-x^{2}\right) \frac{p^{2}}{k^{2}}
+...\right) ;\quad \frac{1}{q^{2}}=\frac{1}{k^{2}}\left( 1+2x\frac{p}{k}
-\left( 1-4x^{2}\right) \frac{p^{2}}{k^{2}}+...\right) .
\label{expansion q and 1/q^2}
\end{equation}
Inserting all this back into (\ref{two solid angle with 1/KS QS-second}) and
performing the integral over $d\Omega _{k}$ which becomes feasible now, we
obtain:
\begin{align}
18T\hspace{-1pt}\sum_{k_{0}}\hspace{-2pt}\int \hspace{-2pt}\frac{d^{3}k}
{4\pi }\,^{\ast }\Delta _{lK}\,^{\ast }\Delta _{lQ}\hspace{-2pt}\int \hspace{
-2pt}\frac{d\Omega _{s_{1}}}{4\pi }\hspace{-2pt}\hspace{-2pt}\int \hspace{
-2pt}\frac{d\Omega _{s_{2}}}{4\pi }\frac{\mathbf{\hat{s}}_{1}.\mathbf{\hat{s}
}_{2}-\mathbf{\hat{s}}_{1}.\mathbf{\hat{p}}\,\mathbf{\hat{s}}_{2}.\mathbf{
\hat{p}}}{PS_{1}\,PS_{2}}\frac{k_{0}q_{0}}{KS_{1}\,QS_{2}}& =\hspace{-1pt}-
\frac{4}{3}T\hspace{-1pt}\sum_{k_{0}}\hspace{-3pt}\int_{\mu }^{+\infty }
\hspace{-4pt}dk\frac{k^{2}}{p_{0}^{2}}\,^{\ast }\Delta _{lK}  \notag \\
& \hspace{-4in}\times \left[ -k_{0}q_{0}k^{2}+\frac{p^{2}}{5p_{0}^{2}}\left[
2k_{0}q_{0}k^{2}+\frac{1}{4}\left( 3+k^{2}+3k_{0}^{2}\right) \left(
3+k^{2}+3q_{0}^{2}\right) \right. \right.  \notag \\
\hspace{-4.3in}& \left. \hspace{-4in}\left. +q_{0}p_{0}k\left(
3+k^{2}+3k_{0}^{2}-2k_{0}p_{0}\right) \partial _{k}-\frac{1}{2}
k_{0}q_{0}p_{0}^{2}k^{2}\partial _{k}^{2}\right] +...\right] \,^{\ast
}\Delta _{l}(q_{0},k).  \label{two solid angle with 1/KS QS-third}
\end{align}

\subsection{Sum over Matsubara}

Everything is ready to perform the sum over the Matsubara frequency $k_{0}$.
It is worth mentioning that the main reason we went through the above steps
is to get $ik_{0}$ (and $iq_{0})$ appear only in the numerator of fractions.
This way, to perform the sum, we only need to use the spectral
representations \cite{pisarski5,pisarski6,le bellac} of the following
quantities:
\begin{align}
^{\ast }\Delta _{t,l}(k_{0},k)& =\int_{0}^{1/T}d\tau \,e^{ik_{0}\tau
}\int_{-\infty }^{+\infty }d\omega \,\rho _{t,l}(k,\omega )\left( 1+n(\omega
)\right) e^{-\omega \tau }\,;  \notag \\
Q_{0}(ik_{0}/k)& =-\frac{1}{2}\int_{0}^{1/T}d\tau \,e^{ik_{0}\tau
}\int_{-\infty }^{+\infty }d\omega \,\Theta (k-\left\vert \omega \right\vert
)\left( 1+n(\omega )\right) e^{-\omega \tau }\,,
\label{spectral representation delta and Q0}
\end{align}
where $n(\omega )=1/\left( e^{\omega /T}-1\right) $ is the Bose-Einstein
distribution and $\rho _{t,l}(k,\omega )$ the spectral densities given in (
\ref{definition rho}) below. After we replace $^{\ast }\Delta _{lK}$ and $
Q_{0k}$ by their spectral representations, we perform the integrals over the
imaginary times. As we mentioned in the text after (\ref{delta-l--1 versus
Qo}), preventing the occurrence of the product of more than two functions
necessitating a spectral decomposition will ensure the occurrence of at most
two (imaginary) time integrals. One such integration yields a delta-function
and the subsequent one an energy denominator. Thus, terms have been arranged
in such a way to yield at most one energy denominator. Only now can we
perform the analytic continuation $ip_{0}\rightarrow \omega _{t}(p)+i0^{+}$.
But just before this, every $e^{\frac{ip_{0}}{T}}$ has to be replaced by $1$
except in the energy denominators. At each time, we are left with two
frequency integrals together with the one over $k$. Terms involving one
integration over imaginary time are real and the imaginary part of those
involving two is obtained using the known relation $1/\left( x+i0^{+}\right)
=\Pr \left( 1/x\right) -i\pi \delta (x)$ applied to the energy denominators; 
$\Pr $ stands for the principal part. This technique is applied to
expressions (\ref{first term in 3gll before spectral decomp.}), (\ref{one
solid angle in 3gll before matsubara}), (\ref{two solid angle term in 3gll})
and (\ref{two solid angle with 1/KS QS-third}). We perform the sum and
obtain after few steps of straightforward algebra: 
\begin{align}
\mathrm{Im}\hspace{1pt}^{\ast }\Pi _{t3gll}(P)& =\frac{g^{2}N_{c}T}{24\pi }
\int_{\mu }^{+\infty }dk\int_{-\infty }^{+\infty }d\omega \int_{-\infty
}^{+\infty }d\omega ^{\prime }\left[ \frac{18k^{2}}{\omega \omega ^{\prime }}
\rho _{l}\rho _{l}^{\prime }+\frac{6\omega ^{2}}{k\omega ^{\prime }}\Theta
\rho _{l}^{\prime }\right.  \notag \\
& +\frac{p^{2}}{5}\left[ \frac{3k^{2}}{2\omega \omega ^{\prime }}\left(
3+44k^{2}-12\omega \omega ^{\prime }\right) \rho _{l}\rho _{l}^{\prime }+
\frac{3k}{2\omega ^{\prime }}\left( 1-\frac{\omega ^{2}}{k^{2}}\right)
\left( 1-9\frac{\omega ^{2}}{k^{2}}\right) \Theta \rho _{l}^{\prime }\right.
\notag \\
& +\frac{4k^{3}}{\omega \omega ^{\prime }}\left( 13+3k^{2}+10\omega -9\omega
^{2}\right) \rho _{l}\hspace{1pt}\partial _{k}\rho _{l}^{\prime }+6\frac{
\omega }{\omega ^{\prime }}\left( 1+2\frac{\omega }{k^{2}}-3\frac{\omega ^{2}
}{k^{2}}\right) \Theta \hspace{1pt}\partial _{k}\rho _{l}^{\prime }  \notag
\\
& \left. \hspace{-2pt}+\hspace{-2pt}\left. \frac{2k^{4}}{\omega \omega
^{\prime }}\hspace{-2pt}\left( 2+\hspace{-2pt}5\omega \right) \rho
_{l}\partial _{k}^{2}\rho _{l}^{\prime }+\hspace{-2pt}\frac{3\omega ^{2}}{
k\omega ^{\prime }}\Theta \partial _{k}^{2}\rho _{l}^{\prime }-\hspace{-2pt}
\frac{54k^{4}}{\omega \omega ^{\prime }}\rho _{l}\rho _{l}^{\prime }\partial
_{\omega }-\frac{18\omega ^{2}}{k\omega ^{\prime }}\Theta \rho _{l}^{\prime
}\partial _{\omega ^{\prime }}\right] \hspace{-2pt}+...\right] \delta .
\label{Im pi_ll-final 1}
\end{align}
The notation is as follows. The quantity $\rho _{l}$ stands for $\rho
_{l}(k,\omega )$ and $\rho _{l}^{\prime }$ for $\rho _{l}(k,\omega ^{\prime
})$. $\Theta $ is $\Theta (k-\left\vert \omega \right\vert )$ and $\delta $
is $\delta (1-\omega -\omega ^{\prime })$. In the above relation, we have
used the approximation $n(\omega ^{(\prime )})\simeq T/\omega ^{(\prime )}$ 
\cite{BPgamt} because only soft values of $\omega ^{(\prime )}$ are to
contribute, as dictated by the HTL-summed perturbation. Also, the last two
terms in (\ref{Im pi_ll-final 1}) come from an expansion of $\delta \left[
\omega _{t}(p)-\omega -\omega ^{\prime }\right] $ in powers of $p$. Finally,
we note that many terms from intermediary steps drop in the end since they
do not carry an imaginary part, i.e., they involve only one imaginary-time
integral as explained. The first two terms in (\ref{Im pi_ll-final 1}) are
those found in \cite{BPgamt}. The terms in the $p^{2}$ contribution are new.

What remains to do is to perform the integrals involved in (\ref{Im
pi_ll-final 1}). How this is done is shown in the next section. Before this,
we comment on the other contributions to $\mathrm{Im}\hspace{1pt}^{\ast }\Pi
_{t}(P)$ and give the corresponding results.

\subsection{The other contributions}

The other contributions to $\mathrm{Im}\hspace{1pt}^{\ast }\Pi _{t}(P)$ are
worked out along similar lines. Concerning the $3g$-contributions, the
intermediary steps of course are much longer. This is because of the
presence of the transverse projector $\left( \delta _{mn}-\hat{k}_{m}\hat{k}
_{n}\right) $ in the three-gluon $lt$ and $tl$ contributions in (\ref{pi
star explicit}) and the additional one $\left( \delta _{rs}-\hat{q}_{r}\hat{q
}_{s}\right) $ in the three-gluon $tt$ contribution. The trigonometry in the
numerator becomes less simple and the expansion in powers of $p$ quite
intricate at times. For example, for the $lt$ contribution, to reduce terms
involving a product of more than two functions requiring a spectral
decomposition, we need to use, in addition to (\ref{delta-l--1 versus Qo}),
the relation: 
\begin{equation}
^{\ast }\Delta _{tK}^{-1}=K^{2}-\frac{3}{2}\left( \frac{k_{0}^{2}}{k^{2}}-
\frac{K^{2}}{k^{2}}\frac{ik_{0}}{k}Q_{0k}\right) .
\label{delta-t--1 versus Qo}
\end{equation}
But in order to avoid dividing by a frequency $k_{0}$ (or $q_{0}$) while
using this relation, something that would spoil the efficiency of the
spectral-decomposition method, the Legendre function $Q_{0k}$ has to appear
in the different expressions in the form $\frac{K^{2}}{k^{2}}\frac{ik_{0}}{k}
Q_{0k}$, which turns out to be not always the case. In contrast, $Q_{0k}$ in
the $3gll$ contribution we just worked out had to always appear only in the
form $\frac{ik_{0}}{k}Q_{0k}$, see (\ref{delta-l--1 versus Qo}). This
latter requirement is less stringent and it was always satisfied there,
which meant there were no additional (technical) difficulties. What we use
to remedy to the situation here is to group different terms together to
obtain a cancellation of undesirable $Q_{0q}$'s. While for the $tl$ and $lt$
contributions this is more or less straightforward, the cancellation is
quite tricky for the $3gtt$ contribution since it necessitates grouping
terms coming from the expansion of the zeroth and first orders when using
(\ref{expansion-1/PS}), terms seemingly not connected together. These
intermediary steps work, but only case by case: we do not have a general
rule regarding this point. These steps are quite lengthy and cumbersome to
report on with any degree of detail. We limit ourselves to giving the final
results, similar to (\ref{Im pi_ll-final 1}). We find for the $tl$ and $lt$
contributions: 
\begin{align}
\mathrm{Im}\hspace{1pt}^{\ast }\Pi _{t3glt}(P)& =\mathrm{Im}\hspace{1pt}^{\ast
}\Pi _{t3gtl}(P)=\frac{g^{2}N_{c}T}{48\pi }\int_{\mu }^{+\infty
}dk\int_{-\infty }^{+\infty }\frac{d\omega }{\omega }\int_{-\infty
}^{+\infty }\frac{d\omega ^{\prime }}{\omega ^{\prime }}\left[ -18\left(
k^{2}-\omega ^{2}\right) ^{2}\rho _{t}\rho _{l}^{\prime }\right.  \notag \\
& -\frac{3\omega }{k^{3}}\left( k^{2}-\omega ^{2}\right) ^{2}\Theta \rho
_{t}^{\prime }+\frac{6\omega }{k}\left( k^{2}-\omega ^{2}\right) ^{2}\Theta
\rho _{l}^{\prime }+\frac{p^{2}}{5}\left[ \left[ 4+93k^{4}+18k^{6}\right.
\right.  \notag \\
& +\left( 8+78k^{4}\right) \omega -\left( +140+78k^{2}+90k^{4}\right) \omega
^{2}+\left( 48-180k^{2}\right) \omega ^{3}  \notag \\
& -\left. \left( 31-126k^{2}\right) \omega ^{4}+102\omega ^{5}-54\omega ^{6} 
\right] \rho _{t}\rho _{l}^{\prime }+\left[ \left( \frac{3}{2k}+3k\right)
\omega -\frac{3}{k}\omega ^{2}\right.  \notag \\
& +\left. \left( \frac{3}{k^{2}}-\frac{15}{k}\right) \omega ^{3}-\frac{6}{
k^{3}}\omega ^{4}-\left( \frac{57}{2k^{5}}-\frac{21}{k^{3}}\right) \omega
^{5}-54\omega ^{6}\right] \Theta \rho _{t}^{\prime }+\left[ -6k\omega +\frac{
24}{k}\omega ^{3}\right.  \notag \\
& -\left. \frac{18}{k^{3}}\omega ^{5}\right] \hspace{-2pt}\Theta \rho
_{l}^{\prime }+\hspace{-2pt}\left[ -12k\hspace{-2pt}-\hspace{-2pt}72k^{5}
\hspace{-2pt}+\hspace{-2pt}\left( -24k-\hspace{-2pt}12k^{3}+48k^{5}\right)
\omega +\hspace{-2pt}\left( -12k+144k^{3}\right) \hspace{-2pt}\omega
^{2}\right.  \notag \\
& +\left. \left( 12k-96k^{3}\right) \omega ^{3}-72k\omega ^{4}+48k\omega
^{5} \right] \rho _{t}\partial _{k}\rho _{l}^{\prime }+\left[
69k+14k^{3}-3k^{5}\right.  \notag \\
& -\left( 54k+16k^{3}-6k^{5}\right) \omega +\left( 2k+14k^{3}\right) \omega
^{2}+\left( 8k-12k^{3}\right) \omega ^{3}  \notag \\
& -\hspace{-2pt}\left. 11k\omega ^{4}\hspace{-2pt}+\hspace{-2pt}6k\omega ^{5}
\right] \hspace{-2pt}\rho _{l}\partial _{k}\rho _{t}^{\prime }\hspace{-2pt}+
\left[ -\frac{9}{2}\omega +9\omega ^{2}\hspace{-2pt}+\frac{3}{k^{2}}\omega
^{3}\hspace{-2pt}-\frac{18}{k^{2}}\omega ^{4}\hspace{-2pt}+\frac{3}{2k^{4}}
\omega ^{5}\hspace{-2pt}+\frac{9}{k^{4}}\omega ^{6}\hspace{-2pt}\right] 
\hspace{-2pt}\Theta \partial _{k}\rho _{t}^{\prime }  \notag \\
& +\left[ 9\omega -18\omega ^{2}-\frac{9}{k^{2}}\omega ^{3}+\frac{18}{k^{2}}
\omega ^{4}\right] \Theta \partial _{k}\rho _{l}^{\prime }-\left[
8k^{2}+16k^{6}+16k^{2}\omega \right.  \notag \\
& +\left. \left( 8k^{2}-32k^{4}\right) \omega ^{2}+16k^{2}\omega ^{4}\right]
\rho _{t}\partial _{k}^{2}\rho _{l}^{\prime }+\left[ 30k^{2}+4k^{4}-2k^{6}-
\left( 24k^{2}+8k^{4}\right) \omega \right.  \notag \\
& -\left. \left( 4k^{2}-4k^{4}\right) \omega ^{2}+8k^{2}\omega
^{3}-2k^{2}\omega ^{4}\right] \rho _{l}\partial _{k}^{2}\rho _{t}^{\prime }-3
\frac{\omega }{k^{3}}\left( k^{2}-\omega ^{2}\right) ^{2}\Theta \partial
_{k}^{2}\rho _{t}^{\prime }  \notag \\
& +6\frac{\omega }{k}\left( k^{2}-\omega ^{2}\right) \Theta \partial
_{k}^{2}\rho _{l}^{\prime }+54k^{2}\left( k^{2}-\omega ^{2}\right) ^{2}\rho
_{t}\rho _{l}^{\prime }\partial _{\omega }+9\frac{\omega }{k^{3}}\left(
k^{2}-\omega ^{2}\right) ^{2}\Theta \rho _{t}^{\prime }\partial _{\omega } 
\notag \\
& \left. -\left. 18\frac{\omega }{k}\left( k^{2}-\omega ^{2}\right) \Theta
\rho _{l}^{\prime }\partial _{\omega }\right] +\dots \right] \delta .
\label{Im pi_lt+tl--final 1}
\end{align}
Note that the two contributions $tl$ and $lt$ are equal as it should be. The
above expression is rather long, and this is partly due to the fact that the
functions corresponding to these two contributions do not benefit from a
symmetry between $\omega $ and $\omega ^{\prime }$, present in the functions
corresponding to the $ll$ and $tt$ contributions. Indeed, though the algebra
for the $tt$ contribution is by far the most tedious, the use of this
symmetry, after the obtainment of expressions like those of (\ref{Im
pi_lt+tl--final 1}) above, renders the corresponding final result relatively
simpler. It reads: 
\begin{align}
\mathrm{Im}\hspace{1pt}^{\ast }\Pi _{t3gtt}(P)& =\frac{g^{2}N_{c}T}{24\pi }
\int_{\mu }^{+\infty }\hspace{-2pt}dk\int_{-\infty }^{+\infty }\hspace{-2pt}
\frac{d\omega }{\omega }\int_{-\infty }^{+\infty }\hspace{-2pt}\frac{d\omega
^{\prime }}{\omega ^{\prime }}\left[ 36\left( k^{2}+\omega \omega ^{\prime
}\right) ^{2}\rho _{t}\rho _{t}^{\prime }-6\frac{\omega ^{3}}{k^{3}}\left(
k^{2}-\omega ^{2}\right) \Theta \rho _{t}^{\prime }\right.  \notag \\
& +\frac{p^{2}}{5}\left[ \left[ -18k^{2}+122k^{4}-\left( 144-174k^{2}\right)
\omega \omega ^{\prime }+\left( \frac{30}{7k^{2}}-138\right) \omega
^{2}\omega ^{\prime 2}\right. \right.  \notag \\
& -\left. \frac{174}{k^{2}}\omega ^{3}\omega ^{\prime 3}\right] \rho
_{t}\rho _{t}^{\prime }-\frac{3\omega }{4k}\left( 1-\frac{\omega ^{2}}{k^{2}}
\right) \left[ 5k^{2}+16\omega -2\omega ^{2}+\frac{21}{k^{2}}\omega ^{4}
\right] \Theta \rho _{t}^{\prime }  \notag \\
& +\left[ 32k^{3}+20k^{5}+\left( 112k+124k^{3}\right) \omega +\left( \frac{48
}{k}-36k-116k^{3}\right) \omega ^{2}\right.  \notag \\
& -\left. \left( \frac{108}{k}+232k\right) \omega ^{3}+\left( \frac{12}{k}
+156k\right) \omega ^{4}+\frac{108}{k}\omega ^{5}-\frac{60}{k}\omega ^{6}
\right] \rho _{t}\partial _{k}\rho _{t}^{\prime }  \notag \\
& -6\frac{\omega ^{2}}{k^{2}}\left( 1-\frac{\omega ^{2}}{k^{2}}\right)
\left( k^{2}+3\omega -3\omega ^{2}\right) \Theta \partial _{k}\rho
_{t}^{\prime }+\left[ 8k^{4}+20k^{2}\left( 1+k^{2}\right) \omega \right. 
\notag \\
& +\left. 12\left( 1+k^{2}\right) \omega ^{2}-\left( 12+32k^{2}\right)
\omega ^{2}-12\omega ^{4}+12\omega ^{5}\right] \rho _{t}\partial
_{k}^{2}\rho _{t}^{\prime }  \notag \\
& -3\frac{\omega ^{3}}{k^{3}}\left( k^{2}-\omega ^{2}\right) \Theta \partial
_{k}^{2}\rho _{t}^{\prime }-108\left( k^{2}+\omega \omega ^{\prime }\right)
^{2}\rho _{t}\rho _{t}^{\prime }\partial _{\omega }  \notag \\
& +\left. \left. 18\frac{\omega ^{3}}{k^{3}}\left( k^{2}-\omega ^{2}\right)
\Theta \rho _{t}^{\prime }\partial _{\omega }\right] +\dots \right] \delta .
\label{Im pi_tt-final 1}
\end{align}

The two $4g$-contributions require less labor but need few additional
ingredients. Consider for example the four-gluon longitudinal contribution.
Using result (\ref{4 gluon HTL vertex}), we have: 
\begin{equation}
^{\ast }\Pi _{t4gl}(P)=-\frac{g^{2}N_{c}}{4}T\sum_{k_{0}}\int \frac{d^{3}k}{
\left( 2\pi \right) ^{3}}\left[ \hspace{-1pt}4-3\hspace{-2pt}\int \hspace{
-2pt}\frac{d\Omega _{s}}{4\pi }\frac{\left( 1-\mathbf{\hat{s}}.\mathbf{\hat{p
}}^{2}\right) }{PS\,KS}\hspace{-2pt}\hspace{-2pt}\left( \frac{i\left(
p_{0}-k_{0}\right) }{\left( P-K\right) S}\hspace{-2pt}-\frac{i\left(
p_{0}+k_{0}\right) }{\left( P+K\right) S}\hspace{-2pt}\right) \hspace{-2pt}
\right] \hspace{1pt}^{\ast }\Delta _{lK}.  \label{pi star 4gl 1st expression}
\end{equation}
In addition to the expansion (\ref{expansion-1/PS}), we use: 
\begin{equation}
\dfrac{1}{\left( P\pm K\right) S}=\frac{\pm 1}{K_{\pm }S}\left[ 1\mp \dfrac{
\mathbf{p}.\mathbf{\hat{s}}}{K_{\pm }S}+\dfrac{\mathbf{p}.\mathbf{\hat{s}}
^{2}}{K_{\pm }S^{2}}\mp ...\right] ;\qquad K_{\pm }=\left( k_{0}\pm p_{0},
\mathbf{k}\right) .  \label{expansion1/(P+-K)S}
\end{equation}
Putting back in (\ref{pi star 4gl 1st expression}), we encounter terms of
the type $1/KS\hspace{1pt}K_{\pm }S^{n}$, with $n=1,2$ and $3$. They are
handled using repeatedly the relation: 
\begin{equation}
\frac{1}{KS\hspace{1pt}K_{\pm }S}=\frac{\pm 1}{ip_{0}}\left[ \frac{1}{K_{\pm
}S}-\frac{1}{KS}\right] .  \label{1/KSK_(+-)S rewritten}
\end{equation}
The rest carries straightforwardly until we obtain: 
\begin{align}
^{\ast }\Pi _{t4gl}(P)& =-\frac{g^{2}N_{c}}{4\pi ^{2}}T\sum_{k_{0}}\int_{\mu
}^{+\infty }dk\hspace{1pt}k^{2}\left[ 2+\frac{1}{p_{0}^{2}}\sum_{\epsilon
=\pm }\left( \frac{ik_{0\epsilon }}{k}Q_{0k_{\epsilon }}-\frac{ik_{0}}{k}
Q_{0k}\right) \right.  \notag \\
& \left. -\frac{p^{2}}{5p_{0}^{2}}\sum_{\epsilon =\pm }\left( \frac{
k_{0\epsilon }}{K_{\epsilon }^{4}}+\frac{2k_{0\epsilon }}{p_{0}K_{\epsilon
}^{2}}+3\frac{ik_{0\epsilon }}{p_{0}^{2}k}Q_{0k_{\epsilon }}-3\frac{ik_{0}}{
p_{0}^{2}k}Q_{0k}\right) +...\right] \hspace{1pt}^{\ast }\Delta _{lK}\hspace{
1pt}.  \label{pi star 4gl 2nd expression}
\end{align}
To perform the Matsubara sum, we need the additional spectral
decompositions: 
\begin{align}
\frac{1}{K^{2}}& =\int_{0}^{1/T}d\tau \,e^{ik_{0}\tau }\int_{-\infty
}^{+\infty }d\omega \left( 1+n(\omega )\right) \,\epsilon \left( \omega
\right) \delta (\omega ^{2}-k^{2})\hspace{1pt}e^{-\omega \tau }\,;  \notag \\
\frac{1}{K^{4}}& =\int_{0}^{1/T}d\tau \,e^{ik_{0}\tau }\int_{-\infty
}^{+\infty }d\omega \,\left( 1+n(\omega )\right) \,\epsilon \left( \omega
\right) \hspace{1pt}\partial _{\omega ^{2}}\delta (\omega ^{2}-k^{2})\hspace{
1pt}e^{-\omega \tau }\,,  \label{spectral representation 1/K^2 1/K^4}
\end{align}
where $\epsilon \left( \omega \right) $ is the sign function. Few steps more
and we obtain: 
\begin{align}
\mathrm{Im}\hspace{1pt}^{\ast }\Pi _{t4gl}(P)& =\frac{g^{2}N_{c}T}{24\pi }
\int_{\mu }^{+\infty }\hspace{-2pt}dk\int_{-\infty }^{+\infty }\hspace{-2pt}
d\omega \int_{-\infty }^{+\infty }\hspace{-2pt}d\omega ^{\prime }\left[ -
\frac{6k}{\omega ^{\prime }}\Theta \rho _{l}^{\prime }+\frac{p^{2}}{5}\left[ 
\frac{18k}{\omega ^{\prime }}\Theta \rho _{l}^{\prime }\right. \right. 
\notag \\
& \hspace{-0.5in}\left. +\left. \frac{24k^{2}}{\omega ^{\prime }}\,\epsilon
\left( \omega \right) \delta \left( k^{2}-\omega ^{2}\right) \rho
_{l}^{\prime }-\frac{12\left\vert \omega \right\vert }{\omega ^{\prime }}
\partial _{\omega }\delta \left( k^{2}-\omega ^{2}\right) \rho _{l}^{\prime
}+\frac{18k}{\omega ^{\prime }}\Theta \rho _{l}^{\prime }\partial _{\omega }
\right] +...\right] \delta .  \label{Im pi_l--final 1}
\end{align}

The transverse $4g$-contribution carries in as much the same way. We obtain
for it:
\begin{align}
\mathrm{Im}\hspace{1pt}^{\ast }\Pi _{t4gt}(P)& =\frac{g^{2}N_{c}T}{24\pi }
\int_{\mu }^{+\infty }\hspace{-2pt}dk\int_{-\infty }^{+\infty }\hspace{-2pt}
d\omega \int_{-\infty }^{+\infty }\hspace{-2pt}d\omega ^{\prime }\left[ 
\frac{6\left( k^{2}-\omega ^{2}\right) }{k\omega ^{\prime }}\Theta \rho
_{t}^{\prime }\right.  \notag \\
& \hspace{-1in}\left. +\frac{p^{2}}{5}\left[ \frac{12\left\vert \omega
\right\vert }{\omega ^{\prime }}\delta \left( k^{2}-\omega ^{2}\right) \rho
_{t}^{\prime }+\frac{6\left( 4\omega -1\right) }{k\omega ^{\prime }}\Theta
\rho _{t}^{\prime }-\frac{18\left( k^{2}-\omega ^{2}\right) }{k\omega
^{\prime }}\Theta \rho _{t}^{\prime }\partial _{\omega }\right] +...\right]
\delta .  \label{Im pi_t--final 1}
\end{align}

\section{the damping rate for transverse gluons}

The damping rate for ultrasoft transverse gluons to lowest order $g^{2}T$ in
HTL-summed perturbation is given by (\ref{definition2 gamma_t}) where the
imaginary part of the HTL-dressed one-loop transverse self-energy is given
by: 
\begin{equation}
\mathrm{Im}\hspace{1pt}^{\ast }\Pi _{t}(P)=\sum_{i,j=l,t}\mathrm{Im}\hspace{1pt}
^{\ast }\Pi _{t3gij}(P)+\sum_{i=l,t}\mathrm{Im}\hspace{1pt}^{\ast }\Pi
_{t4gi}(P).  \label{Im pi_t--contributions}
\end{equation}
The different contributions are given in (\ref{Im pi_ll-final 1}), (\ref{Im
pi_lt+tl--final 1}), (\ref{Im pi_tt-final 1}), (\ref{Im pi_l--final 1}) and (
\ref{Im pi_t--final 1}). These expressions are in the form of triple
integrals which ultimately are going to be performed numerically. The
expressions of the dispersion functions corresponding to the dressed gluon
propagators are as follows \cite{pisarski5,pisarski6,le bellac}:
\begin{equation}
\rho _{t,l}(k,\omega )=\mathfrak{z}_{t,l}(k)\left[ \delta \left( \omega
-\omega _{t,l}(k)\right) -\delta \left( \omega +\omega _{t,l}(k)\right) 
\right] +\beta _{t,l}(k,\omega )\hspace{2pt}\Theta \hspace{-1pt}\left(
k-\left\vert \omega \right\vert \right) ,  \label{definition rho}
\end{equation}
with the residue functions given by: 
\begin{equation}
\mathfrak{z}_{t}(k)=\left. \frac{\omega \left( \omega ^{2}-k^{2}\right) }{
3\omega ^{2}-\left( \omega ^{2}-k^{2}\right) ^{2}}\right\vert _{\omega
=\omega _{t}(k)};\qquad \mathfrak{z}_{l}(k)=\left. -\frac{\omega \left(
\omega ^{2}-k^{2}\right) }{k^{2}\left( 3-\omega ^{2}+k^{2}\right) }
\right\vert _{\omega =\omega _{l}(k)},  \label{residue functions}
\end{equation}
and the cut functions by: 
\begin{align}
\beta _{t}(k,\omega )& =\frac{3\omega \left( k^{2}-\omega ^{2}\right) }{
4k^{3}\left[ \left( k^{2}-\omega ^{2}+\frac{3\omega ^{2}}{2k^{2}}\left( 1+
\frac{k^{2}-\omega ^{2}}{2k\omega }\ln \frac{k+\omega }{k-\omega }\right)
\right) ^{2}+\frac{9\pi ^{2}\omega ^{2}}{16k^{6}}\left( k^{2}-\omega
^{2}\right) ^{2}\right] };  \notag \\
\beta _{l}(k,\omega )& =-\frac{3\omega }{2k\left[ \left( 3+k^{2}-\frac{
3\omega }{2k}\ln \frac{k+\omega }{k-\omega }\right) ^{2}+\frac{9\pi
^{2}\omega ^{2}}{16k^{6}}\right] }\,.  \label{definition cut functions}
\end{align}
Recall that the thermal gluon mass $m_{g}$ is set to one. There are
different types of terms involved in (\ref{Im pi_ll-final 1}), (\ref{Im
pi_lt+tl--final 1}), (\ref{Im pi_tt-final 1}), (\ref{Im pi_l--final 1}) and (
\ref{Im pi_t--final 1}). We will show how we carry through using a generic
term of each type. Because of lengthy expressions, we will use as much
compact a notation as permissible in such a way that the meaning is always
clear from the context.

\subsection{Integration}

The first type of integrals we have to deal with is the one that involves a
$\rho \rho $ contribution. Generically, we consider an integral of the type: 
\begin{equation}
I_{0}=\int_{\mu }^{+\infty }\hspace{-2pt}dk\int_{-\infty }^{+\infty }\hspace{
-2pt}d\omega \int_{-\infty }^{+\infty }\hspace{-2pt}d\omega ^{\prime
}\,\delta \left( 1-\omega -\omega ^{\prime }\right) f(k,\omega ,\omega
^{\prime })\rho _{i}(k,\omega )\rho _{j}(k,\omega ^{\prime }),
\label{integral in rho rho}
\end{equation}
where $i,j=t,l$. The integral over $\omega ^{\prime }$ can be used to
eliminate $\delta \left( 1-\omega -\omega ^{\prime }\right) $. We get: 
\begin{equation}
I_{0}=\int_{\mu }^{+\infty }\hspace{-2pt}dk\int_{-\infty }^{+\infty }\hspace{
-2pt}d\omega f(k,\omega ,1-\omega )\rho _{i}(k,\omega )\rho _{j}(k,1-\omega
).  \label{integral in rho rho-2nd version}
\end{equation}
Using (\ref{definition rho}), there are three kinds of contributions, namely 
$\delta \delta $, $\delta \Theta $ and $\Theta \Theta $ contributions. In
all cases for the indices $i$ and $j$, the $\delta \delta $ contribution is
always zero because of kinematics. Indeed, in order for it to be
nonvanishing, the gluon energies must satisfy $\pm \omega _{i}(k)\pm \omega
_{j}(k)=1$, relations always forbidden by the dispersion relations (\ref
{dispersion relations}). The two $\delta \Theta $ contributions are worked
out as follows. For example, we want to get an expression for $\int_{\mu
}^{+\infty }\hspace{-2pt}dk\int_{-\infty }^{+\infty }\hspace{-2pt}d\omega
f(k,\omega ,1-\omega )\,\mathfrak{z}_{i}(k)\beta _{j}\left( k,1-\omega
\right) \left[ \delta \left( \omega -\omega _{i}\right) -\delta \left(
\omega +\omega _{i}\right) \right] \Theta \hspace{-1pt}\left( k-\left\vert
1-\omega \right\vert \right) $. A non-zero contribution must satisfy $\omega
=\pm \omega _{i}(k)$, together with $1-k\leq \omega \leq 1+k$. It is not
difficult to see that only the case $\omega =\omega _{i}(k)$ is allowed, and
this for all values $k\geq \mu $. The integration over $\omega $ is
straightforward and we obtain for our term the result $\int_{\mu }^{+\infty }
\hspace{-2pt}dkf(k,\omega _{i},1-\omega _{i})\,\mathfrak{z}_{i}\,\beta
_{j}\left( k,1-\omega _{i}\right) $. The other $\delta \Theta $ contribution
is obtained in a similar manner and is equal to $\int_{\mu }^{+\infty }
\hspace{-2pt}dkf(k,1-\omega _{j},\omega _{j})\,\mathfrak{z}_{j}\,\beta
_{i}\left( k,1-\omega _{j}\right) $. The $\Theta \Theta $ contribution does
not vanish when $\omega $ satisfies simultaneously $-k\leq \omega \leq k$
and $1-k\leq \omega \leq 1+k$. This amounts to having $k\geq 1/2$ and
$1-k\leq \omega \leq k$. We write then the $\Theta \Theta $ contribution as
$\int_{1/2}^{+\infty }\hspace{-2pt}dk\int_{1-k}^{k}\hspace{-2pt}d\omega
f(k,\omega ,1-\omega )\beta _{i}(k,\omega )\beta _{j}(k,1-\omega )$. Putting
all these intermediary results together, we have: 
\begin{align}
I_{0}& =\int_{\mu }^{+\infty }\hspace{-2pt}dk\left[ f(k,\omega _{i},1-\omega
_{i})\,\mathfrak{z}_{i}\,\beta _{j}\left( k,1-\omega _{i}\right)
+f(k,1-\omega _{j},\omega _{j})\,\mathfrak{z}_{j}\,\beta _{i}\left(
k,1-\omega _{j}\right) \right]  \notag \\
& +\int_{1/2}^{+\infty }\hspace{-2pt}dk\int_{1-k}^{k}\hspace{-2pt}d\omega
f(k,\omega ,1-\omega )\beta _{i}(k,\omega )\beta _{j}(k,1-\omega ).
\label{integral in rho rho-result}
\end{align}

Another accompanying type of integrals that can straightforwardly be deduced
from $I_{0}$ is the following: 
\begin{align}
I_{0}^{\prime }& =\int_{\mu }^{+\infty }\hspace{-2pt}dk\int_{-\infty
}^{+\infty }\hspace{-2pt}d\omega \int_{-\infty }^{+\infty }\hspace{-2pt}
d\omega ^{\prime }\,\delta \left( 1-\omega -\omega ^{\prime }\right)
f(k,\omega ,\omega ^{\prime })\Theta (k-\left\vert \omega \right\vert )\rho
_{j}(k,\omega ^{\prime })  \notag \\
& =\int_{\mu }^{+\infty }\hspace{-2pt}dk\int_{-\infty }^{+\infty }\hspace{
-2pt}d\omega f(k,\omega ,1-\omega )\Theta (k-\left\vert \omega \right\vert
)\rho _{j}(k,\omega ^{\prime })  \notag \\
& =\int_{\mu }^{+\infty }\hspace{-2pt}dkf(k,1-\omega _{j},\omega _{j})\,
\mathfrak{z}_{j}\,+\int_{1/2}^{+\infty }\hspace{-2pt}dk\int_{1-k}^{k}\hspace{
-2pt}d\omega f(k,\omega ,1-\omega )\beta _{j}(k,1-\omega ).
\label{integral in theta rho}
\end{align}
Indeed, it suffices to replace in (\ref{integral in rho rho-result})
$\mathfrak{z}_{j}(k)$ by zero and $\beta _{j}(k,1-\omega )$ by one.

Only integrals of type $I_{0}$ and $I_{0}^{\prime }$ contribute to the
coefficient of zeroth order in $p$ in the damping rate. With the appropriate
functions $f$, the remaining integrals in (\ref{integral in rho rho-result})
and (\ref{integral in theta rho}) are all finite in the infrared, and so
$\mu $ can safely be taken to zero. However, it is important to mention that
these two types of integrals $I_{0}$ and $I_{0}^{\prime }$ do also intervene
in the second-order coefficient in $p$ and, with the corresponding functions 
$f$, \textit{are} infrared sensitive. We will see this explicitly later in
the text.

The second type of integrals we have to deal with is the following: 
\begin{align}
I_{1}& =\int_{\mu }^{+\infty }\hspace{-2pt}dk\int_{-\infty }^{+\infty }
\hspace{-2pt}d\omega \int_{-\infty }^{+\infty }\hspace{-2pt}d\omega ^{\prime
}\,\delta \left( 1-\omega -\omega ^{\prime }\right) f(k,\omega ,\omega
^{\prime })\rho _{i}(k,\omega )\partial _{k}\rho _{j}(k,\omega ^{\prime }) 
\notag \\
& =\int_{\mu }^{+\infty }\hspace{-2pt}dk\int_{-\infty }^{+\infty }\hspace{
-2pt}d\omega f(k,\omega ,1-\omega )\rho _{i}(k,\omega )\partial _{k}\rho
_{j}(k,1-\omega ).  \label{integral in rho d_k rho}
\end{align}
Here too the discussion has to be carried out contribution by contribution,
using the structure of the spectral functions (\ref{definition rho}). The
first contribution to consider is the one that involves two delta functions,
for example $\int_{\mu }^{+\infty }\hspace{-2pt}dk\int_{-\infty }^{+\infty }
\hspace{-2pt}d\omega \,f\,\mathfrak{z}_{i}\,\delta \left( \omega -\omega
_{i}\right) \partial _{k}\left[ \mathfrak{z}_{j}\delta \left( 1-\omega
-\omega _{j}\right) \right] $. When the derivative over $k$ is applied to $
\mathfrak{z}_{j}(k)$, the supports of the two delta functions do not
intersect and we have zero contribution. But this is also true when $
\partial _{k}$ is applied to $\delta \left( 1-\omega -\omega _{j}(k)\right) $
. This is because whatever the result of the action of $\partial _{k}$, the
two delta functions do not vanish only on the intersection of their
respective supports, which is always empty by kinematics. This can be
checked explicitly by regularizing either the derivative over $k$ (i.e.,
replacing it by a finite difference) or the distribution $\delta \left(
1-\omega -\omega _{j}(k)\right) $. In fact, the same argument applies when
we have a second-order derivative over $k$ and, in general, terms involving $
\delta \left( \omega \pm \omega _{i}\right) \partial _{k}^{n}\left[ \delta
\left( 1-\omega \pm \omega _{j}\right) \right] $ do not contribute.

The second contribution in $I_{1}$ to consider is an integral involving $
\delta \hspace{1pt}\partial _{k}\left( \beta \Theta \right) $. Consider for
example $\int_{\mu }^{+\infty }\hspace{-2pt}dk\int_{-\infty }^{+\infty }
\hspace{-2pt}d\omega f(k,\omega ,1-\omega )\,\mathfrak{z}_{i}\,\delta \left(
\omega \mp \omega _{i}\right) \partial _{k}\left[ \beta _{j}\left(
k,1-\omega \right) \Theta \hspace{-1pt}\left( k-\left\vert 1-\omega
\right\vert \right) \right] $. Applying $\partial _{k}$ to $\beta _{j}$ is
not problematic: only the case $\omega =\omega _{i}(k)$ with $k\geq \mu $
contributes and we obtain for this $\int_{\mu }^{+\infty }\hspace{-2pt}
dkf(k,\omega _{i},1-\omega _{i})\,\mathfrak{z}_{i}\left. \partial _{k}\beta
_{j}\left( k,1-\omega \right) \right\vert _{\omega =\omega _{i}}$. Applying $
\partial _{k}$ to $\Theta \hspace{-1pt}\left( k-\left\vert 1-\omega
\right\vert \right) $ gives $\delta \hspace{-1pt}\left( k-\left\vert
1-\omega \right\vert \right) $ and the support of $\delta \left( \omega \mp
\omega _{i}\right) \delta \hspace{-1pt}\left( k-\left\vert 1-\omega
\right\vert \right) $ is $\omega =1$ and $k=0$. Since $k\geq \mu >0$, this
part of the contribution is zero.

There is another contribution that involves a $\delta $ and a $\Theta $,
namely $\int_{\mu }^{+\infty }\hspace{-2pt}dk\int_{-\infty }^{+\infty }
\hspace{-2pt}d\omega f(k,\omega ,1-\omega )\,\beta _{i}\left( k,\omega
\right) \Theta \hspace{-1pt}\left( k-\left\vert \omega \right\vert \right)
\partial _{k}\left( \mathfrak{z}_{j}\,\delta \left( 1-\omega \mp \omega
_{j}\right) \right) $. We first apply the derivative over $k$ to $\mathfrak{z
}_{i}$ and obtain the piece $\int_{\mu }^{+\infty }\hspace{-2pt}
dkf(k,1-\omega _{j},\omega _{j})\,\mathfrak{z}_{j}^{\prime }\,\beta
_{i}\left( k,1-\omega _{j}\right) $, where $\mathfrak{z}_{j}^{\prime }$
stands for $d\mathfrak{z}_{j}(k)/dk$. We then apply it to $\delta \left(
1-\omega \mp \omega _{j}\right) $. We use the standard rules regulating the
handling of the delta distribution and we always check the results by
regularizing either the derivative $\partial _{k}$ or the delta function
itself. Only $\omega =1-\omega _{j}$ contributes and we get the piece $
\int_{\mu }^{+\infty }\hspace{-2pt}dk\omega _{j}^{\prime }\mathfrak{z}
_{j}\left. \partial _{\omega }\left[ f(k,1-\omega ,\omega )\,\beta
_{i}\left( k,1-\omega \right) \right] \right\vert _{\omega =\omega _{j}}$.

There is one last contribution that we have to consider in $I_{1}$, namely
the double integral $\int_{\mu }^{+\infty }\hspace{-2pt}dk\int_{-\infty
}^{+\infty }\hspace{-2pt}d\omega f(k,\omega ,1-\omega )\,\beta _{i}\left(
k,\omega \right) \Theta \hspace{-1pt}\left( k-\left\vert \omega \right\vert
\right) \partial _{k}\left[ \beta _{j}\left( k,1-\omega \right) \Theta 
\hspace{-1pt}\left( k-\left\vert 1-\omega \right\vert \right) \right] $. The
application of $\partial _{k}$ to $\beta _{j}\left( k,1-\omega \right) $
gives $\int_{1/2}^{+\infty }\hspace{-2pt}dk\int_{1-k}^{k}\hspace{-2pt}
d\omega f(k,\omega ,1-\omega )\,\beta _{i}\left( k,\omega \right) \partial
_{k}\beta _{j}\left( k,1-\omega \right) $ and its application to $\Theta 
\hspace{-1pt}\left( k-\left\vert 1-\omega \right\vert \right) $ gives zero
because of kinematics, as before. Putting all the above contributions
together, we obtain the result: 
\begin{align}
I_{1}& =\int_{\mu }^{+\infty }\hspace{-2pt}dk\left[ f(k,\omega _{i},1-\omega
_{i})\,\mathfrak{z}_{i}\left. \partial _{k}\beta _{j}\left( k,1-\omega
\right) \right\vert _{\omega =\omega _{i}}+f(k,1-\omega _{j},\omega _{j})\,
\mathfrak{z}_{j}^{\prime }\,\beta _{i}\left( k,1-\omega _{j}\right) \right. 
\notag \\
& \hspace{0.6in}\left. +\omega _{j}^{\prime }\,\mathfrak{z}_{j}\left.
\partial _{\omega }\left[ f(k,1-\omega ,\omega )\,\beta _{i}\left(
k,1-\omega \right) \right] \right\vert _{\omega =\omega _{j}}\right]  \notag
\\
& +\int_{1/2}^{+\infty }\hspace{-2pt}dk\int_{1-k}^{k}\hspace{-2pt}d\omega
f(k,\omega ,1-\omega )\,\beta _{i}\left( k,\omega \right) \partial _{k}\beta
_{j}\left( k,1-\omega \right) .  \label{integral in rho d_k rho-result}
\end{align}

Like for $I_{0}$, there is here too an accompanying integral $I_{1}^{\prime
} $ for $I_{1}$ which can be deduced straightforwardly from (\ref{integral
in rho d_k rho-result}) by replacing $\mathfrak{z}_{i}$ by zero and $\beta
_{i}\left( k,\omega \right) $ by one: 
\begin{align}
I_{1}^{\prime }& =\int_{\mu }^{+\infty }\hspace{-2pt}dk\int_{-\infty
}^{+\infty }\hspace{-2pt}d\omega \int_{-\infty }^{+\infty }\hspace{-2pt}
d\omega ^{\prime }\,\delta \left( 1-\omega -\omega ^{\prime }\right)
f(k,\omega ,\omega ^{\prime })\Theta (k-\left\vert \omega \right\vert
)\partial _{k}\rho _{j}(k,\omega ^{\prime })  \notag \\
& =\int_{\mu }^{+\infty }\hspace{-2pt}dk\int_{-\infty }^{+\infty }\hspace{
-2pt}d\omega f(k,\omega ,1-\omega )\Theta (k-\left\vert \omega \right\vert
)\partial _{k}\rho _{j}(k,1-\omega )  \notag \\
& =\int_{\mu }^{+\infty }\hspace{-2pt}dk\left[ f(k,1-\omega _{j},\omega
_{j})\,\mathfrak{z}_{j}^{\prime }\,+\omega _{j}^{\prime }\,\mathfrak{z}
_{j}\left. \partial _{\omega }\left[ f(k,1-\omega ,\omega )\,\right]
\right\vert _{\omega =\omega _{j}}\right]  \notag \\
& +\int_{1/2}^{+\infty }\hspace{-2pt}dk\int_{1-k}^{k}\hspace{-2pt}d\omega
f(k,\omega ,1-\omega )\partial _{k}\beta _{j}\left( k,1-\omega \right) .
\label{integral in theta d_k rho}
\end{align}

The third type of integrals we have to deal with is one that involves a
second derivative in $k$: 
\begin{align}
I_{2}& =\int_{\mu }^{+\infty }\hspace{-2pt}dk\int_{-\infty }^{+\infty }
\hspace{-2pt}d\omega \int_{-\infty }^{+\infty }\hspace{-2pt}d\omega ^{\prime
}\,\delta \left( 1-\omega -\omega ^{\prime }\right) f(k,\omega ,\omega
^{\prime })\rho _{i}(k,\omega )\partial _{k}^{2}\rho _{j}(k,\omega ^{\prime
})  \notag \\
& =\int_{\mu }^{+\infty }\hspace{-2pt}dk\int_{-\infty }^{+\infty }\hspace{
-2pt}d\omega f(k,\omega ,1-\omega )\rho _{i}(k,\omega )\partial _{k}^{2}\rho
_{j}(k,1-\omega ).  \label{integral in rho d_k2 rho}
\end{align}
The steps to treat the different contributions parallel those followed for $
I_{1}$. As explained before, the $\delta \delta $ contribution is zero
because of kinematics. Therefore, the first contribution to look at is $
\int_{\mu }^{+\infty }\hspace{-2pt}dk\int_{-\infty }^{+\infty }\hspace{-2pt}
d\omega f(k,\omega ,1-\omega )\,\mathfrak{z}_{i}\,\delta \left( \omega \mp
\omega _{i}\right) \partial _{k}^{2}\left[ \beta _{j}\left( k,1-\omega
\right) \Theta \hspace{-1pt}\left( k-\left\vert 1-\omega \right\vert \right) 
\right] $. We know that $\partial _{k}^{2}\left( \beta _{j}\Theta \hspace{
-1pt}\right) =\partial _{k}^{2}\beta _{j}\,\Theta \hspace{-1pt}+2\partial
_{k}\beta _{j}\,\partial _{k}\Theta +\beta _{j}\,\partial _{k}^{2}\Theta $.
The first term $\partial _{k}^{2}\beta _{j}\,\Theta \hspace{-1pt}$ yields
simply $\int_{\mu }^{+\infty }\hspace{-2pt}dkf(k,\omega _{i},1-\omega _{i})\,
\mathfrak{z}_{i}\left. \partial _{k}^{2}\beta _{j}\left( k,1-\omega \right)
\right\vert _{\omega =\omega _{i}}$. The second term $2\partial _{k}\beta
_{j}\,\partial _{k}\Theta $ yields zero because the kinematics imposes $k=0$
, which is excluded from the integration region as explained. The same is
true for the term $\beta _{j}\,\partial _{k}^{2}\Theta =\beta _{j}\,\partial
_{k}\delta \left( k-\left\vert 1-\omega \right\vert \right) $ because here
too the support of the delta functions forces $\omega =1$ and $k=0$, which
is excluded. The next contribution is $\int_{\mu }^{+\infty }\hspace{-2pt}
dk\int_{-\infty }^{+\infty }\hspace{-2pt}d\omega f(k,\omega ,1-\omega
)\,\beta _{i}\left( k,\omega \right) \Theta \hspace{-1pt}\left( k-\left\vert
\omega \right\vert \right) \partial _{k}^{2}\left[ \mathfrak{z}_{j}\,\delta
\left( 1-\omega \mp \omega _{j}\right) \right] $. We have $\partial
_{k}^{2}\left( \mathfrak{z}_{j}\,\delta \right) =\mathfrak{z}_{j}^{\prime
\prime }\,\delta +\mathfrak{z}_{j}^{\prime }\,\partial _{k}\delta +\mathfrak{
z}_{j}\,\partial _{k}^{2}\delta $, where $\mathfrak{z}_{j}^{\prime \prime }$
stands for $d^{2}\mathfrak{z}_{j}/dk^{2}$. The term $\mathfrak{z}
_{j}^{\prime \prime }\,\delta $ yields $\int_{\mu }^{+\infty }\hspace{-2pt}
dkf(k,1-\omega _{j},\omega _{j})\,\mathfrak{z}_{j}^{\prime \prime }\beta
_{i}\left( k,1-\omega _{j}\right) $. The two other terms $\mathfrak{z}
_{j}^{\prime }\,\partial _{k}\delta +\mathfrak{z}_{j}\,\partial
_{k}^{2}\delta $ yield together $\int_{\mu }^{+\infty }\hspace{-2pt}dk\left. 
\left[ \left( \mathfrak{z}_{j}\omega _{j}^{\prime \prime }+2\mathfrak{z}
_{j}^{\prime }\omega _{j}^{\prime }\right) \partial _{\omega }+\mathfrak{z}
_{j}\omega _{j}^{\prime 2}\partial _{\omega }^{2}\right] \left[ f(k,1-\omega
,\omega )\,\beta _{i}\left( k,1-\omega \right) \right] \right\vert _{\omega
=\omega _{j}}$. The procedure is to replace $\partial _{k}$ with $\omega
_{j}^{\prime }\partial _{\omega }$ and apply the usual rules regulating the
handling of the delta distribution. Also, integrands that force upon us the
condition $k=0$ are systematically excluded. It remains to look at $
\int_{\mu }^{+\infty }\hspace{-2pt}dk\int_{-\infty }^{+\infty }\hspace{-2pt}
d\omega f(k,\omega ,1-\omega )\,\beta _{i}\left( k,\omega \right) \Theta 
\hspace{-1pt}\left( k-\left\vert \omega \right\vert \right) \partial _{k}^{2}
\left[ \beta _{j}\left( k,1-\omega \right) \Theta \hspace{-1pt}\left(
k-\left\vert 1-\omega \right\vert \right) \right] $. The first term $
\partial _{k}^{2}\beta _{j}\,\Theta \hspace{-1pt}$ yields the piece $
\int_{1/2}^{+\infty }\hspace{-2pt}dk\int_{1-k}^{k}\hspace{-2pt}d\omega
f(k,\omega ,1-\omega )\,\beta _{i}\left( k,\omega \right) \partial
_{k}^{2}\beta _{j}\left( k,1-\omega \right) $ and the second term $2\partial
_{k}\beta _{j}\,\delta $ the piece $2\int_{1/2}^{+\infty }\hspace{-2pt}
dkf(k,1-k,k)\,\beta _{i}\left( k,1-k\right) \left. \partial _{k}\beta
_{j}\left( k,\omega \right) \right\vert _{\omega =k}$, both worked out as
previous similar terms. The third term in this contribution $\int_{\mu
}^{+\infty }\hspace{-2pt}dk\int_{-\infty }^{+\infty }\hspace{-2pt}d\omega
f\,\beta _{i}\beta _{j}\Theta \hspace{-1pt}\partial _{k}\delta =\int_{\mu
}^{+\infty }\hspace{-2pt}dk\int_{-\infty }^{+\infty }\hspace{-2pt}d\omega 
\left[ \partial _{k}\left( f\,\beta _{i}\beta _{j}\Theta \hspace{-1pt}\delta
\right) -\partial _{k}\left( f\,\beta _{i}\right) \beta _{j}\Theta \delta
-f\,\beta _{i}\partial _{k}\beta _{j}\Theta \delta -f\,\beta _{i}\beta
_{j}\delta \hspace{-1pt}\delta \right] $. The support of the $\Theta \hspace{
-1pt}\delta $ distribution is $\omega =1-k$ and $k\geq 1/2$ and that of the $
\delta \hspace{-1pt}\delta $ distribution the point $\omega =k=1/2$. Because 
$\beta _{j}(k,k)=0$ for every $k$, the terms $-\partial _{k}\left( f\,\beta
_{i}\right) \beta _{j}\Theta \delta -f\,\beta _{i}\beta _{j}\delta \hspace{
-1pt}\delta $ are identically equal to zero. The term $-f\,\beta
_{i}\partial _{k}\beta _{j}\Theta \delta $ gives the contribution $
-\int_{1/2}^{+\infty }\hspace{-2pt}dkf(k,1-k,k)\,\beta _{i}\left(
k,1-k\right) \left. \partial _{k}\beta _{j}\left( k,\omega \right)
\right\vert _{\omega =k}$. Generally, the term $\partial _{k}\left( f\,\beta
_{i}\beta _{j}\Theta \hspace{-1pt}\delta \right) $ would be dismissed as a
total derivative. This would be straightforward if we are allowed to
integrate first over $k$ than over $\omega $. But the permutation of the two
integrations is not always permissible and care must be taken; each case has
to be treated individually. Here, one can show that this term gives indeed
zero contribution by either regularizing the delta function or replacing the
derivative by a finite difference. For example, if we use the second method,
we would have: 
\begin{align*}
& \hspace{-1in}\int_{\mu }^{+\infty }\hspace{-2pt}dk\int_{-\infty }^{+\infty
}\hspace{-2pt}d\omega \partial _{k}\left[ g(k,\omega )\Theta \left(
k-\left\vert \omega \right\vert \right) \hspace{-1pt}\delta \left(
k-\left\vert 1-\omega \right\vert \right) \right] \\
& \simeq 1/\epsilon \int_{\mu }^{+\infty }\hspace{-2pt}dk\int_{-\infty
}^{+\infty }\hspace{-2pt}d\omega \left[ g(k^{\prime },\omega )\Theta \left(
k^{\prime }-\left\vert \omega \right\vert \right) \hspace{-1pt}\delta \left(
k^{\prime }-\left\vert 1-\omega \right\vert \right) \right] _{k^{\prime
}=k+\epsilon } \\
& \hspace{0.5in}-1/\epsilon \int_{\mu }^{+\infty }\hspace{-2pt}
dk\int_{-\infty }^{+\infty }\hspace{-2pt}d\omega g(k,\omega )\Theta \left(
k-\left\vert \omega \right\vert \right) \hspace{-1pt}\delta \left(
k-\left\vert 1-\omega \right\vert \right)
\end{align*}
where $\epsilon \rightarrow 0$ and $g(k,\omega )$ stands for $f\,\beta
_{i}\beta _{j}$. This difference is equal to $1/\epsilon \int_{1/2}^{+\infty
}\hspace{-2pt}dk^{\prime }g(k^{\prime },1-k^{\prime })-1/\epsilon
\int_{1/2}^{+\infty }\hspace{-2pt}dkg(k,1-k)=0$. Putting all the above
results together, we obtain: 
\begin{align}
I_{2}& =\int_{\mu }^{+\infty }\hspace{-2pt}dk\left[ f(k,\omega _{i},1-\omega
_{i})\,\mathfrak{z}_{i}\left. \partial _{k}^{2}\beta _{j}\left( k,1-\omega
\right) \right\vert _{\omega =\omega _{i}}+f(k,1-\omega _{j},\omega _{j})\,
\mathfrak{z}_{j}^{\prime \prime }\beta _{i}\left( k,1-\omega _{j}\right)
\right.  \notag \\
& \left. +\left. \left[ \left( \mathfrak{z}_{j}\omega _{j}^{\prime \prime }+2
\mathfrak{z}_{j}^{\prime }\omega _{j}^{\prime }\right) \partial _{\omega }+
\mathfrak{z}_{j}\omega _{j}^{\prime 2}\partial _{\omega }^{2}\right] \left[
f(k,1-\omega ,\omega )\,\beta _{i}\left( k,1-\omega \right) \right]
\right\vert _{\omega =\omega _{j}}\right]  \notag \\
& +\int_{1/2}^{+\infty }\hspace{-2pt}dkf(k,1-k,k)\,\beta _{i}\left(
k,1-k\right) \left. \partial _{k}\beta _{j}\left( k,\omega \right)
\right\vert _{\omega =k}  \notag \\
& +\int_{1/2}^{+\infty }\hspace{-2pt}dk\int_{1-k}^{k}\hspace{-2pt}d\omega
f(k,\omega ,1-\omega )\,\beta _{i}\left( k,\omega \right) \partial
_{k}^{2}\beta _{j}\left( k,1-\omega \right) .
\label{integral in rho d_k^2 rho-result}
\end{align}

As for the two types $I_{0}$ and $I_{1}$, we write for the accompanying
integral $I_{2}^{\prime }$: 
\begin{align}
I_{2}^{\prime }& =\int_{\mu }^{+\infty }\hspace{-2pt}dk\int_{-\infty
}^{+\infty }\hspace{-2pt}d\omega \int_{-\infty }^{+\infty }\hspace{-2pt}
d\omega ^{\prime }\,\delta \left( 1-\omega -\omega ^{\prime }\right)
f(k,\omega ,\omega ^{\prime })\Theta (k-\left\vert \omega \right\vert
)\partial _{k}^{2}\rho _{j}(k,\omega ^{\prime })  \notag \\
& =\int_{\mu }^{+\infty }\hspace{-2pt}dk\int_{-\infty }^{+\infty }\hspace{
-2pt}d\omega f(k,\omega ,1-\omega )\Theta (k-\left\vert \omega \right\vert
)\partial _{k}^{2}\rho _{j}(k,1-\omega ).  \notag \\
& =\int_{\mu }^{+\infty }\hspace{-2pt}dk\left[ f(k,1-\omega _{j},\omega
_{j})\,\mathfrak{z}_{j}^{\prime \prime }+\left. \left( \left( \mathfrak{z}
_{j}\omega _{j}^{\prime \prime }+2\mathfrak{z}_{j}^{\prime }\omega
_{j}^{\prime }\right) \partial _{\omega }+\mathfrak{z}_{j}\omega
_{j}^{\prime 2}\partial _{\omega }^{2}\right) f(k,1-\omega ,\omega
)\,\right\vert _{\omega =\omega _{j}}\right]  \notag \\
& +\int_{1/2}^{+\infty }\hspace{-2pt}\hspace{-2pt}dkf(k,\hspace{-1pt}1
\hspace{-1pt}-\hspace{-1pt}k,\hspace{-1pt}k)\,\hspace{-6pt}\left. \partial
_{k}\beta _{j}\hspace{-2pt}\left( k,\omega \right) \right\vert _{\omega
=k}+\int_{1/2}^{+\infty }\hspace{-2pt}\hspace{-2pt}dk\hspace{-3pt}
\int_{1-k}^{k}\hspace{-2pt}\hspace{-2pt}d\omega f(k,\omega ,1-\omega )\,
\hspace{-3pt}\partial _{k}^{2}\beta _{j}\hspace{-2pt}\left( k,1-\omega
\right) .  \label{integral in theta d_k^2 rho}
\end{align}
Before carrying through, it is important to note that whereas $\beta _{j}
\hspace{-2pt}\left( k,k\right) =0$ for all $k$, the partial derivative $
\left. \partial _{k}\beta _{j}\hspace{-2pt}\left( k,\omega \right)
\right\vert _{\omega =k}$ present in (\ref{integral in rho d_k^2 rho-result}
) and (\ref{integral in theta d_k^2 rho}) is infinite. This divergence comes
from light-cone momenta and a more appropriate notation would be $\left.
\partial _{k}\beta _{j}\hspace{-2pt}\left( k,\omega \right) \right\vert
_{\omega \rightarrow k}$. Such terms would spoil the calculation if we were
not fortunate to have them cancel with similar terms hidden in the\ last
contribution in (\ref{integral in rho d_k^2 rho-result}) and (\ref{integral
in theta d_k^2 rho}). This will be shown later.

The next type of integrals we must handle is: 
\begin{align}
I_{3}& =\int_{\mu }^{+\infty }\hspace{-2pt}dk\int_{-\infty }^{+\infty }
\hspace{-2pt}d\omega \int_{-\infty }^{+\infty }\hspace{-2pt}d\omega ^{\prime
}\,\partial _{\omega }\delta \left( 1-\omega -\omega ^{\prime }\right)
f(k,\omega ,\omega ^{\prime })\rho _{i}(k,\omega )\rho _{j}(k,\omega
^{\prime })  \notag \\
& =-\int_{\mu }^{+\infty }\hspace{-2pt}dk\int_{-\infty }^{+\infty }\hspace{
-2pt}d\omega \,\left. \partial _{\omega }\left[ f(k,\omega ,\omega ^{\prime
})\rho _{i}(k,\omega )\right] \right\vert _{\omega ^{\prime }=1-\omega }\rho
_{j}(k,1-\omega ).  \label{integral in rho rho d_omega delta}
\end{align}
Here too the residue-residue contribution is zero. The residue-cut
contribution is equal to $-\int_{\mu }^{+\infty }\hspace{-2pt}
dk\int_{-\infty }^{+\infty }\hspace{-2pt}d\omega \mathfrak{z}_{i}\partial
_{\omega }\left. \left[ f(k,\omega ,\omega ^{\prime })\delta \left( \omega
\mp \omega _{i}\right) \beta _{j}\left( k,\omega ^{\prime }\right) \Theta 
\hspace{-1pt}\left( k-\left\vert \omega ^{\prime }\right\vert \right) \right]
\right\vert _{\omega ^{\prime }=1-\omega }$. Actually, the easiest way to
handle this contribution is to reconsider in (\ref{integral in rho rho
d_omega delta}) a derivative with respect to $\omega ^{\prime }$ instead of $
\omega $, i.e., replace $\partial _{\omega }\delta \left( 1-\omega -\omega
^{\prime }\right) $ by $\partial _{\omega ^{\prime }}\delta \left( 1-\omega
-\omega ^{\prime }\right) $. Then, the residue-cut contribution would write: 
\begin{align*}
& \hspace{-1.3in}-\int_{\mu }^{+\infty }\hspace{-2pt}dk\int_{-\infty
}^{+\infty }\hspace{-2pt}d\omega \mathfrak{z}_{i}\delta \left( \omega \mp
\omega _{i}\right) \partial _{\omega ^{\prime }}\left. \left[ f(k,\omega
,\omega ^{\prime })\beta _{j}\left( k,\omega ^{\prime }\right) \Theta 
\hspace{-1pt}\left( k-\left\vert \omega ^{\prime }\right\vert \right) \right]
\right\vert _{\omega ^{\prime }=1-\omega } \\
& =-\int_{\mu }^{+\infty }\hspace{-2pt}dk\mathfrak{z}_{i}\left. \partial
_{\omega }\left[ f(k,\omega _{i},\omega )\beta _{j}(k,\omega )\right]
\right\vert _{\omega =1-\omega _{i}},
\end{align*}
where we have implicitly used the usual supports of the delta and theta
functions. The cut-residue contribution is worked out in a similar way and
found to be equal to $-\int_{\mu }^{+\infty }\hspace{-2pt}dk\mathfrak{z}
_{j}\left. \partial _{\omega }\left[ f(k,\omega ,\omega _{j})\beta
_{i}(k,\omega )\right] \right\vert _{\omega =1-\omega _{j}}$. The\ last
contribution, the cut-cut term, writes $-\int_{\mu }^{+\infty }\hspace{-2pt}
dk\int_{-\infty }^{+\infty }\hspace{-2pt}d\omega \partial _{\omega }\left. 
\left[ f(k,\omega ,\omega ^{\prime })\beta _{i}\left( k,\omega \right) \beta
_{j}\left( k,\omega ^{\prime }\right) \Theta \hspace{-1pt}\left(
k-\left\vert \omega \right\vert \right) \Theta \hspace{-1pt}\left(
k-\left\vert \omega ^{\prime }\right\vert \right) \right] \right\vert
_{\omega ^{\prime }=1-\omega }$. Applying the derivative over $\omega $ on $
\Theta \Theta $ implies that either $\omega =k$ or $\omega ^{\prime }=k$. In
both cases $\beta \beta =0$, which means no contribution from here. Hence
the cut-cut contribution is simply equal to $-\int_{1/2}^{+\infty }\hspace{
-2pt}dk\int_{1-k}^{k}\hspace{-2pt}d\omega \beta _{i}\left( k,\omega \right)
\partial _{\omega ^{\prime }}\left. \left[ f(k,\omega ,\omega ^{\prime
})\beta _{j}\left( k,\omega ^{\prime }\right) \right] \right\vert _{\omega
^{\prime }=1-\omega }$. So we get: 
\begin{align}
I_{3}& =-\int_{\mu }^{+\infty }\hspace{-2pt}dk\left[ \mathfrak{z}_{i}\left.
\partial _{\omega }\left[ f(k,\omega _{i},\omega )\beta _{j}(k,\omega )
\right] \right\vert _{\omega =1-\omega _{i}}+\mathfrak{z}_{j}\left. \partial
_{\omega }\left[ f(k,\omega ,\omega _{j})\beta _{i}(k,\omega )\right]
\right\vert _{\omega =1-\omega _{j}}\right]  \notag \\
& -\int_{1/2}^{+\infty }\hspace{-2pt}dk\int_{1-k}^{k}\hspace{-2pt}d\omega
\beta _{i}\left( k,\omega \right) \partial _{\omega ^{\prime }}\left. \left[
f(k,\omega ,\omega ^{\prime })\beta _{j}\left( k,\omega ^{\prime }\right) 
\right] \right\vert _{\omega ^{\prime }=1-\omega }\text{.}
\label{integral in rho rho d_omega delta-result}
\end{align}
The accompanying integral $I_{3}^{\prime }$ where $\rho _{i}(k,\omega )$ is
replaced by $\Theta (k-\left\vert \omega \right\vert )$ is obtained from (
\ref{integral in rho rho d_omega delta-result}) as usual. We have: 
\begin{align}
I_{3}^{\prime }& =\int_{\mu }^{+\infty }\hspace{-2pt}dk\int_{-\infty
}^{+\infty }\hspace{-2pt}d\omega \int_{-\infty }^{+\infty }\hspace{-2pt}
d\omega ^{\prime }\,\partial _{\omega }\delta \left( 1-\omega -\omega
^{\prime }\right) f(k,\omega ,\omega ^{\prime })\Theta (k-\left\vert \omega
\right\vert )\rho _{j}(k,\omega ^{\prime })  \notag \\
& =-\int_{\mu }^{+\infty }\hspace{-2pt}dk\int_{-\infty }^{+\infty }\hspace{
-2pt}d\omega \,\left. \partial _{\omega }\left[ f(k,\omega ,\omega ^{\prime
})\Theta (k-\left\vert \omega \right\vert )\right] \right\vert _{\omega
^{\prime }=1-\omega }\rho _{j}(k,1-\omega )  \notag \\
& =-\int_{\mu }^{+\infty }\hspace{-2pt}dk\mathfrak{z}_{j}\left. \partial
_{\omega }f(k,\omega ,\omega _{j})\right\vert _{\omega =1-\omega _{j}}
\hspace{-2pt}-\hspace{-2pt}\int_{1/2}^{+\infty }\hspace{-2pt}dk\int_{1-k}^{k}
\hspace{-2pt}d\omega \partial _{\omega ^{\prime }}\left. \left[ f(k,\omega
,\omega ^{\prime })\beta _{j}\left( k,\omega ^{\prime }\right) \right]
\right\vert _{\omega ^{\prime }=1-\omega }.
\label{integral in theta rho d_omega delta}
\end{align}

The next type of integrals we have to deal with is: 
\begin{align}
I_{4}& =\int_{\mu }^{+\infty }\hspace{-2pt}dk\int_{-\infty }^{+\infty }
\hspace{-2pt}d\omega \int_{-\infty }^{+\infty }\hspace{-2pt}d\omega ^{\prime
}\,\delta \left( 1-\omega -\omega ^{\prime }\right) f(k,\omega ,\omega
^{\prime })\delta (k^{2}-\omega ^{2})\rho _{j}(k,\omega ^{\prime })  \notag
\\
& =\int_{\mu }^{+\infty }\hspace{-2pt}dk\int_{-\infty }^{+\infty }\hspace{
-2pt}d\omega f(k,\omega ,1-\omega )\delta (k^{2}-\omega ^{2})\rho
_{j}(k,1-\omega ).  \label{integral in delta rho}
\end{align}
It is not difficult to work it out. Using (\ref{definition rho}), the
delta-delta contribution has zero support and hence vanishes. The
delta-theta contribution is easily worked out along the usual lines and we
obtain: 
\begin{equation}
I_{4}=\int_{1/2}^{+\infty }\hspace{-2pt}\frac{dk}{2k}f(k,k,1-k)\beta
_{j}(k,1-k).  \label{integral in delta rho-result}
\end{equation}

There remains one last type of integrals to look at and which too is not
difficult to manipulate. It is: 
\begin{align}
I_{5}& =\int_{\mu }^{+\infty }\hspace{-2pt}dk\int_{-\infty }^{+\infty }
\hspace{-2pt}d\omega \int_{-\infty }^{+\infty }\hspace{-2pt}d\omega ^{\prime
}\,\delta \left( 1-\omega -\omega ^{\prime }\right) f(k,\omega ,\omega
^{\prime })\partial _{\omega }\left[ \delta (k-\omega )-\delta (k+\omega )
\right] \rho _{j}(k,\omega ^{\prime })  \notag \\
& =\int_{\mu }^{+\infty }\hspace{-2pt}dk\int_{-\infty }^{+\infty }\hspace{
-2pt}d\omega f(k,\omega ,1-\omega )\partial _{\omega }\left[ \delta
(k-\omega )-\delta (k+\omega )\right] \rho _{j}(k,1-\omega ).
\label{integral in d_omega delta rho}
\end{align}
Here too the delta-delta contribution is zero because of kinematics. The
delta-theta contribution is treated by turning the derivative onto $f\beta
_{j}\Theta $. Manipulations by now familiar yield: 
\begin{equation}
I_{5}=-\int_{1/2}^{+\infty }\hspace{-2pt}dk\left. \partial _{\omega }\left[
f(k,\omega ,1-\omega )\beta _{j}(k,1-\omega )\right] \right\vert _{\omega
=k}.  \label{integral in d_omega delta rho-result}
\end{equation}

Every integral in the expressions (\ref{Im pi_ll-final 1}), (\ref{Im
pi_lt+tl--final 1}), (\ref{Im pi_tt-final 1}), (\ref{Im pi_l--final 1}) and (
\ref{Im pi_t--final 1}) falls into one of these above types. We are
therefore ready to perform them. However, as we mentioned more than once in
the text, there is occurrence of infrared sensitivity to be aware of and
which we need to treat with care. There are actually two types of
divergences: infrared and light-cone. The first type will persist in the
final result and will be expressed in terms of the magnetic mass $\mu $ we
introduced from the start as a physical infrared regulator. The second type
will not persist as we will see in the sequel.

\subsection{Infrared and light-cone behavior}

It is best to work out few sample examples in some detail, which all come
from the coefficient of $p^{2}$ in the sum (\ref{Im pi_t--contributions}).\
Let us start with the $\rho _{l}\rho _{l}^{\prime }$ contribution. Up to the
factor $g^{2}N_{c}T/120\pi $ for which there is no need writing explicitly,
the corresponding $f$ function is $f(k,\omega ,\omega ^{\prime
})=3k^{2}\left( 3+44k^{2}-12\omega \omega ^{\prime }\right) /2\omega \omega
^{\prime }$. This integral is of type $I_{0}$. From (\ref{integral in rho
rho-result}),\ we can expect infrared divergences coming from the two $
\mathfrak{z}\beta $ terms, which in fact are equal because here $f(k,\omega
,\omega ^{\prime })$ is symmetric in $\omega $ and $\omega ^{\prime }$.
Using the small-$k$ expansions of on-shell energies, the residue and cut
functions given in appendix A, we have the following small-$k$ behavior: 
\begin{align}
f\left( k,1\hspace{-1pt}-\hspace{-1pt}\omega _{l},\omega _{l}\right) 
\mathfrak{z}_{l}\beta _{l}(k,1\hspace{-1pt}-\hspace{-1pt}\omega _{l})& =
\frac{3}{8k}+\left[ \frac{2217}{400}-\frac{27}{3200}\pi ^{2}\right] k  \notag
\\
& -\hspace{-1pt}\left[ \frac{10230949}{1680000}+\frac{134397}{1120000}\pi
^{2}-\frac{243}{1280000}\pi ^{4}\right] \hspace{-2pt}\allowbreak
k^{3}+O(k^{5})\hspace{1pt},  \label{small k for fzb for ll}
\end{align}
which indicates a logarithmic divergence. The corresponding integral can be
written as follows: 
\begin{align}
2\int_{\mu }^{+\infty }dkf\left( k,1-\omega _{l},\omega _{l}\right) 
\mathfrak{z}_{l}(k)\beta _{l}(k,1-\omega _{l})& =\frac{6}{8}\int_{\mu }^{1}
\frac{dk}{k}+\frac{6}{8}\ln k_{l}(\ell)  \notag \\
& \hspace{-1in}+2\int_{0}^{k_{l}(\ell)}dk\left[ f\left( k,1-\omega
_{l},\omega _{l}\right) \mathfrak{z}_{l}(k)\beta _{l}(k,1-\omega _{l})-\frac{
3}{8k}\right]  \notag \\
& \hspace{-1in}+2\int_{\ell}^{1}dxk_{l}^{\prime }(x)\left. f\left(
k,1-\omega _{l},\omega _{l}\right) \mathfrak{z}_{l}(k)\beta _{l}(k,1-\omega
_{l})\right\vert _{k=k_{l}(x)}  \label{integral fzb in ll--1st expression}
\end{align}
What we did is this. We split the original integral in two: one from $\mu $
to a finite value plus one other from that finite value to $+\infty $. The
second integral is finite, and it is (numerically) better for it to change
the integration variable from $k$ to $x\equiv k/\omega _{l}(k)$, which
implies that $k\equiv k_{l}(x)$. The finite value in question that splits
the original integral is chosen in terms of $x$ instead of $k$ and is
denoted by $\ell$. In the integral from $\mu $ to $k_{l}(\ell)$, we subtract 
$6/8k$ from the integrand, which renders it safe in the infrared and so we
can take the lower boundary $\mu \rightarrow 0$. We must add of course the
contribution $\frac{6}{8}\int_{\mu }^{k_{l}(\ell)}\frac{dk}{k}=\frac{6}{8}
\int_{\mu }^{1}\frac{dk}{k}+\frac{6}{8}\ln k_{l}(\ell)$. The finite value $
\ell$ must be chosen small enough in order to make the integral $
\int_{0}^{k_{l}(\ell)}dk\left( f\mathfrak{z}_{l}\beta _{l}-\frac{3}{8k}
\right) $ \textit{numerically} feasible. Indeed, though we are assured of
its finiteness \textit{analytically}, both integrands $f\mathfrak{z}
_{l}\beta _{l}$ and $\frac{3}{8k}$ are still each (here logarithmically)
divergent. However, for a small value of $\ell$, say $\ell=0.1$, we can use
an expansion of $\left( f\mathfrak{z}_{l}\beta _{l}-\frac{3}{8k}\right) $ in
powers of $k$ in order to get a number for the integral. We have pushed the
expansion to $O(k^{15})$ and we start having good convergence for already $
\ell=0.6$. Also, we must (and do) check that the final result does not
depend on a particular value of $\ell$. We finally get: 
\begin{equation}
2\int_{\mu }^{+\infty }dkf\left( k,1-\omega _{l},\omega _{l}\right) 
\mathfrak{z}_{l}(k)\beta _{l}(k,1-\omega _{l})=-\frac{6}{8}\ln \mu
+\allowbreak 5.\allowbreak 46353171.  \label{integral fzb in ll--final}
\end{equation}

It is important to note that the infrared divergence in (\ref{integral fzb
in ll--final}) is not coming from an eventual expansion around ultrasoft $p$
of the dressed propagator $^{\ast }\Delta (P-K)$ or any similar
distribution. Actually, as mentioned in the paragraph after (\ref{integral
in theta rho}), a $\rho _{l}\rho _{l}^{\prime }$ term is present in the
zeroth-order coefficient of the damping rate \cite{BPgamt}, but with a
corresponding function $f$ going sufficiently fast to zero so that the
divergent behavior of $\mathfrak{z}_{l}(k)\beta _{l}(k,1-\omega _{l},)\sim
-1/40k+O(k)$ is neutralized.\ What happens here is that our function $
f\left( k,1-\omega _{l},\omega _{l}\right) \sim -15+O(k^{2})$, a behavior
not fast enough to screen the infrared divergence. Therefore, it is clear
that it is not the expansion of quantities similar to the dressed
propagators $^{\ast }\Delta (P-K)$ around ultrasoft $p$ that is sole
responsible for the occurrence of infrared divergences as one might think.
Note also that the same thing happens to the other contributions to the
imaginary part of the one-loop HTL-dressed self-energy ($tl$, $tt$, etc.).

Returning to the integral we are handling, the corresponding double integral
is $\int_{1/2}^{+\infty }dk\int_{1-k}^{k}d\omega f(k,\omega ,1-\omega )\beta
_{l}(k,\omega )\beta _{l}(k,1-\omega )=2.458381509$ with no problem worth
mentioning. We finally get: 
\begin{equation}
\int \mathcal{D}\frac{3k^{2}}{2\omega \omega ^{\prime }}\left(
3+44k^{2}-12\omega \omega ^{\prime }\right) \rho _{l}\rho _{l}^{\prime }=-
\frac{6}{8}\ln \mu +7.\allowbreak 921913219\hspace{1pt},
\label{rho rho in ll--final}
\end{equation}
where $\int \mathcal{D}$ stands for $\int_{\mu }^{+\infty }\hspace{-3pt}
dk\int_{-\infty }^{+\infty }\hspace{-3pt}d\omega \int_{-\infty }^{+\infty }
\hspace{-3pt}d\omega ^{\prime }\hspace{-2pt}\delta \left( 1-\omega -\omega
^{\prime }\right) $ for short.

Along the same lines and using the type $I_{0}^{\prime }$ given in (\ref
{integral in theta rho}), we get the accompanying result: 
\begin{equation}
\int \mathcal{D}\frac{3k}{2\omega ^{\prime }}\left( 1-\frac{\omega ^{2}}{
k^{2}}\right) \left( 1-9\frac{\omega ^{2}}{k^{2}}\right) \Theta \rho
_{l}^{\prime }=-\frac{9}{4}\ln \mu \allowbreak +0.\allowbreak 611143948.
\label{theta rho in ll--final}
\end{equation}
In terms of the foregoing discussion, convergence is already obtained for $
\ell=0.5$. Note that here too, there is a (logarithmic) infrared divergence
without involving an expansion of the dressed propagators and similar
distributions.

The next generic contribution in (\ref{Im pi_t--contributions}) we consider
is the one that involves $\rho _{l}\hspace{1pt}\partial _{k}\rho
_{l}^{\prime }$, with a corresponding function $f(k,\omega ,\omega ^{\prime
})=4k^{3}\left( 13+3k^{2}+10\omega -9\omega ^{2}\right) /\omega \omega
^{\prime }$. The type of integral we need to use is $I_{1}$ given in (\ref
{integral in rho d_k rho-result}). Let us look at the term $f\mathfrak{z}
_{l}\partial _{k}\beta _{l}$. Using the expression of $f$ and the small-$k$
expansions of $\mathfrak{z}_{l}$ and $\partial _{k}\beta _{l}$ given in
appendix A, we have $f\mathfrak{z}_{l}\partial _{k}\beta _{l}\sim -\frac{14}{
3k}-\left( \frac{682}{225}-\frac{63\pi ^{2}}{200}\right) k+\left( \frac{52167
}{7000}-\frac{149\pi ^{2}}{1000}-\frac{189\pi ^{4}}{16000}\right)
\allowbreak k^{3}+O(k^{5})$, hence a logarithmic divergence. We operate
along similar lines as indicated in the case $\rho _{l}\rho _{l}^{\prime }$
above and obtain $\int_{\mu }^{\infty }dk\mathfrak{z}_{l}(k)f(k,\omega
_{l},1-\omega _{l})\left. \partial _{k}\beta _{l}\left( 1-\omega ,k\right)
\right\vert _{\omega =\omega _{l}}=\frac{14}{3}\ln \mu -0.404833413$. The
convergence is good starting from $\ell=0.3$. Next we look at the term $
\mathfrak{z}_{l}^{\prime }f\beta _{l}$. Its expansion in powers of small $k$
yields $\mathfrak{z}_{l}^{\prime }f\beta _{l}=\frac{-26}{3k}+\left( \frac{
1534}{225}+\frac{39\pi ^{2}}{200}\right) k+O(k^{3})$. A logarithmic
divergence too. We get $\int_{\mu }^{\infty }dk\mathfrak{z}_{l}^{\prime
}f\beta _{l}=\frac{26}{3}\ln \mu +2.193663437$, with a good convergence
starting from $\ell=0.5$. Next is to look at $\omega _{l}^{\prime }\mathfrak{
z}_{l}\partial _{\omega }\left( f\beta _{l}\right) $. We have the expansion $
\omega _{l}^{\prime }\mathfrak{z}_{l}\partial _{\omega }\left( f\beta
_{l}\right) =-\left( \frac{37}{25}+\frac{39\pi ^{2}}{100}\right) k+\left( 
\frac{19177}{26250}+\frac{21647\pi ^{2}}{35000}+\frac{351\pi ^{4}}{20000}
\right) \allowbreak k^{3}+O(k^{5})$. This integrand converges fast enough so
that no infrared singularity is present and we have $\int_{\mu \rightarrow
0}^{\infty }dk\omega _{l}^{\prime }\mathfrak{z}_{l}\partial _{\omega }\left(
f\beta _{l}\right) =-2.222237940$. This finite result also means that not
all contributions are divergent in the infrared, particularly regarding
those coming from an expansion of the effective propagators and indeed, we
encounter many such instances.

We look now at the double integral in $\rho _{l}\partial _{k}\rho
_{l}^{\prime }$, namely $\int_{1/2}^{\infty }dk\int_{1-k}^{k}d\omega
A(k,\omega )$ where $A(k,\omega )$ stands for $\beta _{l}f\partial _{k}\beta
_{l}$. Here care must be taken because the integrand, more precisely $
\partial _{k}\beta _{l}$, is divergent when $\omega \rightarrow 1-k$. To
handle this, let us make the change of variable $y\equiv \omega -1+k$ in
place of $\omega $ and the change of function $G(k,y,Y)/y\equiv A(k,1-k+y)$,
where $Y=1/\ln y$, small when $y$ is small. It is easy to check that $
G(k,y,Y)$ has no pole in $y$. Make then the expansion $
G(k,y,Y)=G_{0}(k,Y)+G_{1}(k,Y)y+O\left( y^{2}\right) $. We obtain the
result: 
\begin{align}
G_{0}(k,Y)& =\frac{27k\left( 56-24k^{2}+32k\right) }{4\left( \left( 3+k^{2}+
\frac{3\left( 1-k\right) }{2k}\ln \left( 2k-1\right) \right) ^{2}+\frac{9}{4}
\pi ^{2}\frac{\left( 1-k\right) ^{2}}{k^{2}}\right) }  \notag \\
& \hspace{1in}\times \frac{\left( \frac{3}{2}\left( Y\ln 2k-1\right) -\left(
k^{2}+3\right) Y\right) Y^{3}}{\allowbreak \left( \left( (3+k^{2})Y-\frac{3}{
2}\left( Y\ln 2k-1\right) \right) ^{2}+\frac{9}{4}\pi ^{2}Y^{2}\right) ^{2}}
\hspace{1pt}.  \label{G_0 in rho_l d_k rho_l}
\end{align}
$G_{1}(k,Y)$ has a lot more complicated expression that we do not need to
display. A small $Y$ expansion of $G_{0}(k,Y)$ yields: 
\begin{equation}
G_{0}(k,Y)=\frac{-2\left( 56k^{3}-24k^{5}+32k^{4}\right) }{k^{2}\left(
\left( 3+k^{2}-\frac{3}{2}\frac{1-k}{k}\ln \frac{1}{2k-1}\right) ^{2}+\frac{9
}{4}\pi ^{2}\frac{\left( 1-k\right) ^{2}}{k^{2}}\right) \allowbreak }
\allowbreak Y^{3}+O(Y^{4})\hspace{1pt}.
\label{G_0 in rho_l d_k rho_l--expansion in Y}
\end{equation}
Therefore, $G_{0}(k,Y)$ is smooth and finite when $Y\rightarrow 0$ and the
divergence of $A(k,\omega \rightarrow k)$ goes only like $1/y$. This
suggests to rewrite: 
\begin{align}
\int_{1/2}^{\infty }dk\int_{1-k}^{k}d\omega A(k,\omega )&
=\int_{0.5^{+}}^{\infty }dk\int_{0^{+}}^{2k-1}\frac{dy}{y}\left[ G\left(
k,y,\ln ^{-1}y\right) -G_{0}\left( k,\ln ^{-1}y\right) \right]  \notag \\
& +\int_{0.5^{+}}^{\infty }dk\int_{1/\ln \left( 2k-1\right) }^{0^{-}}\frac{dY
}{Y^{2}}G_{0}(k,Y)\hspace{1pt}.
\label{double integral in rho_l d_k rho_l--1st expression}
\end{align}
Subtracting $G_{0}\left( k,1/\ln y\right) $ from $G\left( k,y,1/\ln y\right) 
$ in the first integral makes the integrand finite and $y$ can safely go to
zero. It is $G_{0}/y$ itself that is integrated in the second term, but the
integration variable is changed from $y$ to $Y=1/\ln y$. Since $G_{0}(k,Y)$
starts as $Y^{3}$, see (\ref{G_0 in rho_l d_k rho_l--expansion in Y}), the
limit $Y\rightarrow 0$ becomes safe. Therefore, the light-cone divergence
present in $A(k,\omega )=\beta _{l}f\partial _{k}\beta _{l}$ is circumvented
by the above adequate change of variable $y\rightarrow Y$. The first
integral gives $-2.0291277$ with a very good convergence at $k_{\text{max}
}=100$ and $0^{+}=10^{-14}$. The second integral is equal to $0.253206577$.
Therefore: 
\begin{equation}
\int_{1/2}^{\infty }dk\int_{1-k}^{k}d\omega \beta _{l}\left( k,\omega
\right) f(k,\omega ,1-\omega )\partial _{k}\beta _{l}\left( k,1-\omega
\right) =-1.775921122\,.  \label{double integral in rho_l d_k rho_l--final}
\end{equation}
Putting all the above partial results together, we arrive at: 
\begin{equation}
\int \mathcal{D}\frac{4k^{3}}{\omega \omega ^{\prime }}\left(
13+3k^{2}+10\omega -9\omega ^{2}\right) \rho _{l}\hspace{1pt}\partial
_{k}\rho _{l}^{\prime }=\frac{40}{3}\ln \mu -2.209329038\hspace{1pt}.
\label{rho d_k rho in ll--final}
\end{equation}

Before tackling the next example, two comments are in order. First, as we
see, light-cone divergences are brought under control by a suitable change
of variable whereas the infrared divergences are not. This situation is the
same with all the remaining terms. Furthermore, as we see, the handling of
the light-cone divergences does not necessitate the introduction of an
asymptotic gluon mass $m_{\infty }$ as in the improved HTL summation
proposed in \cite{flechsig-rebhan}. However, we should mention that we do
not know if this works for other physical quantities as we do not have a
general proof for it.

The second comment is about $k_{\text{max}}=100$ we mentioned as the upper
bound for $k$ for which the (numerical) integration starts to converge
satisfactorily. Remember that all is in units of $m_{g}$, the soft scale.
Remember also that at dressed one loop, all energies and momenta have to be
soft. One asks then: is $100m_{g}$ still soft? This point is pertinent
because most similar integrations necessitate a $k_{\text{max}}$ in the
hundreds. This observation is an indication that for the HTL approximation
to be applicable, we have to be well in the quark-gluon plasma phase, i.e.
at quite high temperatures so that $100m_{g}$ can still be considered soft.

The next contribution we discuss is $\Theta \partial _{k}\rho $ with $
f(k,\omega ,\omega ^{\prime })=\frac{6\omega }{\omega ^{\prime }}\left[ 1+
\frac{2\omega }{k^{2}}-\frac{3\omega ^{2}}{k^{2}}\right] $. The
corresponding integral is $I_{1}^{\prime }$ given in (\ref{integral in theta
d_k rho}). The first contribution we consider is $f\mathfrak{z}_{l}^{\prime
} $. A small-$k$ expansion yields $f\mathfrak{z}_{l}^{\prime }=9/k^{3}+\frac{
117}{100k}-\frac{2421}{3500}k+O(k^{3})$. Here we should expect a $1/\mu ^{2}$
behavior plus the usual $\ln \mu $. We therefore write $\int_{\mu }^{+\infty
}dkf\mathfrak{z}_{l}^{\prime }=\frac{9}{2\mu ^{2}}-\frac{9}{2k_{l}(\ell)^{2}}
-\frac{117}{100}\ln \mu +\frac{117}{100}\ln
k_{l}(\ell)+\int_{0}^{k_{l}(\ell)}dk\left[ f\mathfrak{z}_{l}^{\prime }-\frac{
9}{k^{3}}-\frac{117}{100k}\right] +\int_{\ell}^{1}dx\mathfrak{z}_{l}^{\prime
}f|_{k=k_{l}(x)}$. Now the integrations are smooth and good convergence and $
\ell-$independence start at $\ell=0.6$. We obtain $\int_{\mu }^{+\infty }dkf
\mathfrak{z}_{l}^{\prime }=\frac{9}{2\mu ^{2}}-\frac{117}{100}\ln \mu
-1.051103551$. The next contribution to look at is $\omega _{l}^{\prime }
\mathfrak{z}_{l}\partial _{\omega }f$. A small-$k$ expansion yields $\omega
_{l}^{\prime }\mathfrak{z}_{l}\partial _{\omega }f=\frac{-36}{25k}+\frac{234%
}{175}k+O(k^{3})$, only logarithmic a divergence. Using the same technique,
we get $\int_{\mu }^{\infty }dk\omega _{l}^{\prime }\mathfrak{z}_{l}\partial
_{\omega }f=\frac{36}{25}\ln \mu +0.3027125835$ with good convergence.

Here too the corresponding double integral has a divergent integrand as $
\omega \rightarrow 1-k$, but the light-cone divergence can be circumvented
by a simple integration by part. Write $d(k,\omega )\equiv f(k,1-\omega
,\omega )=\frac{-1}{\omega (1-\omega )}\allowbreak \left[ 3-\frac{3}{k^{2}}
+\left( 15-\frac{9}{k^{2}}\right) \omega -\left( 12-\frac{9}{k^{2}}\right)
\omega ^{2}-\frac{3}{k^{2}}\allowbreak \omega ^{3}\right] $. We have: 
\begin{align}
\int_{1/2}^{\infty }dk\int_{1-k}^{k}d\omega d(k,\omega )\partial _{k}\beta
_{l}\left( k,\omega \right) & =-\int_{1/2}^{\infty }dk\int_{1-k}^{k}d\omega
\partial _{k}d(k,\omega )\,\beta _{l}\left( k,\omega \right)  \notag \\
& \hspace{-2in}-\int_{1/2}^{\infty }dkd(k,1-k)\,\beta _{l}\left(
k,1-k\right) +\lim \left. \int_{1-k}^{k}d\omega d(k,\omega )\,\beta
_{l}\left( k,\omega \right) \right\vert _{k\rightarrow 1/2}^{k\rightarrow
\infty }.  \label{double in theta d_k rho-ll}
\end{align}
The first integral is finite. Indeed, write it as $\int_{0}^{2}d\kappa
\int_{1-1/k}^{1/k}d\omega \partial _{\kappa }d(1/\kappa ,\omega )\,\beta
_{l}\left( 1/\kappa ,\omega \right) $ where $\kappa \equiv 1/k$ and it
yields $-0.2338436626$. The second one is also finite and rewritten as $
-\int_{0}^{2}\frac{d\kappa }{\kappa ^{2}}d(1/\kappa ,1-1/\kappa )\,\beta
_{l}\left( 1/\kappa ,1-1/\kappa \right) $, it yields exactly $-0.24$. Note
that we have no independent analytic check of this exact value. The third
term can be rewritten as $\lim \left. \int_{1-1/\kappa }^{1/\kappa }d\omega
d(1/\kappa ,\omega )\,\beta _{l}\left( 1/\kappa ,\omega \right) \right\vert
_{\kappa \rightarrow 2}^{\kappa \rightarrow 0}$ and the integral can easily
be shown to go to zero in both limits $\kappa \rightarrow 0$ and $\kappa
\rightarrow 2$. The light-cone divergence is thus circumvented. Putting all
the above partial results together, we get: 
\begin{equation}
\int \mathcal{D}\frac{6\omega }{\omega ^{\prime }}\left[ 1+2\frac{\omega }{
k^{2}}-3\frac{\omega ^{2}}{k^{2}}\right] \Theta \hspace{1pt}\partial
_{k}\rho _{l}^{\prime }=\frac{9}{2\mu ^{2}}+\frac{27}{100}\ln \mu
-1.\allowbreak 22223463\,.  \label{theta d_k rho in ll--final}
\end{equation}

The next contribution we consider is $\rho _{l}\partial _{k}^{2}\rho _{l}$.
The corresponding integral is of type $I_{2}$\ given in (\ref{integral in
rho d_k^2 rho-result}) and the function $f(k,\omega ,\omega ^{\prime })=
\frac{2k^{4}}{\omega \omega ^{\prime }}\hspace{-2pt}\left( 2+\hspace{-2pt}
5\omega \right) $. The treatment is as before. The first term to consider is 
$\mathfrak{z}_{l}^{\prime \prime }f\beta _{l}$. A small-$k$ expansion yields 
$\mathfrak{z}_{l}^{\prime \prime }f\beta _{l}=\frac{2}{k}+\left( -\frac{461}{
150}-\frac{9\pi ^{2}}{200}\right) k+O(k^{3})$, hence a logarithmic
divergence. Using the same techniques as above, we get $\int_{\mu }^{\infty
}dk\mathfrak{z}_{l}^{\prime \prime }f\beta _{l}=-2\ln \mu -1.435815399$,
with a very good stability starting from $\ell=0.5$. The next term to
consider is $\int_{\mu }^{\infty }dk\mathfrak{z}_{l}f\partial _{k}^{2}\beta
_{l}$. The small-$k$ expansion yields $\mathfrak{z}_{l}f\partial
_{k}^{2}\beta _{l}=\frac{7}{3k}+\left( \frac{81}{50}-\frac{63\pi ^{2}}{200}
\right) k+O(k^{3})$. We obtain $\int_{\mu }^{\infty }dk\mathfrak{z}
_{l}f\partial _{k}^{2}\beta _{l}=-\frac{7}{3}\ln \mu +0.731445450$, here too
with very good stability starting from $\ell=0.5$. Next we look at $
2\int_{\mu }^{\infty }dk\omega _{l}^{\prime }\mathfrak{z}_{l}^{\prime
}\partial _{\omega }(f\beta _{l})$. The small-$k$ expansion yields $2\omega
_{l}^{\prime }\mathfrak{z}_{l}^{\prime }\partial _{\omega }(f\beta
_{l})=\left( \frac{46}{25}+\frac{3\pi ^{2}}{25}\right) k+O(k^{3})$. This
term is finite and we have $2\int_{0}^{\infty }dk\omega _{l}^{\prime }
\mathfrak{z}_{l}^{\prime }\partial _{\omega }(f\beta _{l})=1.347361846$,
with a good convergence. The next contribution is $\mathfrak{z}_{l}\left(
\omega _{l}^{\prime 2}\partial _{\omega }^{2}+\omega _{l}^{\prime \prime
}\partial _{\omega }\right) (f\beta _{l})$. A small-$k$ expansion yields $
\left( \frac{1}{50}-\frac{9\pi ^{2}}{100}\right) k+O(k^{3})$, finite. We
therefore have $\int_{\mu }^{\infty }dk\mathfrak{z}_{l}\left( \omega
_{l}^{\prime 2}\partial _{\omega }^{2}+\omega _{l}^{\prime \prime }\partial
_{\omega }\right) (f\beta _{l})=-0.1176374160$, with good convergence too.

It remains to handle $\int_{1/2}^{\infty }dkf\beta _{l}\partial _{k}\beta
_{l}+\int_{1/2}^{\infty }dk\int_{1-k}^{k}d\omega f\beta _{l}\partial
_{k}^{2}\beta _{l}$. We have already remarked that $\left. \partial
_{k}\beta _{l}\left( k,\omega \right) \right\vert _{\omega \rightarrow k}$
is infinite. First a suitable change of variable from $\omega $ to $1-\omega 
$. Then perform the second integral by parts to get: 
\begin{align}
\int_{1/2}^{\infty }dkf\beta _{l}\partial _{k}\beta _{l}+\int_{1/2}^{\infty
}dk\int_{1-k}^{k}d\omega f\beta _{l}\partial _{k}^{2}\beta _{l}& =\left.
\int_{1-k}^{k}d\omega d(k,\omega )\partial _{k}\beta _{l}\left( k,\omega
\right) \right\vert _{k\rightarrow 0.5}^{k\rightarrow \infty }  \notag \\
& \hspace{-2.5in}-\int_{1/2}^{\infty }dkd(k,1-k)\left. \partial _{k}\beta
_{l}\left( k,\omega \right) \right\vert _{\omega =1-k}-\int_{1/2}^{\infty
}dk\int_{1-k}^{k}d\omega \partial _{k}d(k,\omega )\partial _{k}\beta
_{l}\left( k,\omega \right) ,  \label{double in rho d_k^2 rho}
\end{align}
where we used the definition $d(k,\omega )\equiv f(k,1-\omega ,\omega )\beta
_{l}\left( k,1-\omega \right) $. The first term is equal to zero (the
integral over $\omega $ as a function of $k$ vanishes in the two limits $
k\rightarrow \infty $ and $k\rightarrow 1/2$). The second term is also zero
because $\beta _{l}\left( k,k\right) =0$. The integrand of the third term
diverges both in the limit $\omega \rightarrow k$ and $\omega \rightarrow
1-k $. This can be handled in the following manner. Define $A(k,\omega
)\equiv \partial _{k}d(k,\omega )\partial _{k}\beta _{l}\left( k,\omega
\right) $; $F_{1}(k,y)\equiv yA(k,k-y)$ and $F_{2}(k,y)\equiv yA(k,1-k+y)$.
Define for $i=1,2$ the function $G_{i}(k,y,Y)\equiv F_{i}(k,y)$ with $
Y=1/\ln y$. An expansion in powers of small $y$ yields $G_{i}(k,y,Y)=
\allowbreak G_{i0}(k,Y)+O\left( y\right) \allowbreak $ and an expansion of $
\allowbreak G_{i0}(k,Y)$ in small $Y$ yields $\allowbreak
G_{10}(k,Y)=K_{1}(k)Y^{3}+O(Y^{4})$ and $G_{20}(k,Y)=K_{2}(k)Y^{2}+O(Y^{3})$
with $K_{i}(k)$ a (complicated) function of $k$ that does not need to be
displayed. Therefore, we can write: 
\begin{equation}
-\int_{1/2}^{\infty }dk\int_{1-k}^{k}d\omega A(k,\omega
)=-\int_{0.5^{+}}^{\infty }dk\int_{0^{+}}^{\left( k-0.5\right) }\frac{dy}{y}
F_{1}(k,y)-\int_{0.5^{+}}^{\infty }dk\int_{0^{+}}^{\left( k-0.5\right) }
\frac{dy}{y}F_{2}(k,y),
\label{double in rho_l d_k^2 rho_l--expression of A in F_i}
\end{equation}
with: 
\begin{align}
-\int_{0.5^{+}}^{\infty }dk\int_{0^{+}}^{k-0.5}\frac{dy}{y}F_{i}(k,y)&
=-\int_{0.5^{+}}^{\infty }dk\int_{0^{+}}^{k-0.5}\frac{dy}{y}\left[
F_{i}(k,y)-G_{i0}(k,\ln ^{-1}y)\right]  \notag \\
& -\int_{0.5^{+}}^{\infty }dk\int_{1/\ln \left( k-0.5\right) }^{0^{-}}\frac{
dY}{Y^{2}}G_{i0}(k,Y).  \label{F_i in double rho_l d_k^2 rho_l}
\end{align}
Both integrals ($i=1,2$) are finite and the logic behind this technique has
been discussed previously, around (\ref{double integral in rho_l d_k
rho_l--1st expression}). We obtain $-0.0059819227$ for $i=1$ and $
0.917793484 $ for $i=2$. Hence, in all we get $\int_{1/2}^{\infty }dkf\beta
_{l}\partial _{k}\beta _{l}+\int_{1/2}^{\infty }dk\int_{1-k}^{k}d\omega
f\beta _{l}\partial _{k}^{2}\beta _{l}=0.911811562$. Putting all these
partial results together, we get: 
\begin{equation}
\int \mathcal{D}\frac{2k^{4}}{\omega \omega ^{\prime }}\hspace{-2pt}\left( 2+
\hspace{-2pt}5\omega \right) \rho _{l}\partial _{k}^{2}\rho _{l}^{\prime
}=\allowbreak -\frac{13}{3}\ln \mu +1.437166043.
\label{rho_l d_k^2 rho_l--final}
\end{equation}

The accompanying term to work out is $\Theta \partial _{k}^{2}\rho _{l}$.
The corresponding function is $f\left( k,\omega ,\omega ^{\prime }\right) =
\frac{3(2k^{2}-\omega ^{2})}{k\omega ^{\prime }}$ and the type of integral
we need is $I_{2}^{\prime }$, given in (\ref{integral in theta d_k^2 rho}).
The first term to look at is $f\mathfrak{z}_{l}^{\prime \prime }$. A small-$
k $ expansion yields $f\mathfrak{z}_{l}^{\prime \prime }=\frac{-18}{k^{3}}+
\frac{621}{100k}-\frac{8073}{3500}k+O(k^{3})$, hence a $1/\mu ^{2}$ plus a
logarithmic divergence. Using the same techniques, we obtain $\int_{\mu
}^{+\infty }dkf\mathfrak{z}_{l}^{\prime \prime }=-\frac{9}{\mu ^{2}}-\frac{
621}{100}\ln \mu +3.32815131$. The other contributions are worked out
similarly and yield only a logarithmic divergence. We have $\int_{\mu
}^{+\infty }\hspace{-2pt}dk\left( \left( \mathfrak{z}_{l}\omega _{l}^{\prime
\prime }+2\mathfrak{z}_{l}^{\prime }\omega _{l}^{\prime }\right) \partial
_{\omega }+\mathfrak{z}_{l}\omega _{l}^{\prime 2}\partial _{\omega
}^{2}\right) f=\allowbreak \frac{297}{50}\ln \mu +0.127063871\allowbreak 2$.
Next is to look at the double integral. We have: 
\begin{align}
\int_{1/2}^{\infty }dk\int_{1-k}^{k}d\omega f\partial _{k}^{2}\beta
_{l}+\int_{1/2}^{\infty }dkf\partial _{k}\beta _{l}& =\left.
\int_{1-k}^{k}d\omega f\partial _{k}\beta _{l}\right\vert _{k\rightarrow
0.5}^{k\rightarrow \infty }-\int_{1/2}^{\infty }dkf(k,1-k,k)\partial
_{k}\beta _{l}  \notag \\
& -\int_{1/2}^{\infty }dk\int_{1-k}^{k}d\omega \partial _{k}f\partial
_{k}\beta _{l}.  \label{double in theta d_k^2 rho_l}
\end{align}
The first term is zero. The second term is analytically finite but must
numerically be handled with care because it does not converge easily in the
limit $k\rightarrow 0.5$. A change of variable $k\rightarrow 1/k$, usually
useful, is not of any help here. The best way we find to treat this is to
introduce a small $\epsilon $ such that the integral $\int_{0.5+\epsilon
}^{\infty }dk$ carries with no problem, and handle $\int_{0.5}^{0.5+\epsilon
}dk$ using an expansion of the integrand. The first term in the expansion
has been sufficient. Thus we find this second term (exactly) equal to $
-0.645 $. Note that here too, this simple and exact number is obtained
numerically and has not been shown analytically. The third term has a
light-cone divergence and is handled in the $\left( y,Y\right) $ variables.
We find it equal to $-0.936156340$. Putting everything together, we get: 
\begin{equation}
\int \mathcal{D}\frac{3\omega ^{2}}{k\omega ^{\prime }}\Theta \partial
_{k}^{2}\rho _{l}^{\prime }=-\frac{9}{\mu ^{2}}-\frac{27}{100}\ln \mu
+1.874058841.  \label{theta d_k^2 rho_l--final}
\end{equation}

The next term we work out explicitly is $\rho _{l}\rho _{l}^{\prime
}\partial _{\omega }\delta $ with a function $f\left( k,\omega ,\omega
^{\prime }\right) =-\hspace{-2pt}\frac{54k^{4}}{\omega \omega ^{\prime }}$.
The corresponding integral is of the type $I_{3}$, given in (\ref{integral
in rho rho d_omega delta}). We consider first the contribution $-2\mathfrak{z
}_{l}\partial _{\omega }\left( f\beta _{l}\right) $, the factor of 2 coming
from the fact that two terms in (\ref{integral in rho rho d_omega delta})
are equal. The small-$k$ expansion yields $\left( -\frac{54}{5}+\frac{27\pi
^{2}}{20}\right) k+O(k^{3})$, which means no infrared divergent behavior
analytically, but practically, the integration does not converge easily when 
$k\rightarrow 0$. We then split the integral into two pieces, one piece
equal to $\int_{0}^{k_{l}(\ell)}dk\left[ \left( -\frac{54}{5}+\frac{27\pi
^{2}}{20}\right) k+\left( \frac{12339}{875}-\frac{23607\pi ^{2}}{14000}-
\frac{243\pi ^{4}}{4000}\right) \allowbreak k^{3}+O(k^{5})\right] $ where we
perform an expansion of the integrand, and a second piece equal to $
-2\int_{\ell}^{1}dxk_{l}^{\prime }\mathfrak{z}_{l}\partial _{\omega }\left(
f\beta _{l}\right) $ which is integrated smoothly. We get in all $
0.00866285165$, $\,$with very good stability and $\ell-$independence already
from $\ell=0.2$. The corresponding double integral presents a light-cone
divergence that is handled in the $\left( y,Y\right) $ variables. We find it
equal to $-0.825017233$. In all, we have: 
\begin{equation}
-\int \mathcal{D\,}\frac{54k^{4}}{\omega \omega ^{\prime }}\rho _{l}\rho
_{l}^{\prime }\partial _{\omega }\delta \left( 1-\omega -\omega ^{\prime
}\right) =-0.\allowbreak 816354381\allowbreak 4.
\label{rho_l rho_l d_omega delta--final}
\end{equation}

The last term we discuss is $\Theta \rho _{l}^{\prime }\partial _{\omega
}\delta $ with a vanishing corresponding $f$. Hence: 
\begin{equation}
\int \mathcal{D\,}\hspace{-2pt}f\left( k,\omega ,\omega ^{\prime }\right)
\Theta \rho _{l}^{\prime }\partial _{\omega }\delta \left( 1-\omega -\omega
^{\prime }\right) =0.  \label{theta rho_l d_omega delta--final}
\end{equation}

These are all the sample terms contributing to the second coefficient in $
p^{2}$ of $\mathrm{Im}\hspace{1pt}^{\ast }\Pi _{t}$ we wanted to discuss in
the main text. All the remaining contributions are treated along parallel
lines and no new subtleties not touched upon here do occur. The only new
feature is the occurrence of divergences of order $1/\mu ^{4}$. We give in
appendix B all the partial results for each contribution.

The coefficient of order zero in $\mathrm{Im}\hspace{1pt}^{\ast }\Pi _{t}$ is
finite and is already determined in \cite{BPgamt}. Putting everything
together and restoring back $m_{g}$, we obtain: 
\begin{equation}
\mathrm{Im}\hspace{1pt}^{\ast }\Pi _{t}(P)=\frac{g^{2}N_{c}T}{24\pi }\left[
13.086+\left( 90.064/\bar{\mu}^{4}+2.7/\bar{\mu}^{2}+\frac{44}{15}\ln \bar{
\mu}+\allowbreak 1.5192728\right) \bar{p}^{2}+\mathcal{O}\left( \bar{p}
^{4}\right) \right] .  \label{pi star_t--final}
\end{equation}
where $\bar{\mu}=\mu /m_{g}$, a dimensionless parameter, and $\bar{p}
=p/m_{g} $. Using (\ref{definition2 gamma_t}) , we finally find: 
\begin{equation}
\gamma _{t}(p)=\frac{g^{2}N_{c}T}{24\pi }\left[ 6.543+\left( 45.032/\bar{\mu}
^{4}-314.21/\bar{\mu}^{2}-507.96\ln \bar{\mu}+\allowbreak 1014.65\right) 
\bar{p}^{2}+\mathcal{O}\left( \bar{p}^{4}\right) \right] .
\label{gamma_t--final}
\end{equation}
As we have claimed, when the damping rate $\gamma _{t}(p)$ for transverse
gluons is calculated in the sole context of HTL-summed perturbation theory
in an expansion in powers of $p^{2}$, the second coefficient is divergent in
the infrared. We discuss this result in the following and last section.

\section{Discussion}

Using the hard-thermal-loop perturbative framework, we have determined the
damping rate $\gamma _{t}(p)$ for ultrasoft transverse gluons in
high-temperature QCD to its lowest order $g^{2}T$. The calculation requires
the obtainment of an expression for the imaginary part of the fully
HTL-dressed one-loop-order gluon self-energy. We have determined this
quantity in the form of a series in powers of the external momentum $\bar{p}
=p/m_{g}$, a small parameter when $p$ is ultrasoft. The aim of this section is to
carry a critical discussion of the results (\ref {pi star_t--final} )
and (\ref {gamma_t--final}), and the method used to obtain them.

\subsection{Magnetic mass as an infrared regulator}

The presence of 1/$\bar{\mu}^{4}$, $1/\bar{\mu}^{2}$ and $\ln \bar{\mu}$ (to
a lesser degree) in the coefficient of $\bar{p}^{2}$ in (\ref{gamma_t--final}
) is troublesome to some extent because it suggests a divergence when $\bar{
\mu}\rightarrow 0$. However, we may regard the issue from a different
perspective. Let us suppose that the HTL-summed perturbation is \textit{all}
of massless QCD at high temperature. Since, on physical grounds, we \textit{
must} have $\gamma _{l}(0)=\gamma _{t}(0),$ we can impose this condition on (
\ref{gaml-0}) and obtain thus a (crude) estimate of the magnetic mass:
\begin{equation}
\mu =0.32259\,m_{g}+\dots\,.  \label{estimate-mu}
\end{equation}
We can put this value back in (\ref{gamma_t--final}) and obtain the
following expression for $\bar{\gamma}_{t}\equiv \gamma _{t}/m_{g}$:
\begin{equation}
\bar{\gamma}_{t}(\bar{p})=\left[ 6.543+\allowbreak 2728.3\,\,\bar{p}^{2}+
\mathcal{O}\left( \bar{p}^{4}\right) \right] \,\bar{g}\,+\dots,
\label{gamma_t-with-estimate-mu}
\end{equation}
where $\bar{g}\equiv \frac{3N_{c}}{24\pi \sqrt{N_{c}+N_{f}/2}}g$. Note that $
\bar{g}$ is a better expansion parameter than $g$ because it is much
smaller. Indeed, for $N_{c}=3$ and $N_{f}=1$, we have $\bar{g}=0.0638\,g$,
and for $N_{c}=3$ and $N_{f}=3$, $\bar{g}=0.0563\,g$.

Result (\ref{gamma_t-with-estimate-mu}) is quite acceptable, even with a
relatively large coefficient for $\bar{p}^{2}$. Remember that in massless
high-temperature QCD, there is a hierarchy of scales $\bar{g}^{n}T$, with $n$
a nonnegative integer \cite{aurenche-gelis-kobes-petitgirard1}. Consider for
instance the quantity $\bar{\Omega}_{t}\left( \bar{p}\right) \equiv \Omega
_{t}\left( p\right) /m_{g}$, where $\Omega _{t}\left( p\right) $\ is the
full (complex) on-shell energy of the ultrasoft transverse gluon. In an
expansion in powers of the coupling $\bar{g}$, we anticipate the following:
\begin{equation}
\bar{\Omega}_{t}\left( \bar{p}\right) =\bar{\omega}_{t}(\bar{p})+\left( \,
\bar{\omega}_{t}^{\left( 1\right) }(\bar{p})-i\bar{\gamma}_{t}(\bar{p}
)\right) +\left( \bar{\omega}_{t}^{\left( 2\right) }(\bar{p})-i\bar{\gamma}
_{t}^{\left( 2\right) }(\bar{p})\right) +\dots ,
\label{expansion-OmegaBar-gBar}
\end{equation}
where $\bar{\omega}_{t}(\bar{p})\equiv $ $\omega _{t}(p)/m_{g}=1+3/5\,\bar{p}
^{2}-9/35\,\bar{p}^{4}+\dots $ from (\ref{dispersion relations}) and is of
order one (lowest-order contribution); $\bar{\omega}_{t}^{\left( 1\right) }(
\bar{p})$ and $\bar{\gamma}_{t}(\bar{p})$ are functions of $\bar{p}$ times $
\bar{g}$ (first-order corrections); $\bar{\omega}_{t}^{\left( 2\right) }(
\bar{p})$ and $\bar{\gamma}_{t}^{\left( 2\right) }(\bar{p})$ are functions
of $\bar{p}$ times $\bar{g}^{2}$ (second-order corrections), etc. Remember
that the very definition (\ref{definition1 gamma_t}) of $\gamma _{t}(p)$ is
based on the feasibility of the expansion (\ref{expansion-OmegaBar-gBar}).
The full complex energy $\Omega _{t}$ is a solution to the full dispersion
relation:
\begin{equation}
-\Omega _{t}^{2}+p^{2}-\Pi _{t}(-i\Omega _{t},p)=0,
\label{full-dispersion-relation}
\end{equation}
where $\Pi _{t}$ is the \textit{full} transverse-gluon self-energy, a
fortiori expected also to admit an expansion in powers of $\bar{g}$:
\begin{equation}
\Pi _{t}=\Pi _{t}^{\left( 0\right) }+\Pi _{t}^{\left( 1\right) }+\Pi
_{t}^{\left( 2\right) }+\dots \,.  \label{expansion-full-self-energy}
\end{equation}
The logic behind the hard-thermal-loop perturbation in $g$ or equivalently
in $\bar{g}$ can be understood as follows. In order to get the lowest-order
term $\bar{\omega}_{t}$ of $\bar{\Omega}_{t}$, one uses only the
lowest-order term $\Pi _{t}^{\left( 0\right) }=\delta \Pi _{t}$ of $\Pi _{t}$
in the dispersion relation (\ref{full-dispersion-relation}). The
contribution $\delta \Pi _{t}$ is obtained by summing naive-perturbation
one-loop graphs with \textit{hard} internal momenta, more precisely momenta
ranging from the soft scale $gT$, or preferably $m_{g}$, to $+\infty $. In
order to get the first-order corrections $\bar{\omega}_{t}^{\left( 1\right)
} $ and $\bar{\gamma}_{t}$ to $\bar{\Omega}_{t}$, one assumes that all of
the next-to-leading-order contribution $\Pi _{t}^{\left( 1\right) }$ is $
^{\ast }\Pi _{t}$ and adds it to\ $\delta \Pi _{t}$ in (\ref
{full-dispersion-relation}). It is evaluated as a sum of HTL-dressed
one-loop graphs with\ \textit{soft} momenta ranging from $g^{2}T$ to $gT$
or, more preferably from a physical standpoint, from the magnetic scale $\mu 
$ to the electric scale $m_{g}$. In order to get the second-order
corrections $\bar{\omega}_{t}^{\left( 2\right) }$ and $\bar{\gamma}
_{t}^{\left( 2\right) }$ to $\bar{\Omega}_{t}$, one adds $\Pi _{t}^{\left(
2\right) }$ to the dispersion relation, assuming that all of it is a set of
HTL-dressed two-loop graphs with \textit{ultrasoft} loop-momenta ranging
from the scale $g^{3}T$ to the magnetic scale $\mu $. And so on. At the
present time, there is no known physical process that involves in a natural
way the scale $g^{3}T$ and explicit calculations in HTL-dressed perturbation
can be carried with much hardship only to first order.

Hence, it is quite admissible to use the magnetic mass $\mu $ as an
infrared regulator in HTL-dressed one-loop calculations, and this we did in
the present work. The problem that arises is how to estimate $\mu $ itself.
The common belief is that chromomagnetic screening is a non-perturbative
effect \cite{braaten-nieto1,rebhan3}. The somewhat heuristic argument
leading to (\ref{estimate-mu}) may be useful, but is probably a temporary
way out. It assumes that all of massless QCD at high temperature is in the
HTL-summed perturbation. This is certainly true to lowest order, and some
would argue that it is also true to next-to-leading order. But there is no
certainty beyond. Recall that there is no generating functional that yields
the whole series of HTL perturbation. The only available generating
functionals are those generating the hard thermal loops only.

Still, the result (\ref{gamma_t--final}) depends strongly on the infrared
regulator and for the term in $\bar{p}^{2}$ to be regarded as a correction
to the lowest-order coefficient, we must have $p\lesssim 0.049\,m_{g}$. A
different perspective is to look at the full complex energy $\overline{
\Omega }_{t}\left( \bar{p}\right) $ in an expansion in powers of $\bar{p}
=p/m_{g}$. This must be allowed for ultrasoft $p$ on physical grounds. Write
then:
\begin{equation}
\overline{\Omega }_{t}\left( \bar{p}\right) =\overline{\Omega }_{t}^{\left(
0\right) }\left( \bar{g}\right) +\overline{\Omega }_{t}^{\left( 1\right)
}\left( \bar{g}\right) \,\bar{p}^{2}+\overline{\Omega }_{t}^{\left( 2\right)
}\left( \bar{g}\right) \,\bar{p}^{4}+\dots  \label{Omaga in powers of p}
\end{equation}
Where the different $\overline{\Omega }_{t}^{\left( i\right) }\left( \bar{g}
\right) $, $i=0,1,\dots $ can be written in powers of $\bar{g}$. For
example, $\overline{\Omega }_{t}^{\left( 0\right) }\left( \bar{g}\right) $
is the \textit{full} zero-momentum complex energy for the transverse gluon
(in units of $m_{g}$), and we have:
\begin{equation}
\overline{\Omega }_{t}^{\left( 0\right) }\left( \bar{g}\right) =1+\left[ 
\bar{\omega}_{t}^{\left( 1\right) }(0)/\bar{g}-6.543\,i\right] \bar{g}+
\mathcal{O}\left( \bar{g}^{2}\right) .  \label{lowest order in Omega}
\end{equation}
The quantity $\omega _{t}^{\left( 1\right) }(0)$ is the next-to-leading
order contribution to the gluon thermal mass, and assuming that both the
longitudinal and transverse gluons are the same at zero momentum, we do have
a number for it from \cite{schulz} in the case of a pure gluonic theory ($
N_{f}=0$). In our notation, $\bar{\omega}_{t}^{\left( 1\right) }(0)/\bar{g}
=-22.6\sqrt{N_{c}}$. We also have, with the estimate (\ref{estimate-mu}) for 
$\mu $:
\begin{equation}
\overline{\Omega }_{t}^{\left( 1\right) }\left( \bar{g}\right) =\frac{3}{5}+
\left[ \bar{\omega}_{t1}^{\left( 1\right) }/\bar{g}-2728.3\,i\right] +
\mathcal{O}\left( \bar{g}^{2}\right) ,  \label{first order in Omega}
\end{equation}
where $\bar{\omega}_{t1}^{\left( 1\right) }$ is the coefficient of $\bar{p}
^{2}$ in the expansion of $\bar{\omega}_{t}^{\left( 1\right) }(\bar{p})$,
and $3/5$ comes from the expansion of $\bar{\omega}_{t}(\bar{p})=1+3/5\,\bar{
p}^{2}-9/35\,\bar{p}^{4}+\dots $ . We do not have an expression for $\bar{
\omega}_{t1}^{\left( 1\right) }$, but for the perturbation to be reliable,
we should at least have $2728.3\,\bar{g}\lesssim 3/5$, which means $\bar{g}
\lesssim 0.0002$, or $g\lesssim 0.003$ for $N_{c}=3$ and $N_{f}=1$, and $
g\lesssim 0.004$ for $N_{c}=3$ and $N_{f}=3$. This implies that the
temperature\ $T$ must be quite high so that the (running) coupling constant
be so small (asymptotic freedom). At very high temperature, classical
behavior dominates and quantum effects are small. It seems therefore that
the HTL-summed perturbation may be reliable well into the quark-gluon plasma
phase, but if we try to lower the temperature, this perturbation may become
less useful.

Another interesting rough inference from this `orders-of-magnitude'
discussion is the relation of the magnetic mass $\mu $ to the magnetic scale 
$g^{2}T$. Indeed, let us write $\mu =a\,g^{2}T=\frac{24\pi }{N_{c}}a\bar{g}
\,m_{g}=0.32259\,m_{g}$. This implies that $a\bar{g}=0.32259\frac{N_{c}}{
24\pi }=0.012835$, and with $\bar{g}\lesssim 0.0002$, we have the lower
limit $a=\mu /g^{2}T\gtrsim 64.18$. This is between one to two orders of
magnitude larger than the magnetic scale $g^{2}T$ and adds water to the
suggestion that not only simple integer powers of the coupling $g$ play a
role in this perturbation \cite{aurenche-gelis-kobes-petitgirard1}.

\subsection{A Slavnov-Taylor identity}

Away from these orders-of-magnitude estimates, we focus attention on a
Slavnov-Taylor identity for the gluon polarization tensor derived in \cite
{dirks-niegawa-okano} in the Coulomb gauge and applied to the
next-to-leading-order contribution. The identity recovers already known
transversality conditions, and when applied to the imaginary part of
next-to-leading order gluon self-energy, it implies the equality between the
damping rates for non-moving transverse and longitudinal gluons. This result
is important and worth discussing in the present context. We first recall
the main steps of that work with a slight change in the notation suitable
for our purposes. The Slavnov-Taylor identity derived in \cite
{dirks-niegawa-okano} is:
\begin{equation}
P^{\nu }\Pi _{\mu \nu }\left( P\right) =-\left[ \delta _{\mu }^{\nu
}P^{2}-P_{\mu }P^{\nu }+\Pi _{\mu }^{\nu }\left( P\right) \right] \Pi _{
\text{g}\,\nu }\left( P\right) ,  \label{slavnov-taylor}
\end{equation}
where $\Pi _{\mu \nu }\left( P\right) $ is the \textit{full} gluon
self-energy and $\Pi _{\text{g}\,\nu }\left( P\right) $ is related to the
Coulomb-ghost self-energy $\Pi _{\text{g}}$ via the relation $\Pi _{\text{g}
}\left( P\right) =p^{i}\,\Pi _{\text{g}\,i}\left( P\right) $. Note that
identity (\ref{slavnov-taylor}) is exact and valid for \textit{all} momenta $
P$.

In the sequel, concentrate on the soft momentum region and consider the
expansion of the full $\Pi _{\mu \nu }$ in powers of the coupling, relation (
\ref{expansion-full-self-energy}). Again, $\Pi _{\mu \nu }^{\left( 0\right)
} $ is the full leading-order contribution of order $g^{2}T^{2}$, $\Pi _{\mu
\nu }^{\left( 1\right) }$ the full next-to-leading-order contribution of
order $g\left( g^{2}T^{2}\right) $, and so on. When reading the identity (
\ref{slavnov-taylor}) in perturbation in powers of the coupling, we can
deduce the two transversality conditions:
\begin{equation}
P^{\nu }\Pi _{\mu \nu }^{\left( 0\right) }\left( P\right) =0;\qquad P^{\mu
}P^{\nu }\Pi _{\mu \nu }^{\left( 1\right) }\left( P\right) =0,
\label{transversality conditions}
\end{equation}
both relations known to be satisfied by the hard thermal loop $\delta \Pi $
and the HTL-summed one-loop-order self-energy $^{\ast }\Pi $ respectively.
For our purposes, we need the Coulomb-ghost self-energy only to lowest
order, and it is derived in \cite{dirks-niegawa-okano} to read:
\begin{equation}
\Pi _{\text{g}\,\mu }^{\left( 0\right) }\left( P\right) =-\frac{g^{2}N_{c}T}{
16}\delta _{\mu }^{i}\,\hat{p}^{i},  \label{pi ghost lowest order}
\end{equation}
which is already of order $g^{2}T$ and has the limit $\Pi _{\text{g}\,\mu
}^{\left( 0\right) }\left( \omega ,\mathbf{p}=\mathbf{0}\right) =0$. From
this we get the relation \cite{dirks-niegawa-okano}:
\begin{equation}
P^{\nu }\Pi _{\nu \mu }^{\left( 1\right) }\left( P\right) =\frac{g^{2}N_{c}T
}{16}\frac{\omega }{p}\,^{\ast }\Delta _{l}^{-1}\left( P\right) \left( 1,
\frac{\omega }{p}\hat{p}\right) _{\mu },
\label{relation pi one and Delta star}
\end{equation}
where $^{\ast }\Delta _{l}$ is the longitudinal gluon propagator to lowest
order given in (\ref{gluon propagator}). The above relation was checked in 
\cite{dirks-niegawa-okano} to be satisfied by $^{\ast }\Pi $ as defined in (
\ref{definition pi star}). Since the right-hand side of (\ref{relation pi
one and Delta star}) is real, we immediately obtain:
\begin{equation}
P^{\nu }\mathrm{Im}\Pi _{\nu \mu }^{\left( 1\right) }\left( P\right) =0.
\label{ImPi one contracted is zero}
\end{equation}
From here, it is straightforward to show that \cite{dirks-niegawa-okano}:
\begin{equation}
\gamma _{l}\left( 0\right) =\gamma _{t}\left( 0\right) ,
\label{gam_l equal gam_t}
\end{equation}
with the same definition of the damping rates as the one we adopt. One thing
pertinent in the result (\ref{gam_l equal gam_t}) is that, since $^{\ast
}\Pi $ as defined in (\ref{definition pi star}) is shown explicitly to
satisfy (\ref{relation pi one and Delta star}), it too would lead to the
same equality, independently of the issue we raise whether $^{\ast }\Pi $ is
the full $\Pi ^{\left( 1\right) }$.

We have emphasized at several occasions that at zero momentum, the
longitudinal and transverse gluon damping rates have to be equal, and that
any consistent computational scheme ought to yield this equality. The result
of \cite{dirks-niegawa-okano} is an additional confirmation of this.
However, all calculations in \cite{BPgamt} leading to (\ref{gamt-0}) and in 
\cite{AAB,AA} leading to (\ref{gaml-0}), and even those of \cite
{kobes-kunstatter-mak,braaten-pisarski (quarks)} leading to (\ref{gamq-0})
for the quarks are \textit{explicit}, whereas those in \cite
{dirks-niegawa-okano} are formal: the loop-integral/Matsubara-sum are never
done explicitly; no infrared cutoff is introduced to regulate the infrared
sector and there is no mention of what region should be kept in the
integral/sum. All these issues are at the heart of any explicit calculation
and have to be dealt with. Different sets of rules can and do lead to
different results, and the important point to emphasize is that when one
`goes' explicit, so to speak, one encounters problems.

There is no doubt that any explicit calculation should be handled with care,
and in any case, should be in accordance with general rules and checks \cite
{dirks-niegawa-okano}. But in our specific situation, relations (\ref
{relation pi one and Delta star}) and (\ref{ImPi one contracted is zero})
are not of any direct help. Indeed, since we have an explicit on-shell
expressions for $\mathrm{Im}\,^{\ast }\Pi _{l}$ and $\mathrm{Im}\,^{\ast }\Pi
_{t}$ to order $p^{2}$, from \cite{AA} and (\ref{pi star_t--final})
respectively, we could hope to obtain a constraint from (\ref{ImPi one
contracted is zero}) that we try to check whether it is satisfied.
Unfortunately this is not the case. Indeed, because $^{\ast }\Pi _{\mu \nu }$
satisfies the second relation of (\ref{transversality conditions}), it
depends on three independent functions $^{\ast }\Pi _{l},^{\ast }\Pi _{t}$
and $^{\ast }\Pi _{l}^{\prime }$ such that \cite{BPgamt}:
\begin{eqnarray}
^{\ast }\Pi _{00} &=&\,^{\ast }\Pi _{l};\qquad ^{\ast }\Pi _{0i}=-\frac{
p^{0}p^{i}}{p^{2}}\left( ^{\ast }\Pi _{l}+\,^{\ast }\Pi _{l}^{\prime
}\right) ;  \notag \\
^{\ast }\Pi _{ij} &=&\left( \delta ^{ij}-\hat{p}^{i}\hat{p}^{j}\right)
\,^{\ast }\Pi _{t}+\hat{p}^{i}\hat{p}^{j}\left( ^{\ast }\Pi _{l}+2\,^{\ast
}\Pi _{l}^{\prime }\right) .  \label{structure Pi star}
\end{eqnarray}
What happens is that constraint (\ref{ImPi one contracted is zero}), when
applied to $\mathrm{Im}^{\ast }\Pi _{\mu \nu }$, implies only the condition:
\begin{equation}
\mathrm{Im}\,^{\ast }\Pi _{l}^{\prime }\left( P\right) =0.
\label{ImPi_l prime is zero}
\end{equation}
Recall that the choice of the strict Coulomb gauge was precisely made to
avoid having to deal with $^{\ast }\Pi _{l}^{\prime }$ \cite{BPgamt}, which
is hence of no direct physical significance, as much is therefore (\ref
{ImPi_l prime is zero}). It would still be of academic interest though to
check whether indeed an explicit calculation, through similar \textit{
explicit} steps as in this work, would satisfy (\ref{ImPi_l prime is zero}).
This would take us way beyond the scope of the present article.

\subsection{Other explicit calculations}

Besides, it is important to emphasize that there are other explicit
estimates of the gluon damping rates where approximation schemes have been
used, away from formal manipulations. One such estimate is that of \cite
{pisarski4} where result (\ref{gam glu qua-nonzero p-pisarski}) is derived.
We have discussed in detail in \cite{ABD1} the difference between this
result and the one we carry here: (\ref{gam glu qua-nonzero p-pisarski}) is
obtained by integrating ultrasoft loop momenta $0\leq k<\mu $ while keeping $
p$ just soft, and the infrared behavior is regulated via the introduction by
hand of a magnetic mass directly in the static magnetic propagator. Result (
\ref{gam glu qua-nonzero p-pisarski}) cannot be carried as it is to the
limit $p\rightarrow 0$ since it would give zero whereas the dampings there
are finite.

Interesting also is the result (\ref{gam soft p-flechsig-rebhan-schulz})
from \cite{flechsig-rebhan-schulz}, which is the same as (\ref{gam glu
qua-nonzero p-pisarski}) except that the constant $a$ is determined to be
equal to $N_{c}/4\pi $. It is argued in \cite{flechsig-rebhan-schulz} that
when the external momentum $p$ of the quasi-particle is of order $gT$, the
dominant term in the next-to-leading order self-energies comes from loop
integrals involving the dressed transverse-gluon propagator $^{\ast }\Delta
_{t}$, and it is argued that such dominant terms come proportional to the
function: 
\begin{equation}
S_{i}\left( p\right) =T\sum\limits_{k_{0}}\int \frac{d^{3}k}{\left( 2\pi
\right) ^{3}}\,^{\ast }\Delta _{t}\left( K\right) \,\left. ^{\ast }\Delta
_{i}\left( P-K\right) \right\vert _{p_{0}=-i\omega _{t}\left( p\right)
+0^{+}}\,,  \label{function S_i of flechsig-rebhan-schulz}
\end{equation}
where $i$ indicates any of the quasi-particles. Using the structure of the
spectral function corresponding to $^{\ast }\Delta _{t}\left( K\right) $, it
is argued that the dominant contribution to the Matsubara sum in (\ref
{function S_i of flechsig-rebhan-schulz}) is the `static' contribution $
k_{0}=0$. The integration over the spatial loop-momentum is then done for
this static contribution with the magnetic mass $\mu $ taken as the infrared
cutoff, i.e., $\int_{0}^{+\infty }dk\rightarrow \int_{\mu }^{+\infty }dk$.
The logarithmic infrared sensitivity $\ln \left( m_{g}/\mu \right) $ is
obtained as a consequence of the fact that the calculation is performed on
mass-shell, and result (\ref{gam soft p-flechsig-rebhan-schulz}) is said to
be `mass-shell' infrared sensitive.

But it is stressed that the leading infrared sensitive result (\ref{gam soft
p-flechsig-rebhan-schulz}) should not be carried from the soft region $p\sim
gT$ where it is determined to the ultrasoft region $p<\mu \sim g^{2}T$
because there, the logarithmic sensitivity of $S_{i}\left( P\right) $ is
softened and the result becomes finite. It is also argued that in the
ultrasoft region, formally higher-order terms in HTL-summed perturbation may
contribute to the self-energies with the same magnitude as the
next-to-leading-order one we are concerned with in this work.

It is clear that here too our result (\ref{gamma_t--final}) is not in any
way in contradiction with (\ref{gam soft p-flechsig-rebhan-schulz}) or in
fact with any other result available in the literature: we carry the
calculation directly in the ultrasoft region and sum soft loop momenta in
the context of \textit{full} one-loop-order HTL-dressed perturbation.

\subsection{Momentum expansion in scalar QED}

But we may still question (\ref{gamma_t--final}) since it depends strongly
on the manner we have carried\ out the calculation. A crucial step in this
latter is to perform a momentum expansion before the sum over Matsubara
frequencies and analytic continuation to real energies are done. Given the
fact that the final result is significantly infrared sensitive, it is
legitimate to question the validity of this method. It may also be true that
our calculations do not provide the equality between the damping rates of
non-moving longitudinal and transverse gluons because intricate intermediary
steps, dealing with potential infinite terms, are not handled with
sufficient care \cite{dirks-niegawa-okano}. We have put forward arguments in
previous works regarding this issue \cite{AA,ABD1}, but probably the best
way to show that the method used has little to do with the infrared
sensitivity of the damping rate is to evaluate some physical quantity with
no momentum expansion, then perform the expansion and compare the result
with that where the expansion is done at an early stage, before Matsubara
sum and analytic continuation to real energies are done. It would be most
suitable to make this comparison directly in massless QCD, but unfortunately
the calculations are forbiddingly difficult. However, the structure of
scalar quantum electrodynamics (scalar QED) is manageable enough that it is
able to offer such an opportunity.

The HTL-summed photon self-energies are\ evaluated for all $\omega $ and $p$
in hot scalar QED without recourse to any momentum expansion \cite
{kraemmer-rebhan-schulz}. Let us focus on the longitudinal case and quickly
review the corresponding results, with a slight editing in the notation from 
\cite{kraemmer-rebhan-schulz}. Three regions in $\omega $ and $p$ are to be
distinguished: $\omega ^{2}<p^{2}$ (region I), $p^{2}<\omega
^{2}<4m_{e}^{2}+p^{2}$ (region II) and \ $4m_{e}^{2}+p^{2}<\omega ^{2}$
(region III), where $m_{e}=eT/2$ is the scalar thermal mass and $e$ the
coupling constant. The longitudinal next-to-leading order HTL-summed photon
self-energy is found to have the following expression:
\begin{equation}
\delta \,^{\ast }\Pi _{l}\left( \omega ,p\right) =\dfrac{e^{2}T}{8\pi }
\dfrac{\omega ^{2}-p^{2}}{p^{2}}\left[ 4m_{e}+2i\varepsilon -i\dfrac{\omega
^{2}}{p}\ln \left( \dfrac{2m_{e}-i\left( \varepsilon +p\right) }{
2m_{e}-i\left( \varepsilon -p\right) }\right) \right] ,
\label{HTL-summed photon self-energy}
\end{equation}
where the prefix $\delta $ indicates here that the known leading
hard-thermal-loop contribution has been subtracted for ultraviolet
convergence, and $\varepsilon =\Theta $ in regions I and III and $
\varepsilon =i\left\vert \Theta \right\vert $ in region II, with $\Theta
^{2}=\omega ^{2}\left( \omega ^{2}-p^{2}-4m_{e}^{2}\right) /\left( \omega
^{2}-p^{2}\right) $. The longitudinal photon dispersion relation is:
\begin{equation}
\Omega ^{2}=p^{2}+\Pi _{l}\left( \Omega ,p\right) ,
\label{full-dispersion-in-SQED}
\end{equation}
and, up to next-to-leading order, writes:
\begin{equation}
\Omega ^{2}\left( p\right) =\omega _{l}^{2}\left( p\right) +\dfrac{\delta
^{\ast }\Pi _{l}\left( \omega _{l},p\right) }{1-\partial _{\omega
_{l}^{2}}\delta \Pi _{l}\left( \omega _{l},p\right) },
\label{next-to-leading-dispersion}
\end{equation}
where $\delta \Pi _{l}\left( \omega ,p\right) $ is the hard thermal loop
given by \cite{kraemmer-rebhan-schulz}:
\begin{equation}
\delta \Pi _{l}\left( \omega ,p\right) =3m_{p}^{2}\left( 1-\frac{\omega ^{2}
}{p^{2}}\right) \left( 1-\frac{\omega }{2p}\ln \frac{\omega +p}{\omega -p}
\right) ,  \label{photon-htl}
\end{equation}
with $m_{p}=eT/3$ the photon thermal mass, and $\omega _{l}\left( p\right) $
the on-shell longitudinal photon energy, solution to (\ref
{full-dispersion-in-SQED}) to lowest order, where only the hard thermal loop
is kept in the self-energy. Using (\ref{photon-htl}), we can rewrite (\ref
{next-to-leading-dispersion}) as:
\begin{equation}
\Omega ^{2}\left( p\right) =p^{2}+\delta \Pi _{l}\left( \omega _{l},p\right)
+\dfrac{2\omega _{l}^{2}}{3m_{p}^{2}+p^{2}-\omega _{l}^{2}}\delta ^{\ast
}\Pi _{l}\left( \omega _{l},p\right) .
\label{next-to-leading-dispersion-rewritten}
\end{equation}

Remember that all these results involve no expansion in $p$ and the
Matsubara sum and analytic continuation to real energies are done. Now we
expand. The region of interest to us is region II, where we are allowed to
perform the following expansion: 
\begin{equation}
\omega _{l}\left( p\right) =1+\frac{3}{10}\bar{p}^{2}-\frac{3}{280}\bar{p}
^{4}+\mathcal{O}\left( \bar{p}^{6}\right) ;\qquad \bar{p}=p/m_{p}.
\label{expansion-omaga-SQED}
\end{equation}
Using this, we perform the expansion of (\ref
{next-to-leading-dispersion-rewritten}) for soft $p$. We find:
\begin{equation}
\Omega ^{2}\left( p\right) =m_{p}^{2}\left[ \left( 1-0.368\,e+\dots\right)
+\left( \frac{3}{5}-0.0536\,e+\dots\right) \bar{p}^{2}+\mathcal{O}(\bar{p}
^{4})\right] .  \label{Omega-expanded}
\end{equation}
Here we have together the leading and next to leading orders in the coupling 
$e$. What remains to do is to perform the expansion in powers of $\bar{p}$, 
\textit{before} the Matsubara sum and analytic continuation to real energies
are done, in a way similar to what we have done in this work for the
transverse gluon damping rate. This we do and we obtain the \textit{same}
result (\ref{Omega-expanded}). This clearly indicates that the calculation
method we use is likely to have little to do with the infrared sensitivity
of the results.

\subsection{Final comments}

Trying to determine the damping rates for quasi-particles is essentially a
mean of trying to understand the (analytic) structure of QCD in the infrared
at finite temperature. Already a $1/\bar{\mu}^{4}$ appears at order $\bar{p}
^{2}$. If by some mean we are able to manage an expression for the
coefficient of $\bar{p}^{4}$, should we expect a $1/\bar{\mu}^{6}$? This is
not certain. Indeed, if we look at the damping rates $\gamma _{\pm }(p)$ for
quarks in an expansion in powers of $p$, the zeroth-order and first-order
coefficients are free from infrared divergences; only the second-order
coefficient, that of $p^{2}$, is infrared sensitive, and only
logarithmically \cite{ADB}:
\begin{equation}
\gamma _{\pm }\left( p\right) =\frac{g^{2}C_{f}T}{16\pi }\left[ 5.705672\mp
1.056796\,\tilde{p}+\left( \allowbreak 8.553573-5.\allowbreak
968072\allowbreak \ln \tilde{\mu}\right) \tilde{p}^{2}+\mathcal{O}\left( 
\tilde{p}^{3}\right) \right] ,  \label{result for quarks}
\end{equation}
where $\tilde{p}=p/m_{f}$ and $\tilde{\mu}=\mu /m_{f}$ with $m_{f}=\sqrt{
C_{f}/8}\,gT$ the quark thermal mass. Recall that $C_{f}=\left(
N_{c}^{2}-1\right) /2N_{c}$ and here, $N_{c}=N_{f}=3$. The same behavior is
obtained for other values of $N_{c}$ and $N_{f}$. We are not able to explain
why there is such a difference in the infrared sensitivity between the
quarks and the gluons.

Once the work on the dampings is finished, it would be interesting to look
at $\omega ^{\left( 1\right) }(p)$, the correction to the energy at
next-to-leading order in HTL-summed perturbation. One would have to deal
with the real part of the HTL-dressed one-loop-order self-energy, something
much more difficult than dealing with the imaginary part. There is already
the work \cite{schulz}, where the next-to-leading order correction to $m_{g}$
is estimated in the pure gluonic case using the longitudinal one-loop
HTL-summed self-energy. The result is: 
\begin{equation}
\omega _{0}^{2}=m_{g}^{2}\left( 1-1.8N_{c}g+\dots \right) ,
\label{correction to m_g--schulz}
\end{equation}
and is gauge independent within the class of covariant gauges. The quantity $
\omega _{0}$ is the long-wavelength limit of the frequency spectrum and it
is this result we have used in the text after (\ref{lowest order in Omega}).
As one can see, the result is finite and the HTL-summation scheme poses no
problem. But how about trying to calculate the energy for moving ultrasoft
quasi-particles? Is there going to be infrared sensitivity as for the
damping rates?

\appendix

\section{Expansion in powers of momentum}

In the main text in section four, small-$k$ expansions have been frequently
used. They relied partly on the expansion of the gluon on-shell energies,
the residue and cut functions together with their first and second
derivatives. We give in this appendix the first terms of these expansions.
We note that in actual calculations, it very often occurs that the actual
expansion has to be pushed quite further, especially for the residue
functions and their derivatives. We do not show these extra terms here to
avoid saturation of the display.

Small-$k$ gluon energies are given by:
\begin{align}
\omega_{l}(k) & =1+\frac{3\hspace{1pt}k^{2}}{10}-\frac{3\hspace{1pt}k^{4}}{
280}+\frac{k^{6}}{6000}+\frac{489\hspace{1pt}k^{8}}{43120000}-\frac {79
\hspace{1pt}k^{10}}{509600000}-\frac{22129\hspace{1pt}k^{12}}{
543\allowbreak312000000}+...\hspace{2pt};  \notag \\
\omega_{t}(k) & =1+\frac{3\hspace{1pt}k^{2}}{5}-\frac{9\hspace{1pt}k^{4}}{35}
+\frac{88\hspace{1pt}k^{6}}{375}-\frac{91617\hspace{1pt}k^{8}}{336875}+\frac{
706662\hspace{1pt}k^{10}}{1990625}-\frac{264087844\hspace {1pt}k^{12}}{
530578125}+...\hspace{2pt}.  \label{omega_l,t}
\end{align}
Their first derivatives are given by the following expansions: 
\begin{align}
\omega_{l}^{\prime}\left( k\right) & =\frac{3\hspace{1pt}k}{5}-\frac{3
\hspace{1pt}k^{3}}{70}\hspace{-2pt}+\frac{k^{5}}{1000}+\frac {489\hspace{1pt}
k^{7}}{5390000}-\frac{79\hspace{1pt}k^{9}}{50960000}-\frac{22129\hspace{1pt}
k^{11}}{45\allowbreak276000000}  \notag \\
& -\frac{2677\hspace{1pt}k^{13}}{381180800000}+...\hspace{2pt};  \notag \\
\omega_{t}^{\prime}\left( k\right) & =\frac{6k}{5}-\frac{36k^{3}}{35}+\frac{
176k^{5}}{125}-\frac{732936k^{7}}{336875}+\frac{1413324k^{9}}{398125}-\frac{
1056351376k^{11}}{176859375}  \notag \\
& +\frac{3815200504k^{13}}{372246875}+...\hspace{2pt},
\label{omega^prime_l,t}
\end{align}
and their second derivatives by the following expansions: 
\begin{align}
\omega_{l}^{\prime\prime}(k) & =\frac{3}{5}-\frac{9k^{2}}{70}+\frac{k^{4}}{
200}+\frac{489k^{6}}{770000}-\frac{711k^{8}}{50960000}-\frac{22129k^{10}}{
4116000000}-\frac{2677k^{12}}{29\allowbreak321600000}+...\hspace {2pt}; 
\notag \\
\omega_{t}^{\prime\prime}(k) & =\frac{6}{5}\hspace{-3pt}-\frac{108k^{2}}{35}
\hspace{-2pt}+\frac{176k^{4}}{25}\hspace{-2pt}-\frac{732936k^{6}}{48125}
\hspace{-2pt}+\frac{12719916k^{8}}{398125}\hspace{-2pt}-\frac {
1056351376k^{10}}{16\,078125}\hspace{-2pt}  \notag \\
& +\frac{3815200504k^{12}}{28634375}+...\hspace{2pt}.
\label{omega^second_l,t}
\end{align}

The residue functions are expanded as follows: 
\begin{align}
\mathfrak{z}_{l}(k) & =\frac{-1}{2k^{2}}\left[ 1-\frac{3k^{2}}{10}+\frac{
9k^{4}}{280}-\frac{k^{6}}{1200}-\frac{489k^{8}}{6160000}+\frac {711k^{10}}{
509600000}+\frac{22129k^{12}}{49\allowbreak392000000}+...\right] ;  \notag \\
\mathfrak{z}_{t}(k) & =\frac{1}{2}\hspace{-3pt}-\frac{2k^{2}}{5}\hspace {-3pt
}+\frac{19k^{4}}{35}\hspace{-3pt}-\frac{314k^{6}}{375}\hspace{-3pt}+\frac{
736k^{8}}{539}\hspace{-3pt}-\frac{13717342k^{10}}{5971875}\hspace {-3pt}+
\frac{697\,037162k^{12}}{176859375}\hspace{-3pt}  \notag \\
& -\frac{10718685944k^{14}}{1563436875}\hspace{-3pt}+\hspace{-3pt}...\hspace{
2pt}.  \label{residue_l,t}
\end{align}
Their first derivatives are given by: 
\begin{align}
\mathfrak{z}_{l}^{\prime }(k) & =\frac{1}{k^{3}}\left[ 1-\frac{9k^{4}}{280}+
\frac{k^{6}}{600}+\frac{1467k^{8}}{6160000}-\frac{711k^{10}}{127400000}-
\frac{22129k^{12}}{9878400000}+...\right] ;  \notag \\
\mathfrak{z}_{t}^{\prime}(k) & =-\frac{4k}{5}+\frac{76k^{3}}{35}-\frac{
628k^{5}}{125}+\frac{5888k^{7}}{539}-\frac{27434684k^{9}}{1194375}+\frac{
2788148648k^{11}}{58953125}  \notag \\
& -\frac{21437371888k^{13}}{223348125}+...\hspace{2pt},
\label{residue^prime_l,t}
\end{align}
and their second derivatives by: 
\begin{align}
\mathfrak{z}_{l}^{\prime\prime}\hspace{-2pt}(k) & =-\frac{3}{k^{4}}-\frac {9
}{280}+\frac{k^{2}}{200}+\frac{1467k^{4}}{1232000}-\frac{711k^{6}}{18200000}-
\frac{22129k^{8}}{1097600000}-\frac{8031k^{10}}{18\allowbreak 659200000}+...
\hspace{2pt};\allowbreak  \notag \\
\mathfrak{z}_{t}^{\prime\prime}\hspace{-2pt}(k) & =-\frac{4}{5}+\frac{
228k^{2}}{35}-\frac{628k^{4}}{25}+\frac{5888k^{6}}{77}-\frac {82304052k^{8}}{
398125}+\frac{2788148648k^{10}}{5359375}  \notag \\
& -\frac{21\allowbreak437371888k^{12}}{17180625}+...\hspace{2pt}.
\label{residue^second_l,t}
\end{align}

The longitudinal cut function expands as follows: 
\begin{align}
\beta _{l}(1-\omega _{l},k)& =\frac{1}{20}k-\left[ \frac{1097}{42000}+\frac{
9\pi ^{2}}{8000}\right] k^{3}+\left[ \frac{118511}{12600000}+\frac{6807\pi
^{2}}{5600000}+\frac{81\pi ^{4}}{3200000}\right] \allowbreak k^{5}  \notag \\
& -\left[ \frac{1065078779}{363\allowbreak 825000000}+\frac{3029563\pi ^{2}}{
3920000000}+\frac{92907\pi ^{4}}{2240000000}\allowbreak +\frac{729\pi ^{6}}{
1280000000}\right] \allowbreak k^{7}  \notag \\
& +\left[ \frac{16\,801\allowbreak 915872097}{19864845000000000}+\frac{
85\allowbreak 916963857\pi ^{2}}{226380000000000}\allowbreak +\frac{
59428197\pi ^{4}}{1568\allowbreak 000000000}\right.  \notag \\
& +\left. \frac{160137\pi ^{6}}{128000000000}\allowbreak +\frac{6561\pi ^{8}
}{512000000000}\right] \allowbreak k^{9}+...\hspace{2pt};  \notag \\
\beta _{l}(1-\omega _{t},k)& =\frac{1}{10}k-\left[ \frac{197}{5250}+\frac{
9\pi ^{2}}{1000}\right] k^{3}-\left[ \frac{451}{28125}-\frac{1857\pi ^{2}}{
175000}-\frac{81\pi ^{4}}{100000}\right] k^{5}  \notag \\
& +\left[ \frac{120421003}{3789843750}-\frac{146863\pi ^{2}}{30625000}-\frac{
28107\allowbreak \pi ^{4}}{17500000}-\frac{729\pi ^{6}}{10000000}\right]
\allowbreak k^{7}  \notag \\
& -\left[ \frac{2454132521881}{77597050781250}+\frac{85705817\pi
^{2}}{110537109375}-\frac{580509\pi ^{4}}{382812500}\right. 
\notag \\
& -\left. \frac{50787\pi ^{6}}{250000000}-\frac{6561\pi ^{8}}{1000000000}
\right] k^{9}+...\hspace{2pt}.  \label{cut_l}
\end{align}
In actual calculations, we had to expand the above cut functions till
$\mathcal{O}(k^{15})$. The same is true for the following transverse cut
functions: 
\begin{align}
\beta _{t}(1-\omega _{l},k)& =-\frac{40}{9\pi ^{2}k}-\left[ \frac{176}
{315\pi ^{2}}-\frac{516\,128}{3645\pi ^{4}}\right] k-\left[ \frac{
33298514048}{7381125\pi ^{6}}-\frac{827024}{30375\pi ^{4}}+\frac{
24721}{992250\pi ^{2}}\right]k^{3} \notag \\
& +\left[ \hspace{-2pt}\frac{2148286932320768}{14946778125\pi
^{8}}-\frac{1508396467072}{1291696875\pi ^{6}}\hspace{-2pt}+\hspace{-2pt}
\frac{3742632736}{1674421875\pi ^{4}}\hspace{-2pt}+\hspace{-2pt}
\frac{1025393}{1910081250\pi ^{2}}\hspace{-2pt}\right] \hspace{-2pt}k^{5}
\notag \\
& -\left[ \frac{138598879725606668288}
{30267225703125\pi ^{10}}-\frac{24468480689779456}
{523137234375\pi ^{8}}+\frac{1394874080310776}
{10172112890625\pi ^{6}}\right.  \notag \\
& -\left. \frac{38108673112}{644652421875\pi ^{4}}-
\frac{52966144687}{625742617500000\pi ^{2}}\right]
k^{7}+...\hspace{2pt};  \notag \\
\beta _{t}(1-\omega _{t},k)& =-\frac{20}{9\pi ^{2}k}+\left[  
\frac{173056}{3645\pi ^{4}}-\frac{184}{105\pi ^{2}}\right] k-\left[ \frac{
7487094784}{7381125\pi ^{6}}-\frac{3381248}{70875\pi ^{4}}-\frac{255916}
{496125\pi ^{2}}\right] k^{3}  \notag \\
& +\left[ \frac{323921668734976}{14946778125\pi ^{8}}-\frac{
533311520768}{430565625\pi ^{6}}-\frac{6589211648}{1674421875\pi
^{4}}-\frac{153921104}{318346875\pi ^{2}}\right] k^{5} \notag \\
& -\left[ \frac{14014147076150001664}{30267225703125\pi ^{10}}-
\frac{5431917214171136}{174379078125\pi ^{8}}+
\frac{1167210299260928}{10172112890625\pi ^{6}}\right.  \notag \\
& -\left. \frac{1529299484672}{214884140625\pi ^{4}}+ 
\frac{11073473445148}{19554456796875\pi ^{2}}\right]
k^{7}+...\hspace{2pt}. \label{cut_t}
\end{align}

The expansion of the first derivatives of the longitudinal cut functions
writes: 
\begin{align}
\left. \partial _{k}\beta _{l}(k,\omega )\right\vert _{\omega =1-\omega
_{l}}& =-\frac{1}{20}-\left[ \frac{2459}{42000}-\frac{27\pi ^{2}}{8000}
\right] k^{2}+\left[ \frac{20027}{360000}+\frac{183\pi ^{2}}{1120000}-\frac{
81\pi ^{4}}{640000}\right] k^{4} \notag \\
& -\left[ \frac{3467595349}{121275000000}+\frac{10718959\pi ^{2}}{3920000000}
-\frac{176499\pi ^{4}}{2240000000}-\frac{5103\pi ^{6}}{1280000000}\right] k^{6}
\notag \\
& +\left[ \frac{229778221708927}{19864845000000000}+
\frac{203974481679\pi ^{2}}{75460000000000}+
\frac{82788327\pi ^{4}}{1568000000000}\right. \notag \\
& -\left. \frac{627183\pi ^{6}}{128000000000}-\frac{59049\pi ^{8}}
{512000000nn000}\right] k^{8}+...\hspace{2pt}; \notag \\
\left. \partial _{k}\beta _{l}(k,\omega )\right\vert _{\omega =1-\omega
_{t}}& =-\frac{1}{10}-\left[ \frac{1259}{5250}-\frac{27\pi ^{2}}{1000}\right]
k^{2}+\left[ \frac{7937}{39375}+\frac{633\pi ^{2}}{35000}-\frac{81\pi ^{4}}
{20000}\right] k^{4} \notag \\
& -\left[ \frac{53836759}{1624218750}+\frac{1811059\pi ^{2}}{30625000}-
\frac{22599 \pi ^{4}}{17500000}-\frac{5103\pi ^{6}}{10000000}\right] k^{6}
\notag \\
& -\left[ \frac{4091313116071}{77597050781250}-\frac{1804152701 \pi^{2}}
{36845703125}-\frac{2970819 \pi^{4}}{382812500}\right. \notag \\
& -\left. \frac{153333 \pi^{6}}{250000000}-\frac{59049\pi ^{8}}{1000000000}
\right] k^{8}+...\hspace{2pt}. \label{d_k cut_l}
\end{align}

It is important to note that these expressions are not obtained by simply
deriving (\ref{cut_l}). One has first to calculate $\partial _{k}\beta
_{l}(k,\omega )$, then replace $\omega $ by $1-\omega _{l,t}$, and only then
perform the expansion. The derivative of the transverse cut function is
given by: 
\begin{align}
\left. \partial _{k}\beta _{t}(k,\omega )\right\vert _{\omega =1-\omega _{l}}
\hspace{-3pt}& =\hspace{-2pt}\frac{-40}{9\pi ^{2}k^{2}}\hspace{-2pt}+\hspace{
-2pt}\frac{2735072}{3645\pi ^{4}}\hspace{-2pt}+\hspace{-2pt}\frac{76}{315\pi
^{2}}\hspace{-2pt}-\hspace{-2pt}\hspace{-2pt}\left[ \hspace{-2pt}\frac{
319613296256}{7381125\pi ^{6}}-\frac{24183296}{212625\pi ^{4}}-
\frac{89813}{992250\pi ^{2}}\hspace{-2pt}\right]\hspace{-2pt}k^{2} \notag \\
& \hspace{-3pt}+\hspace{-3pt}\left[ \hspace{-2pt}\frac{5971222733143552}
{2989355625\pi ^{8}}\hspace{-2pt}-\hspace{-2pt}\frac{2602609504448}
{258339375\pi ^{6}}\hspace{-2pt}+\hspace{-2pt}
\frac{1874876804}{334884375\pi ^{4}}\hspace{-2pt}+\hspace{-2pt}
\frac{3814589}{764032500\pi ^{2}}\hspace{-2pt}\right] \hspace{-5pt}k^{4} \notag \\
& -\left[ \frac{2522063079101393782784}
{30267225703125\pi ^{10}}-\frac{45494694191332864}
{74733890625\pi ^{8}}\right.  \notag \\
& +\hspace{-3pt}\left. \frac{10313692441736168}{10172112890625\pi ^{6}}
\hspace{-2pt}+\hspace{-2pt}\frac{98656172204}{644652421875\pi ^{4}}\hspace{-2pt}
+\hspace{-2pt}\frac{145623897509}{625742617500000\pi ^{2}}\hspace{-2pt}\right]
\hspace{-3pt}k^{6}\hspace{-2pt}+\hspace{-2pt}...\hspace{2pt}; 
\notag \\
\left. \partial _{k}\beta _{t}(k,\omega )\right\vert _{\omega =1-\omega_{t}}
& =\frac{-20}{9\pi ^{2}k^{2}}+\frac{492544}{3645\pi ^{4}}-\frac{16}{105\pi ^{2}}
-\left[ \frac{35131752448}{7381125\pi ^{6}}-\frac{3223552}{23625\pi ^{4}}
-\frac{487252}{496125\pi ^{2}}\right]k^{2}  \notag \\
& +\hspace{-3pt}\left[ \hspace{-3pt}\frac{423589874499584}
{2989355625\pi ^{8}}-\frac{565829435392}{86113125\pi ^{6}}
-\frac{5720261632}{334884375\pi ^{4}}-\frac{45637336}
{63669375\pi ^{2}}\hspace{-3pt}\right] \hspace{-3pt}k^{4} \notag \\
& -\left[ \frac{117503233176950013952}{30267225703125\pi ^{10}}
-\frac{41116968270430208}{174379078125\pi ^{8}}\right. \label{d_k cut_t} \\
& +\left. \frac{8361475700424704}{10172112890625\pi ^{6}}
-\frac{482750388224}{23876015625\pi ^{4}}-
\frac{15508324754764}{19554456796875\pi ^{2}}\right]k^{6}+...\hspace{2pt}. \notag
\end{align}

The derivative over $\omega$ of the longitudinal cut function is
expanded as: 
\begin{align}
\left. \partial_{\omega}\beta_{l}(k,\omega)\right| _{\omega=1-\omega_{l}} & =
\frac{-1}{6k}+\left[ \frac{19}{900}+\frac{9\pi^{2}}{800}\right] k+\left[ 
\frac{4063}{252000}-\frac{507\pi^{2}}{56000}-\frac{27\pi^{4}}{64000}\right] k^{3}
\notag \\
& -\left[ \frac{245274179}{19845000000}-\frac{1487147\pi^{2}}{392000000}-
\frac{65529\pi^{4}}{112000000}-\frac{1701\pi^{6}}{128000000}
\right] k^{5} \notag \\
& +\left[ \frac{371842066997}{65488500000000}-\frac{973840051\pi
^{2}}{1078000000000}-\frac{68681091\pi^{4}}{156800000000}\right. 
\notag \\
& -\left. \frac{166293\pi^{6}}{6400000000}-\frac{19683\pi^{8}}{51200000000}
\right] k^{7}+...\hspace{2pt};  \notag \\
\left. \partial_{\omega}\beta_{l}(k,\omega)\right| _{\omega=1-\omega_{t}} & =
\frac{-1}{6k}-\left[ \frac{56}{225}-\frac{9\pi^{2}}{200}\right] k+\left[ 
\frac{1709}{7875}+\frac{33\pi^{2}}{3500}-\frac{27\pi^{4}}{4000}\right]
k^{3}  \notag \\
& -\left[ \frac{11507054}{310078125}+\frac{404403\pi^{2}}{6125000}
-\frac{4077\pi^{4}}{875000}-\frac{1701}{2000000}\right] k^{5}
\notag \\
& -\left[ \frac{26289763331}{511628906250}-\frac{764943847\pi^{2}}
{12632812500}-\frac{536523\pi^{4}}{76562500}\right.  \notag \\
& +\left. \frac{32643\pi^{6}}{25000000}+\frac{19683\pi^{8}}{200000000}\right]
k^{7}+...\hspace{2pt};  \label{d_omega cut_l}
\end{align}
The corresponding results for the transverse cut function are: 
\begin{align}
\left. \partial_{\omega}\beta_{t}(k,\omega)\right| _{\omega=1-\omega_{l}} & =
\frac{-400}{27\pi^{2}k^{3}}+\left[ \frac{739648}{729\pi^{4}}+\frac {52}
{189\pi^{2}}\right] \frac{1}{k}-\left[ \frac{219717754112}{4428675\pi^{6}}
-\frac{16061392}{127575\pi^{4}}\right.  \notag \\
& \left. -\frac{95188}{297675\pi^{2}}\right] k+\hspace {-2pt}
\left[ \hspace{-2pt}\frac{3822935800822784}{1793613375\pi^{8}}\hspace{-2pt}
-\hspace{-2pt}\frac{547835807552}{51667875\pi^{6}}\hspace{-2pt}-\hspace{-2pt}
\frac{400625132}{200930625\pi^{4}}\right.  \notag \\
& \left. \hspace{-2pt}+\frac{4350391}{152806500\pi^{2}}\right]\hspace{-2pt}k^{3}
-\hspace{-2pt}\left[ \hspace{-2pt}\frac{172430102335794122752}{2017815046875\pi^{10}}
-\hspace{-2pt}\frac{194970192021158656}{313882340625\pi^{8}}\right.  \notag \\
& +\hspace{-2pt}\left. \frac{1555758096434912}{2034422578125\pi^{6}}\hspace{-2pt}
+\hspace{-2pt}\frac {80143211084}{55255921875\pi^{4}}\hspace{-2pt}-\hspace{-2pt}
\frac{2836963147}{62574261750000\pi^{2}}\right] \hspace{-2pt}k^{5}
\hspace{-2pt}+...\hspace{2pt};  \notag \\
\left. \partial_{\omega}\beta_{t}(k,\omega)\right| _{\omega=1-\omega_{t}} & =
\frac{-100}{27\pi^{2}k^{3}}\hspace{-2pt}+\hspace{-2pt}\left[ \hspace {-2pt}
\frac{53248}{729\pi^{4}}\hspace{-2pt}-\hspace{-2pt}\frac{116}{63\pi^{2}}
\hspace{-2pt}\right] \hspace{-2pt}\frac{1}{k}-\hspace{-2pt}\left[ 
\hspace{-2pt}\frac{6335234048}{4428675\pi^{6}}-\hspace{-2pt}\frac {2425856}
{42525\pi^{4}}-\frac{683552}{297675\pi^{2}}\right] \hspace {-2pt} k \notag \\
& +\hspace{-2pt}\left[ \frac{49834102882304}{1793613375\pi^{8}}-
\frac{26537099264}{17222625\pi^{6}}-\frac{6961211392}
{200930625\pi^{4}}-\frac{14887772}{12733875\pi^{2}}\right] k^{3} 
\notag \\
& -\left[ \frac{1078011313550000128}{2017815046875\pi^{10}}
-\frac{576925575675904}{14946778125\pi^{8}}-\frac{695975917125632}{2034422578125\pi^{6}}\right.
\notag \\
& -\hspace{-2pt}\left. \frac{1435899206656}{128930484375\pi^{4}}-
\frac{5003479782988}{3910891359375\pi^{2}}\right]
k^{5}+...\hspace{2pt}.  \label{d_omega cut_t}
\end{align}

We turn to the second derivatives. We have for the longitudinal cut function: 
\begin{align}
\left. \partial_{k}^{2}\beta_{l}(k,\omega)\right| _{\omega=1-\omega_{l}} & =
\frac{1}{10k}+\left[ \frac{731}{7000}-\frac{27\pi^{2}}{2000}\right] k+\left[ 
\frac{144943}{1260000}-\frac{2157\pi^{2}}{280000}+\frac{243\pi^{4}}{320000}
\right] k^{3}  \notag \\
& -\left[ \frac{4333821991}{25987500000}-\frac{4008843\pi^{2}}{980000000}
-\frac{25029\pi^{4}}{1120000000}+\frac{5103\pi^{6}}{160000000}
\right] k^{5}  \notag \\
& +\left[ \frac{8237191315679}{73573500000000}+\frac{28393782023\pi^{2}}
{3773000000000}-\frac{76104117\pi^{4}}{156800000000}\right.  \notag \\
& +\left. \frac{53217\pi^{6}}{3200000000}+\frac{59049\pi^{8}}
{51200000000}\right] k^{7}+...\hspace{2pt};  \notag \\
\left. \partial_{k}^{2}\beta_{l}(k,\omega)\right| _{\omega=1-\omega_{t}} & =
\frac{1}{5k}+\left[ \frac{681}{875}-\frac{27\pi^{2}}{250}\right] k+
\left[ \frac{17678}{39375}-\frac{1977\pi^{2}}{8750}+\frac{243\pi^{4}}{10000}\right]
k^{3}  \notag \\
& -\left[ \frac{5151227623}{5684765625}-\frac{1033443\pi^{2}}{7656250}
-\frac{231579\pi^{4}}{8750000}+\frac{5103\pi^{6}}{1250000}\right]
k^{5}  \notag \\
& +\left[ \frac{979623985177}{2586\,568359375}
+\frac {1714903248\pi^{2}}{7369140625}-\frac{2354859\pi^{4}}{38281250}\right.  \notag \\
& -\left. \frac{5103\pi^{6}}{6250000}+\frac{59049\pi^{8}}{100000000}
\right] k^{7}+...\hspace{2pt}.  \label{d_k^2 cut_l}
\end{align}
The corresponding results for the transverse cut function are: 
\begin{align}
\left. \partial_{k}^{2}\beta_{t}(k,\omega)\right| _{\omega=1-\omega_{l}} & = 
\left[ \frac{4841152}{1215\pi^{4}}-\frac{4}{5\pi^{2}}\right] \frac{1}{k}-
\left[ \frac{1034\allowbreak972897792}{2460375\pi^{6}}-\frac {41030672}
{70875\pi^{4}}+\frac{353}{875\pi^{2}}\right] k  \notag \\
& +\hspace{-2pt}\left[ \hspace{-2pt}\frac{9324383097777152}
{332150625\pi^{8}}\hspace{-2pt}-\hspace{-2pt}\frac{8040766589504}
{86113125\pi^{6}}\hspace{-2pt}+\hspace{-2pt}\frac{4285921012}
{111628125\pi^{4}}\hspace{-2pt}-\hspace{-2pt}\frac{8089}{157500\pi^{2}}
\hspace{-2pt}\right] \hspace{-2pt}k^{3}  \notag \\
& -\left[ \frac{15447111103341084491776}
{10089075234375\pi^{10}}-\frac{479468923614041344}{58126359375\pi^{8}}\right.  \notag \\
& +\left. \frac{31407182749531952}{3390704296875\pi^{6}}-\frac{399195734764}
{214884140625\pi^{4}}+\frac{463459}
{336875000\pi^{2}}\right] k^{5}+...\hspace{2pt};  \notag \\
\left. \partial_{k}^{2}\beta_{t}(k,\omega)\right| _{\omega=1-\omega_{t}} & = 
\left[ \frac{763904}{1215\pi^{4}}-\frac{8}{5\pi^{2}}\right] \frac{1}{k}
-\left[ \frac{83111182336}{2460375\pi^{6}}-\frac{60286976}
{70875\pi^{4}}+\frac{2424}{875\pi^{2}}\right] k  \notag \\
& +\hspace{-2pt}\left[ \hspace{-2pt}\frac{433173355823104}{332150625\pi^{8}}
\hspace{-2pt}-\hspace{-2pt}\frac{1589288370176}{28704375\pi^{6}}\hspace{-2pt}
+\hspace{-2pt}\frac{4023629824}{111628125\pi^{4}}\hspace{-2pt}+\hspace{-2pt}
\frac{27848}{39375\pi^{2}}\hspace{-2pt}\right] \hspace{-2pt}k^{3}  \notag \\
& -\left[ \frac{438501832773261590528}{10089075234375\pi^{10}}-\frac{20721255879540736}
{8303765625\pi^{8}}\right.  \notag \\
& +\hspace{-2pt}\left. \frac{41103339808096256}{3390704296875\pi^{6}}\hspace{-2pt}
-\hspace{-2pt}\frac{16084864550912}
{214884140625\pi^{4}}\hspace{-2pt}+\hspace{-2pt}\frac{3965392}{6015625\pi^{2}}
\hspace{-2pt}\right] \hspace{-2pt}k^{5}\hspace{-2pt}+...\hspace{2pt}. \label{d_k^2 cut_t}
\end{align}

We also need the second derivatives with respect to $\omega$. We have for
the longitudinal cut function: 
\begin{align}
\partial_{\omega}^{2}\left. \beta_{l}\left( \omega,k\right) )\right|
_{\omega=1-\omega_{l}} & =\left[ \frac{3}{5}-\frac{3\pi^{2}}{40}\right] 
\frac{1}{k}-\left[ \frac{51}{175}-\frac{71\pi^{2}}{5600}-\frac{9\pi^{4}}{1600}\right] k
+\left[ \frac{395161}{5250000}+\frac {17151\pi^{2}}{700000}\right.  \notag \\
& \hspace{-0.5in}-\left. \frac{15327\pi^{4}}{2800000}-\frac{1701\pi^{6}}
{6400000}\right] k^{3}-\left[ \frac{22213679651\pi^{2}}{970200000000}
-\frac{4137597\pi^{4}}{1960000000}\right.  \notag \\
& \hspace{-0.5in}-\left. \frac{274023\pi^{6}}{640000000}-\frac{6561\pi^{8}}{640000000}\right]
k^{5}-\left[ \frac{192681536863}{36786750000000}-\frac{4104150013\pi^{2}}
{339026688000}\right.  \notag \\
& \hspace{-0.5in}-\hspace{-2pt}\left. \frac{4045427293\pi^{4}}{22638000000000}
\hspace{-2pt}+\hspace{-2pt}\frac{3871503\pi^{6}}{11200000000}\hspace{-2pt}+\hspace{-2pt}
\frac{115911\pi^{8}}{5120000000}\hspace{-2pt}+\hspace{-2pt}\frac{72171\pi^{10}}
{204800000000}\hspace{-2pt}\right] \hspace{-2pt} k^{7}\hspace{-2pt}+\hspace{-2pt}...
\hspace{2pt};  \notag \\
\left. \partial_{\omega}^{2}\beta_{l}(k,\omega)\right| _{\omega=1-\omega
_{t}} & =\hspace{-2pt}\left[ \hspace{-2pt}\frac{6}{5}\hspace{-2pt}-\hspace{-2pt}
\frac{3\pi^{2}}{20}\hspace{-2pt}\right] \hspace{-2pt}\frac {1}{k}\hspace{-2pt}
+\hspace{-2pt}\left[ \hspace{-2pt}\frac{162}{175}\hspace{-2pt}
-\hspace{-2pt}\frac{319\pi^{2}}{700}\hspace{-2pt}+\hspace {-2pt}\frac{9\pi^{4}}{200}
\hspace{-2pt}\right] \hspace{-2pt}k\hspace {-2pt}-\hspace{-2pt}
\left[ \frac{69704}{46875}-\frac{2554\pi^{2}}{9375}-\frac{4923\pi^{4}}{87500}\right.  \notag \\
& \hspace{-0.5in}+\left. \frac{1701\pi^{6}}{200000}\right] k^{3}
+\left[ \frac{1256517454}{1894921875}+\frac{925954283\pi^{2}}{2526562500}-\frac{2028603\pi^{4}}
{15312500}-\frac{5427\pi^{6}}{5000000}\right.  \notag \\
& \hspace{-0.5in}+\left. \frac{6561\pi^{8}}{5000000}\right] k^{5}
-\left[ \frac{6968609678}{287396484375}-\frac{11304813868\pi^{2}}{19159765625}
+\frac{1591770343\pi^{4}}{44214843750}\right. \notag \\
& \hspace{-0.5in}+\left. \frac{2226393\pi^{6}}{87500000}-\frac{2187\pi^{8}}{2000000}
-\frac{72171\pi^{10}}{400000000}\right] k^{7}+...\hspace{2pt}.  \label{d_omega^2 cut_l}
\end{align}
For the transverse cut function we have: 
\begin{align}
\left. \partial_{\omega}^{2}\beta_{t}(k,\omega)\right| _{\omega=1-\omega
_{l}} & =\frac{-8000}{81\pi^{2}k^{5}}-\left[ \frac{2000}{189\pi^{2}}-\frac{27904000}
{2187\pi^{4}}\right] \frac{1}{k^{3}}-\left[ \frac{178084}{59535\pi^{2}}-\frac{188938432}
{76545\pi^{4}}\right.  \notag \\
& \hspace{-1.2in}+\left. \frac{748658115584}{885735\pi^{6}}\right]
\frac{1}{k}-\left[ \frac{14773972}{22920975\pi^{2}}-\frac {5463228688}{17222625\pi^{4}}
+\frac{6853501867264}{31000725\pi^{6}}\right. 
\notag \\
& \hspace{-1.2in}-\left. \frac{48565749544951808}{1076168025\pi^{8}}\right] k
-\left[ \frac{278379176783}{4693069631250\pi^{2}}-\frac{3300542659852}
{77358290625\pi^{4}}\right.  \notag \\
& \hspace{-1.2in}+\left. \frac{37302948014309056}{1220653546875\pi^{6}}
-\frac{309677336393249792}{20925489375\pi^{8}}+\frac{2605511136700461694976}
{1210689028125\pi^{10}}\right] k^{3}+...\hspace{2pt};  \notag \\
\left. \partial_{\omega}^{2}\beta_{t}(k,\omega)\right| _{\omega=1-\omega_{t}}
& =\frac{-1000}{81\pi^{2}k^{5}}-\left[ \frac{1000}{63\pi^{2}}-\frac{1088000}
{2187\pi^{4}}\right] \frac{1}{k^{3}}-\left[ \frac {233368}{59535\pi^{2}}
-\frac{29452288}{25515\pi^{4}}\right.  \notag \\
& \hspace{-1.2in}+\left. \frac{13866631168}{885735\pi^{6}}\right]
\frac{1}{k}+\left[ \frac{41076832}{7640325\pi^{2}}+\frac{86458213888}{120558375\pi^{4}}
-\frac{456797978624}{10333575\pi^{6}}\right.  \notag \\
& \hspace{-1.2in}+\left. \frac{476299021778944}{1076168025\pi^{8}}\right] k
+\left[ \frac{7827935167072}{2346534815625\pi^{2}}+\frac{4077064460288}
{25786096875\pi^{4}}\right.  \notag \\
& \hspace{-1.2in}-\left. \frac{43774954685857792}{1220653546875\pi^{6}}
+\frac{3273422343766016}{2325054375\pi^{8}}-\frac{14262918917738463232}
{1210689028125\pi^{10}}\right] k^{3}+...\hspace{2pt}. \label{d_omega^2 cut_t}
\end{align}

\section{The other terms}

In the main text, we have shown how to determine the $3gll$ contribution to
the coefficient of order $p^{2}$ in $\gamma _{t}\left( p\right) $. Here we
give the expressions of the other contributions. Their derivation follows
the same steps as in the main text. Whenever there is an infrared
divergence, it is extracted as explained there. All light-cone divergences
are brought under control as explained too. Recall that all is in unit of
$m_{g}$ and we use the same notation as in the main text, except that
$\bar{\mu}=\mu /m_{g}$ is directly used instead of $\mu $.

We have: 
\begin{align}
& \hspace{-3.5in}\int \mathcal{D}\frac{1}{\omega \omega ^{\prime }}\left[
4+93k^{4}+18k^{6}+(8+78k^{4})\omega -(140+78k^{2}+90k^{4})\omega
^{2}+(48-180k^{2})\omega ^{3}\right.  \notag \\
\hspace{0.5in}-\left. (31-126k^{2})\omega ^{4}+102\omega ^{5}-54\omega ^{6}
\right] \rho _{t}\rho _{l}^{\prime }& =-\frac{0.7505272862}
{\bar{\mu}^{4}}+\frac{6.404458463}{\bar{\mu}^{2}}  \notag \\
& +97.0603739\ln \bar{\mu}+73.01499652\hspace{2pt}.
\label{rho_t rho_l--final}
\end{align}
The infrared behavior here is $1/\bar{\mu}^{4}$ together with $1/\bar{\mu}
^{2}$ and $\ln \bar{\mu}$. Because of this, we have good convergence
starting from $\ell =0.2$ but the convergence is lost from $\ell =0.025$
downwards. This is not a problem because $\ell $ does not have to be so
small. The corresponding double integral is finite. We make the change of
variable $\kappa =1/k$. To get the number, we have first to push the
boundary $\omega =1/\kappa $ to $\omega =1/\kappa -\epsilon $ and the limit
$\kappa =2$ to $\kappa =2-\epsilon ^{\prime }$with the condition $\epsilon
^{\prime }>2\epsilon $ (in order for the arguments of the logarithms in the
cut functions to remain positive). With this, there is good convergence to
the number $-4.96229365...$ starting already at $\epsilon =10^{-6}$.

The next term is: 
\begin{align}
& \hspace{-3in}\int \mathcal{D}\frac{1}{\omega \omega ^{\prime }}\left[
-18k^{2}+122k^{4}-(144-174k^{2})\omega \omega ^{\prime }+\left( \frac{30}
{k^{2}}-138\right) \omega ^{2}\omega ^{\prime 2}\right.  \notag \\
\hspace{2in}-\left. \frac{174}{k^{2}}\omega ^{3}\omega ^{\prime 3}\right]
\rho _{t}\rho _{t}^{\prime }& =-\frac{880}{3\pi ^{2}}\ln \bar{\mu}
-9.45862819\hspace{2pt}.  \label{rho_t rho_t--final}
\end{align}
Here $\ell =0.2,0.1$ is sufficient. There is in fact good convergence: we
have a good number already for $\ell =0.3$. We note here that we have to
replace in $\int_{\ell }^{1}dxk_{t}^{\prime }\mathfrak{z}_{t}\beta _{t}f$
the upper boundary $1\,$by $0.9999...$ because the integral does not
converge numerically in the ultraviolet (analytically it is convergent).
However, for fixed $\ell $, say $\ell =0.1$, there is stability of the
integral when adding nines. The corresponding double integral does not pose
a problem.

Next we have: 
\begin{align}
& \hspace{-4in}\int \mathcal{D}\frac{1}{\omega ^{\prime }}\left[
-\allowbreak \left( \frac{9}{2k}+\frac{3}{4}k\right) +\frac{9}{k}\omega
+\left( \frac{3}{k^{3}}+\frac{33}{4k}\right) \omega ^{2}+\frac{6}{k^{3}}
\omega ^{3}-\allowbreak \left( \frac{57}{4k^{3}}+\frac{57}{2k^{5}}\right)
\omega ^{4}\right.  \notag \\
\hspace{2in}+\left. \frac{9}{k^{5}}\omega ^{5}+\frac{27}{4k^{5}}\omega ^{6}
\right] \Theta \rho _{t}^{\prime }& =\frac{9}{4}\ln \bar{\mu}-4.904610308\hspace{2pt}.
\label{theta rho_t--final}
\end{align}
For the double integral, one has to regularize at $\omega =1-k$: one shifts
the boundary to $\omega =1-k-\epsilon $. We have good convergence starting
at $\epsilon =0.0001$.

The next contribution is: 
\begin{equation}
\int \mathcal{D} \frac{1}{\omega ^{\prime }}\left[ -\frac{9}{2}
+3\omega -\frac{15}{k^{2}}\omega ^{2}+\dfrac{6}{k^{2}}\omega ^{3}+\frac{39}
{2k^{4}}\omega ^{4}-\frac{9}{k^{4}}\omega ^{5}\right] \Theta \hspace{1pt}
\partial _{k}\rho _{t}^{\prime }=4.464952656\hspace{2pt}. \label{theta d_k rho_t--final}
\end{equation}
This contribution is finite. The corresponding double integral has initially
an apparent light-cone divergence which is handled as explained in the text.

The next term is: 
\begin{align}
& \hspace{-2.5in}\int \mathcal{D} \frac{1}{\omega \omega ^{\prime}}
\left[ 69k+14k^{3}-3k^{5}-\left( 54k+16k^{3}-6k^{5}\right) \omega +\left(
2k+14k^{3}\right) \omega ^{2}+\left( 8k-12k^{3}\right) \omega ^{3}\right. 
\notag \\
\hspace{0.8in}\left. -11k\omega ^{4}+6k\omega ^{5}\right] \rho _{l}\hspace{1pt}
\partial _{k}\rho _{t}^{\prime }& = -\frac{1000}{27\pi ^{2}\bar{\mu}^{4}}
+\left( \frac{34120}{189\pi ^{2}}+\frac{54701440}{2187\pi ^{4}}
\right) \frac{1}{2\bar{\mu}^{2}}  \notag \\
& \hspace{-1.5in}+\left( \frac{4277584}{59535\pi ^{2}}+\frac{1938884128}
{76545\pi ^{4}}+\frac{1278453185024}{885735\pi ^{6}}\right) \ln \bar{\mu}+1477.282103
\hspace{2pt}. \label{rho_l d_k rho_t--final}
\end{align}
Here too there is a $1/\bar{\mu}^{4}$. But there is no light-cone divergence.

The next term is: 
\begin{align}
& \hspace{-2.5in}\int \mathcal{D}\frac{1}{\omega \omega ^{\prime}}
\left[ -12k-72k^{5}-\left( 24k+12k^{3}-48k^{5}\right) \omega -\left(12k-144k^{3}
\right) \omega ^{2}+(12k-96k^{3})\omega ^{3}\right.  \notag \\
\hspace{0.5in}-\left. 72k\omega ^{4}+48k\omega ^{5}\right] \rho _{t}\hspace{1pt}
\partial _{k}\rho _{l}^{\prime }& =-\frac{400}{3\pi ^{2}\bar{\mu}^{4}}
+\left( \frac{12136}{63\pi ^{2}}+\frac{8566912}{729\pi ^{4}}\right)
\frac{1}{\bar{\mu}^{2}}  \notag \\
& +\hspace{-1.9in}\left( 5+\frac{317576216}{99225\pi ^{2}}+\frac{2290058432}
{127575\pi ^{4}}+\frac{1411647241216}{1476225\pi ^{6}}\right) \ln \bar{\mu}
+1109.862748\hspace{2pt}. \label{rho_t d_k rho_l--final}
\end{align}
The stability band here is a little narrow: $\ell =0.1,0.045$. Also, the
double integral has an apparent light-cone divergence and is treated as
usual. We note that in order to have a convergence in the double integral,
$k_{\text{max}}=400$.

The next term to discuss is: 
\begin{align}
& \hspace{-4in}\int \mathcal{D}\frac{1}{\omega \omega ^{\prime }}\left[
32k^{3}+20k^{5}+\left( 112k+124k^{3}\right) \omega +\left( \frac{48}{k}
-36k-116k^{3}\right) \omega ^{2}\allowbreak -\left( \frac{108}{k}
+232k\right) \omega ^{3}\right.  \notag \\
\hspace{0.85in}+\left. \left( \frac{12}{k}+156k\right) \omega ^{4}+\frac{108}
{k}\omega ^{5}-\frac{60}{k}\omega ^{6}\right] \rho _{t}\partial _{k}\rho _{t}^{\prime }
& =-\frac{928}{27\pi ^{2}}\ln \bar{\mu}-1.97446447\hspace{2pt}. 
\label{rho_t d_k rho_t--final}
\end{align}
There are no light-cone divergences here.

The next integral is: 
\begin{equation}
\int \mathcal{D}\frac{1}{\omega ^{\prime }} \left( -3k+\frac{3}{k}\omega ^{2}\right)
\Theta \hspace{1pt}\partial _{k}^{2}\rho _{t}^{\prime}=-3.245787103\hspace{2pt}.
\label{theta d_k^2 rho_t--final}
\end{equation}
No infrared and no light-cone divergences.

The next contribution is: 
\begin{align}
& \hspace{-0.1in}\int \mathcal{D}\frac{1}{\omega \omega ^{\prime }}\left[
30k^{2}+4k^{4}-2k^{6}-\left( 24k^{2}+8k^{4}\right) \omega -\left(
4k^{2}-4k^{4}\right) \omega ^{2}+8k^{2}\omega ^{3}-2k^{2}\omega ^{4}\right]
\rho _{l}\hspace{1pt}\partial _{k}^{2}\rho _{t}^{\prime }  \notag \\
& =\hspace{-2pt}\left[ \hspace{-2pt}\frac{19364608}{729\pi ^{4}}\hspace{-2pt}
-\hspace{-2pt}\frac{16}{3\pi ^{2}}\hspace{-2pt}\right] \hspace{-2pt}\frac{1}
{\bar{\mu}^{2}}\hspace{-2pt}+\hspace{-2pt}\left[ \hspace{-2pt}
\frac{176}{25\pi ^{2}}\hspace{-2pt}+\hspace{-2pt}\frac{6906341632}{127575\pi^{4}}
-\frac{8279783182336}{1476225\pi ^{6}}\hspace{-2pt}\right]\hspace{-2pt}
\ln \bar{\mu}\hspace{-2pt}+\hspace{-2pt}6810.290071\hspace{2pt}.
\label{rho_l d_k^2 rho_t--final}
\end{align}
The double integral has a light-cone divergence that is treated as usual.

The next contribution is: 
\begin{align}
\int \mathcal{D}\frac{-1}{\omega \omega ^{\prime }}\left[
8k^{2}+16k^{6}+16k^{2}\omega +\left( 8k^{2}-32k^{4}\right) \omega
^{2}+16k^{2}\omega ^{4}\right] \rho _{t}\partial _{k}^{2}\rho _{l}^{\prime}
& =\frac{5600}{9\pi ^{2}\bar{\mu}^{4}}  \notag \\
& \hspace{-4in}-\left[ \frac{123376}{189\pi ^{2}}+\frac{170390272}{2187\pi ^{4}}\right]
\frac{1}{\bar{\mu}^{2}}-\left[ 8+\frac{515581456}{99225\pi ^{2}}
+\frac{11866583750656}{1476225\pi ^{6}}\right.  \notag \\
& \hspace{-2.5in}\left. +\frac{4181643904}{42525\pi ^{4}}\right] \ln \bar{\mu }
-8137.840681\hspace{2pt}.  \label{rho_t d_k^2 rho_l--final}
\end{align}
Because of the presence of 1/$\bar{\mu}^{4}$ and because we are requiring
more precision, the stability band for the single integral is narrow: $\ell
=0.1,0.05$. The double integral is light-cone divergent and handled as usual.

The next contribution is: 
\begin{align}
& \hspace{-2.1in}\int \mathcal{D}\frac{1}{\omega \omega ^{\prime }}\left[
8k^{4}+\left( 20k^{2}+20k^{4}\right) \omega +\left( 12+12k^{2}\right) \omega ^{2}
-\left( 12+32k^{2}\right) \omega ^{3}-12\omega ^{4}+12\omega ^{5}\right]
\rho _{t}\partial _{k}^{2}\rho _{t}^{\prime }  \notag \\
\hspace{2in}& =-\frac{4400}{27\pi ^{2}}\ln \bar{\mu}-10.70561628\hspace{2pt}.
\label{rho_t d_k^2 rho_t--final}
\end{align}
Generally, the transverse terms are `smoother'. The double integral has no
light-cone divergence.

The next contribution is: 
\begin{equation}
\int_{\bar{\mu}}^{+\infty }dk\int_{-\infty }^{+\infty }d\omega \int_{-\infty
}^{+\infty }d\omega ^{\prime }\frac{-9\left( k^{2}-\omega ^{2}\right) ^{2}}
{k^{3}\omega ^{\prime }}\Theta \hspace{1pt}\rho _{t}^{\prime }\hspace{1pt}
\partial _{\omega }\delta \left( 1-\omega -\omega ^{\prime }\right)
=4.65989517\hspace{2pt}. \label{theta rho_t d_omega delta--final}
\end{equation}
No infrared or light-cone divergences, though this term is highly `uncommon'.

The next contribution is: 
\begin{equation}
\int_{\bar{\mu}}^{+\infty }\hspace{-2pt}dk\int_{-\infty }^{+\infty }\hspace{-2pt}
d\omega \int_{-\infty }^{+\infty }\hspace{-2pt}d\omega ^{\prime }
\hspace{-2pt}\frac{54k^{2}}{\omega \omega ^{\prime }}\left( k^{2}-\omega
^{2}\right) ^{2}\rho _{t}\rho _{l}^{\prime }\partial _{\omega }\delta \left(
1-\omega -\omega ^{\prime }\right) =-\frac{8000}{3\pi ^{2}}\ln \bar{\mu}
-268.9616004.  \label{rho_t rho_l d_omega delta--final}
\end{equation}
The double integral has a light-cone divergence which is treated as usual.

The next contribution is: 
\begin{equation}
\int_{\bar{\mu}}^{+\infty }\hspace{-2pt}dk\int_{-\infty }^{+\infty }\hspace{-6pt}
d\omega \int_{-\infty }^{+\infty }\hspace{-6pt}d\omega ^{\prime }
\frac{-108}{\omega \omega ^{\prime }}\left( k^{2}+\omega
\omega ^{\prime }\right) ^{2}\rho _{t}\rho _{t}^{\prime }\partial _{\omega}
\delta \left( 1-\omega -\omega ^{\prime }\right) =-\frac{1600}{3\pi ^{2}}
\ln \bar{\mu}-34.46857932. \label{rho_t rho_t d_omega delta--final}
\end{equation}
The double integral has no light-con divergence and converges fairly well.

The next contributions are: 
\begin{equation}
\int \mathcal{D}\frac{-24k^{3}}{\omega \omega ^{\prime }}\delta \left(
\omega ^{2}-k^{2}\right) \rho _{l}=1.8\hspace{1pt};\hspace{0.5in}
\int \mathcal{D}\frac{12\omega }{\omega ^{\prime }}\epsilon (\omega )\delta
\left( \omega ^{2}-k^{2}\right) \rho _{t}=3.0\hspace{1pt}.
\label{delta rho--finals}
\end{equation}
\begin{equation}
\int \mathcal{D}\frac{6k\omega }{\omega ^{\prime }}\epsilon (\omega )\rho
_{l}\partial _{\omega }\left[ \delta \left( k-\omega \right) -\delta \left(
k+\omega \right) \right] =-3.0\hspace{1pt}.
\label{rho_l d_omega delta--final}
\end{equation}
There are no light-cone problems with these integrals. Note that the values
here are `exact' though the integration is done numerically. We have no
analytical check of this.

These are all the terms that intervene in the determination of the one-loop
HTL-dressed transverse-gluon self-energy. When added to the one discussed
with more detail in the main text, we obtain the result displayed in 
(\ref {pi star_t--final}).

\begin{acknowledgments}
a.a acknowledges the kind hospitality of LPT-Orsay. He warmly thanks Asmaa
Abada, Francois Gelis, Olivier P\`{e}ne and Dominique Schiff for useful
discussions during his visit.
\end{acknowledgments}


\begin{thebibliography}{99}
\bibitem{aurenche-gelis-kobes-petitgirard1} Allowing $n$ to be only an
integer may be too restrictive because it is indicated in P. Aurenche, F.
Gelis, R. Kobes and E. Petitgirard, Z.~Phys.~\textbf{C}75 (1996) 315 that
intermediary scales between $gT$ and $T$ may also be relevant. Here we
assume $T$ to be the largest scale in the problem.
\end{thebibliography}
\end{document}